# Knowledge Gaps and Research Needs for Modeling $CO_2$ Mineralization in the Basalt-$CO_2$-Water System: A Review of Laboratory Experiments

Peng Lu[1,2,*], John Apps[3], Guanru Zhang[4], Alexander Gysi[6,5], Chen Zhu[*2]

[1]EXPEC Advanced Research Center, Saudi Aramco, Dhahran, 31311, Saudi Arabia.

[2]Department of Earth and Atmospheric Sciences, Indiana University, Bloomington, IN 47405, USA

[3]Earth and Environmental Sciences Area, Lawrence Berkeley National Laboratory, Berkeley, CA 94705, USA.

[4]College of Ecology and Environment, Chengdu University of Technology, Chengdu 610059, China

[5]New Mexico Bureau of Geology and Mineral Resources, New Mexico Institute of Mining and Technology, Socorro, NM 87801, USA

[6]Department of Earth and Environmental Science, New Mexico Institute of Mining and Technology, Socorro, NM 87801, USA

# ABSTRACT

Carbon capture and storage (CCS) in basalt is being actively investigated as a scalable climate change mitigation option. Accurate geochemical modeling prediction of the extent and rate of $CO_2$ mineralization is a critical component in assessing the local and global feasibility and efficacy of this strategy. In this study, we review basalt-$CO_2$-water interaction experimental studies conducted during the last two decades to determine whether they provide useable information for geochemical modeling. Most of the cited experiments generate data on the temporal evolution of water composition, and a few provide identification of secondary precipitates and their compositions, offering empirical and semi-quantitative information about the reactivity of basalts and the likelihood of secondary carbonate mineralization at various temperatures, pHs, and $pCO_2$ conditions. However, most experiments provide insufficient information on the properties and quantity of secondary minerals formed, prohibiting accurate mass balance calculations and hence more quantitative geochemical modeling studies. Primary Ca, Mg, and Fe-bearing minerals in basalt control the availability of major ions released into aqueous solution for carbonate precipitation, and many secondary minerals, i.e., smectites, Ca-Mg-Fe carbonates, and zeolites, provide sinks for the same major ions, some of which are difficult to quantify experimentally. Thus, we have a "multi-source and multi-sink" inverse mass balance problem with insufficient constraints on the bulk system in which the temporal evolution of major ions does not provide sufficient information on which mineral(s) dissolve or the sequence of dissolution and precipitation reactions.

Going forward, this review highlights the need for providing better-constrained experimental data to calibrate basalt-$CO_2$-water geochemical models. Hence, we propose that future experimental work should focus on trace elements and multiple isotopic tracers as well as emphasize the need for better characterization of the solid reaction products with modern analytical instruments.

## 1. INTRODUCTION

Carbon dioxide mineralization (i.e., turning $CO_2$ into carbonate minerals) can potentially be used to capture and store gigatons of $CO_2$ per year (National Academies of Sciences Engineering and Medicine, 2019; IPCC, 2021; US Department of State, 2021). Among the most effective are those processes involving mafic and ultramafic rocks, as their capacity to mineralize $CO_2$ is greater than those of other rock types such as sandstones (Schaef et al., 2010;). Natural rock weathering of these rocks may be enhanced by milling and spreading the comminuted product on croplands and forests, marine beaches and in oceans (National Academies of Sciences Engineering Medicine, 2019; 2021); or by injecting $CO_2$ into aquifers in the most reactive mafic rocks, i.e., extrusive basalts, both on land and offshore (McGrail et al., 2006; Snæbjörnsdóttir et al., 2020 ) (**Figure 1**). However, as the US long-term strategy report (US Department of State, 2021) stated, techniques of "enhanced mineralization, are still in nascent stages of research and development, so the potential magnitude of reductions and the time frames over which these technologies might deliver reductions is unknown". This is particularly pertinent to knowledge gaps in basic geochemical kinetics, and underscores the urgency of resolving these stubborn scientific challenges to meet society's current need for climate mitigation.

The knowledge gaps in geochemical kinetics hinder the successful global implementation of these actionable and promising climate mitigation strategies. For example, the safety of $CO_2$ storage in basalt aquifers—unlike $CO_2$ storage in sedimentary aquifers that rely on the integrity of the cap rocks—counts on the rapid fixation of secondary carbonate minerals, basalt formations are not always bounded by caprocks. Furthermore, the potential of removing gigatons of $CO_2$ from the atmosphere by spreading comminuted basalt on croplands and forests can become a reality only if the geochemical reaction kinetics is favorable and continues to be favorable across time frames of decades to a millennium. How many megatons of soil-spread comminuted basalt are required to capture so many gigatons of $CO_2$ in the first year, the 10th year, and the 100th year? The basic science information that we discuss in this communication will feed into techno-economic analyses which will then assess whether global implantation of a climate mitigation strategies is economically feasible. In other words, the quantitative answers on a climate change mitigation over a human time scale (years, decades) require a higher level of knowledge of geochemical kinetics than is currently practiced.

Basalts are rich in divalent cations and typically contain about 7-10 wt.% CaO, 5-6 wt.% MgO, and 7-13 wt.% $FeO_{tot}$ (Philpotts & Ague, 2022). Furthermore, basaltic glass and minerals (i.e., olivine, pyroxene, and plagioclase) are quite reactive, potentially expediting mineralization of the injected $CO_2$ as stable carbonate minerals (McGrail et al., 2017; Gunnarsson et al., 2018; Pogge von Strandmann et al., 2019; Snæbjörnsdóttir et al., 2020). The objective of this paper is to delineate the critical knowledge gained through a review of the literature on laboratory experiments of basalt-$CO_2$-water interaction and identify the knowledge gaps and prospective research needs to advance our knowledge for the technology development of mineral carbonation in subsurface basaltic formations. Satisfying this objective could have broader implications relating to both enhanced atmospheric $CO_2$ and natural $CO_2$ capture from other sources by basalts, as discussed in the **DISCUSSION** section.

Over the last half century, numerous basalt-water interaction experiments have been conducted because of their potential significance controlling on the global cycling of elements (Kump et al., 2000; Holland, 2005; Gislason et al., 2006; Jones et al., 2012), the long-term chemistry of the ocean and oceanic crusts (Hart, 1973; Wolery and Sleep, 1976; Mottl and Wheat, 1994), and impacts on the global carbon cycle (Berner et al., 1983; Coogan and Gillis, 2013). Since the early 2000s, many studies on basalt-$CO_2$-water experiments have been carried out in response to interest in climate change mitigation. Indeed, we identified 49 such studies in the literature (Brady and Gíslason, 1997; McGrail et al., 2006; Matter et al., 2007; McGrail et al., 2009; Prasad et al., 2009; Schaef and McGrail, 2009; Schaef et al., 2010; Gudbrandsson et al. 2011; Schaef et al., 2011; Stockmann et al., 2011; Wolff-Boenisch et al., 2011; Gysi and Stefánsson, 2012a, b, c; Jones et al., 2012; Rosenbauer et al., 2012; Adam et al., 2013; Rani et al., 2013; Schaef et al., 2013; Shibuya et al., 2013; Stockmann et al., 2013; Galeczka et al., 2014; Schaef et al., 2014; Sissmann et al., 2014; Peuble et al., 2015; Guha Roy et al., 2016; Shrivastava et al., 2016; Adeoye et al., 2017; Hellevang et al., 2017; Kanakiya et al., 2017; Kumar et al., 2017; Luhmann et al., 2017a, b; Meysman and Montserrat, 2017; Xiong et al., 2017a, b; Marieni et al., 2018; Menefee et al., 2018; Rigopoulos et al., 2018; Wolff-Boenisch and Galeczka, 2018; Xiong et al., 2018; Clark et al., 2019; Kelland et al., 2020; Marieni et al., 2021; Menefee and Ellis, 2021; Phukan et al., 2021a,b; Voigt et al., 2021; Delerce et al., 2023). In this paper, we review these experiments and evaluate what is critical to the assessment of the feasibility of basalt reservoirs as a source of Ca, Mg, and Fe, and as a sink for carbon, which together are essential to constrain geochemical models and support the design and operation of field-scale $CO_2$ injection.

# 2. SUMMARY OF THE FINDINGS FROM THE EXPERIMENTS IN THE LITERATURE

The 49 experimental studies that we have reviewed here cover a wide range of experimental conditions: temperature (room temperature to 350 °C), pH (acidic to basic), $pCO_2$ (0.3-310 bars; some are at supercritical conditions), starting basalt compositions (basaltic glass and crystalline basalt), duration (up to 1,387 days), and chemical composition and ionic strength of the experimental solutions (seawater, brine, or water). These conditions are described in **Appendix A**. Basalt is a rock with complex mineralogical compositions; at least five silicate mineral groups (olivine, pyroxene, amphibole, plagioclase, glass, and alkali-feldspar) are involved. Each mineral group has numerous publications on their dissolution experiments. Therefore, experiments on a single mineral, e.g., forsterite, are not the focus of this review.

## 2.1. Types of Experiments

Four types of experiments have been used to investigate basalt-$CO_2$-water reactions, including batch-type, mixed flow (or flow-through) reactors, core flooding, and column leaching experiments (**Appendix A**).

### 2.1.1 Batch-type experiments

Batch-type experiments usually contains a gas (e.g., $CO_2$) input, reactor vessel, and sampling system. The reactor vessel hosts the rock powders and fluids and the mixture is typically stirred continuously. The experimental duration is relatively long (months to years). Batch-type experiments are used to monitor reaction progress in closed systems. The advantage of this method is the ability to conduct experiments at variable fluid/rock ratios, approach to equilibrium, and *in situ* sampling of the reacted solutions, as well as analysis of the solid reaction products. The measured parameters are typically pH and major ions (e.g., major cations such as K, Na, Mg, Ca, Fe, Si, and Al, and anions such as $HCO_3^-/CO_3^{2-}$, $Cl^-$, $SO_4^{2-}$, etc.) as a function of time. *In situ* pH for experiments at elevated temperature and pressure is calculated from geochemical speciation and measured solution chemistry, and/or including pH values measured at room temperature. Measurements of minor elements (e.g., Al) may be less accurate due to (1) very low concentration and insufficient sensitivity of the analytical method; (2) the filtration process fails to remove small particles (Barnes, 1975). The concepts of alkalinity and dissolved inorganic carbon (DIC) have been misapplied interchangeably in many studies. Carbonate alkalinity is defined as Alkalinity = $mHCO_3^- + 2mCO_3^{2-} + mOH^- - mH^+$ (Stumm and Morgan, 1995; Zhu and Anderson, 2003), while DIC = $mHCO_3^- + mCO_3^{2-} + mCO_2(aq)$.

Secondary minerals have been identified using X-ray diffraction (XRD), electron microprobe analysis (EMPA), scanning electron microscope (SEM), energy dispersive X-ray spectroscopy (EDS), Fourier-transform infrared spectroscopy (FTIR), and transmission electron microscopy (TEM) (**Appendix A**).

Data and knowledge gained through these experiments can be used to improve reaction path models, even though the primary purpose may not necessarily have been to provide a basis for modeling: (1) type and compositions of secondary minerals formed and how they change upon reaction progress (e.g., Gysi and Stefansson, 2012b; c); (2) comparison of water chemistry evolution between the modeling results and experimental data (currently good for $CO_2$, Ca, Mg and relatively poor for Fe, Si, Al) (e.g., Gysi and Stefansson, 2012a; c; Voigt et al., 2021). However, information from these experiments is still insufficient for more accurate models. For example, (1) multicomponent, multiphase systems have limits on how much they can feed our models to improve the accuracy of simulated mineral properties and their compositions. Using fluid chemistry data to infer the mineralogy changes is an undetermined inverse problem that an element concentration change may stem from the contributions of dissolution and precipitation of different minerals, which is here called "multi-sources, multi-sinks problem". For example, calcium can be sourced from multiple reactants (e.g., plagioclase, glass, and pyroxene), and sink into multiple secondary minerals (e.g., clay and carbonate minerals), with variable compositions as a function of reaction progress between basalt and $CO_2$-charged waters. (2) despite more or less controlled fluid/rock ratios and reactive surface area (RSA), challenges still remain as to how to model reaction kinetics that is strongly influenced by the RSA of primary and secondary minerals, solid solutions, rates, rate laws for dissolution, nucleation, and growth, as well as coupled dissolution and reprecipitation reactions.

Nevertheless, data from batch experiments are valuable as we can model the chemical evolution with our current knowledge of thermodynamics and kinetics without having to consider the complexity of transport processes. Future experimental studies on this topic could be devoted to (1) the use of isotope tracers and trace elements to increase mass balance constraints from fluid chemistry, (2) more targeted simple experiments on pure minerals to develop thermodynamic solid solution models, and (3) dedicated experiments for reaction kinetic (dissolution/precipitation rates, nucleation and growth) of relatively simple systems (one primary mineral with one or two secondary minerals).

*2.1.2 Mixed flow-through reactors*

Mixed single flow-through reactor type experiments usually consist of a high-pressure column filled with crushed solid material through which fluids are pumped via an inlet/outlet system. The system permits sampling in time and space and allows modified flow rates. Experiments are done with relatively low fluid-rock ratios compared to batch experiments. These experiments can be used to measure quasi steady-state major ion concentrations and outlet solution pH. These measurements are typically used to determine the dissolution rate constants and how specific parameters affect reaction rates (e.g., temperature, pressure, ionic strength, pH, catalyzing and inhibiting species, and $pCO_2$). This type of experiment has been conducted for natural basaltic glass (Oelkers and Gislason, 2001; Gislason and Oelkers, 2003; Delerce et al., 2023), crystalline basalts (Gudbrandsson et al., 2011), and basalt derived single minerals (e.g., Daval et al., 2010; Gudbrandsson et al, 2014; Xiong and Giammar, 2014; Montserrat et al., 2017; Mesfin et al., 2023).

Basaltic glass and rock dissolution rates (typically based on steady-state Si release rates) can be derived from these experiments (e.g., Oelkers and Gislaslon, 2001; Gislason and Oelkers, 2003; Schaef and McGrail, 2009; Stockmann et al., 2011; Wolff-Boenisch et al., 2011; Galeczka et a. 2014), and sometimes the aggregate activation energies of basalt dissolution reactions can be calculated based on Si or Ca release rates at different temperatures (Schaef and McGrail, 2009). However, as pointed out by Galeczka et al. (2014), even for a relatively simple system with only basaltic glass as the primary reactant, published modeling results differ substantially from observations. Both experimental setups and interpretation must be improved before the latter can be used confidently for designing and operating of field-scale projects. SEM and XRD analyses on the reaction products (usually at the end of the experiments) are also conducted. However, due to the difficulties of secondary mineral characterization, quantitative analysis of the reaction products and their evolution remains a challenge.

*2.1.3. Reactive core-flooding experiments*

In some studies, basalt core flooding experiments under simulated reservoir conditions are performed to monitor time- and space-dependent permeability, porosity, mineral surface changes and variations in aqueous fluid concentrations at the outlet (Peuble et al., 2015; Luhmann et al., 2017a; Luhmann et al., 2017b; Menefee et al., 2018; Menefee and Ellis, 2021). This type of experiment is similar to the mixed flow-through reactor experiment. However, instead of using the crushed materials, a core holder hosting half or whole core plugs (e.g., with 25.4 mm diameter) is used in an attempt to mimic an advection-dominated flow condition (*ibid.*). SEM and XRD analyses are commonly performed on the sectioned core samples after the experiments. Permeability can be calculated using Darcy's law and using variable flow rates. Porosity changes in the reacted basaltic rock cores are determined using small-angle and ultra-small-angle neutron scattering or X-ray computed tomography (xCT) scans pre- and post-experiments. Flooding experiments with core artificially fractured into two parts, misleadingly referred to as "half-core" experiments, have been conducted to study the carbonate mineral precipitation in the fractures (e.g., Menefee et al., 2018). However, it would be more useful to study the performance of naturally fractured rocks in "whole-core" experiments because the artificially generated fractures may not represent the heterogeneity and complexity of natural fractures.

Other issues arise when using the core-flooding experimental data for modeling. Thus, the effluent fluid chemistry represents an integrated respnse over the entire core sample, the mineralogical characterizations are conducted mainly locally either on a plane perpendicular to the main axis of the core (for sliced core) or at the end of the core. The variabilities of reactive surface area and kinetics (due to heterogeneity) for this type of experiment are higher than mixed flow reactor experiments using rock powders. If the flow rate of the experiments is too high, the system will be in a transport controlled regime that is only relevant to the vicinity of the borehole. Although the core flooding experiments are conducted under "*in situ*" conditions, they cannot fully represent the reservoir conditions due to the complexity of the subsurface environment. Upscaling core flooding experimental data for field-scale models is still challenging due to rock heterogeneities and the consequent infeasible scale of a representative elementary volume, proper characterization of the rock, the controlling regime (kinetic/transport/mixed), and boundary conditions.

## 2.2. Secondary Mineral Precipitation

When a solution is supersaturated with respect to a mineral, this mineral has the tendency to precipitate. The precipitated minerals are secondary, and differ from the starting, i.e., primary minerals in the system.

The saturation index (SI) is a measure of the saturation state of a mineral. SI>0, =0, or <0 denote the solution is supersaturated, at equilibrium, or undersaturated with respect to a mineral of interest, respectively. SI is defined as SI = log(IAP/$K_{sp}$), where IAP denotes ion activity product and $K_{sp}$ is the equilibrium solubility product. Considering calcite precipitation reaction $Ca^{2+} + CO_3^{2-} = CaCO_3$ (calcite), the IAP for this reaction is actual ($aCa^{2+}$) ($aCO_3^{2-}$), while $K_{sp}$ is equilibrium ($aCa^{2+}$) ($aCO_3^{2-}$). $a$ is the activity of the species, a measure of the "effective concentration" of a species in a mixture.

The wide range of discrete basaltic reactants and reported experimental conditions have produced a bewildering range of secondary mineral precipitates. (**Appendix B**). Calcite was the most common and abundant secondary carbonate mineral over a wide range of temperature, $pCO_2$, and pH conditions relevant to carbon capture and storage (CCS) in basalt (**Figure 2**).

11Saudi Aramco: Public

Aragonite was the second most abundant carbonate. At room temperature and atmospheric $p$CO$_2$, only calcite and aragonite formed, whereas other carbonate minerals formed only at elevated $p$CO$_2$ and temperature (**Figure 2**).

Of other secondary carbonate minerals, the pure endmembers of siderite, ankerite, dolomite, and magnesite rarely precipitated in the experiments; either solid solutions or their poorly crystalline precursors formed instead, which necessitated crystallographic and/or chemical analyses for their characterization. These minerals were identified as (Ca)-Fe-Mg carbonates (Gysi and Stefánsson, 2012a, b, c; Adam et al., 2013; Xiong et al., 2017b), Mg-bearing siderite (Adeoye et al., 2017), siderite (Rani et al., 2013; Shrivastava et al., 2016; Kumar et al., 2017; Luhmann et al., 2017a; Luhmann et al., 2017b; Xiong et al., 2017a; Xiong et al., 2017b; Clark et al., 2019; Voigt et al., 2021), ankerite (McGrail et al., 2006; Prasad et al., 2009; Schaef et al., 2013; Kanakiya et al., 2017; Kumar et al., 2017; Menefee et al., 2018; Voigt et al., 2021), magnesite (Rani et al., 2013; Schaef et al., 2013; Sissmann et al., 2014; Wolff-Boenisch and Galeczka, 2018; Menefee and Ellis, 2021; Voigt et al., 2021), and Fe-magnesite (Rosenbauer et al., 2012). It should be noted, however, that there is presently no consensus regarding the defined stoichiometry of ankerite. In some papers, these secondary carbonates are poorly characterized, due in part to technical challenges.

The dolomite-ankerite solid-solution series (CaMg(CO$_3$)$_2$-CaFe(CO$_3$)$_2$) are commonly found in clastic reservoirs in a mesogenetic diagenetic environment and as the weathering product of mafic and ultramafic rocks. Fe-dolomite, siderite, ankerite, and magnesite as a group, were abundant secondary carbonates, primarily forming at elevated temperature, slightly elevated $p$CO$_2$, and mildly acidic pH. Other carbonates, kutnohorite (Ca(Mn, Mg, Fe$^{2+}$)(CO$_3$)$_2$), rhodochrosite ($MnCO_3$), end-member dolomite, and huntite ($CaMg_3(CO_3)_4$) were rarely observed, as they formed primarily at temperatures ≥90 °C and at elevated $pCO_2$ (**Figure 2**).

Smectite is the most abundant non-carbonate mineral precipitated in the $CO_2$-basalt-water interaction experiments (**Figure 3**). Gysi and Stefánsson (2012b, b, c) found mixed Ca-Mg-Fe smectites in their 40 °C and 150 °C experiments with elevated $pCO_2$ (11-19 bars) using SEM with EDS analysis and EMPA. Nontronite was identified at 80-150 °C with 0.01 M $Na_2CO_3$ solution and lower $pCO_2$ (0.5-0.76 bars) (Hellevang et al., 2017). Saponite was detected using XRD, SEM-EDS, and Raman spectroscopy by Kumar et al. (2017) and Voigt et al. (2021) in their 100 °C and 5-10 bar $pCO_2$ and 130 °C and 2.75 bar $pCO_2$ experiments. Either smectite or clay was ambiguously mentioned, but not verified in some studies (Rani et al., 2013; Schaef et al., 2013; Shibuya et al., 2013; Sissmann et al., 2014; Kanakiya et al., 2017; Luhmann et al., 2017a; Luhmann et al., 2017b; Menefee et al., 2018; Wolff-Boenisch and Galeczka, 2018). In nature, smectite is one of the most abundant secondary clay minerals in basaltic rocks at temperatures <100 °C (Schiffman and Fridleifsson, 1991).

Amorphous silica ($SiO_{2(a)}$), zeolites, chlorite, Fe-bearing oxides, and sulfides (pyrite, marcasite) were also abundant as secondary minerals in experiments (**Figure 3**). $SiO_{2(a)}$ was found as a product in numerous experiments between 75-250 °C, 11.2 – 310 bars $pCO_2$, and under acidic to neutral pH (Prasad et al., 2009; Schaef et al., 2011; Gysi and Stefánsson, 2012a, c; Rosenbauer et al., 2012; Peuble et al., 2015; Adeoye et al., 2017; Xiong et al., 2017b; Menefee et al., 2018; Xiong et al., 2018). Due to its amorphous nature and indefinite shape (e.g., milky coating; Menefee et al., 2018), its occurrence could have been underestimated or ignored in many studies.

Zeolites precipitated at elevated temperatures (70-150 °C) and near-neutral to basic pH in the experiments (**Table 1**; **Figure 3**). Several different species of zeolite were recognized, for example, chabazite (Kumar et al., 2017; Wolff-Boenisch and Galeczka, 2018), analcime (Shibuya et al., 2013), mesolite and scolecite (Stockmann et al., 2011), and stilbite (Phukan et al. 2021a). Zeolite group minerals are mentioned based on morphology in some studies, although their composition could not be verified (Gysi and Stefánsson, 2012a, c; Kanakiya et al., 2017; Voigt et al., 2021).

Chlorite also precipitated at elevated temperatures (90-250 °C) and near-neutral pH in some experiments (**Figure 3**). However, usually, only the generic name was mentioned without specifying its chemical composition.

Pyrite/marcasite formed in the experiments under the reducing environments (when $O_2(g)$ was removed in both the solution and the gas) and where 1-1.5% $H_2S(g)$ was often found in the gas cap (Schaef et al., 2009; Schaef et al., 2010; Gysi and Stefánsson, 2012c; Schaef et al., 2013; Luhmann et al., 2017a; Luhmann et al., 2017b; Marieni et al., 2018). They required elevated temperature (60-250 °C) and acidic to neutral pH (**Figure 3**).

Iron oxides formed in the experiments when $O_2(g)$ was not purged from either the solution or the gas (Menefee et al., 2018; Rigopoulos et al., 2018; Wolff-Boenisch and Galeczka, 2018). The Schikorr reaction (Schikorr, 1929) involving hydrolysis by ferrous ion with the formation of magnetite cannot be ruled out but would require the detection of secondary hydrogen. Because Fe oxides commonly form as an amorphous coating, their occurrence could have been underestimated or neglected in many studies. Magnetite, hematite, anhydrite, and albite are rare secondary minerals, usually forming at relatively high temperatures (≥90 °C; **Figure 3**).

At elevated temperatures, secondary smectite and chlorite competed with carbonate minerals for $Fe^{2+}$ and $Mg^{2+}$ (Gysi and Stefánsson, 2012a). At 150 °C and 250 °C, most Mg and Fe were incorporated into smectites and/or chlorite and only ~20% of secondary minerals were carbonates (mainly calcite).

## 2.3. Influence of geochemical variables

### 2.3.1. pH

In the basalt-$CO_2$-water systems, pH may vary widely from acidic to basic for different scenarios. The typical pH range of injected solutions containing dissolved $CO_2$ (e.g., CarbFix) or soil amendments is near-neutral to basic (Snæbjörnsdóttir et al., 2017; Kelland et al., 2020), but in a basaltic aquifer subject to the injection of supercritical $CO_2$, the pH may range from acidic to basic (Schaef and McGrail., 2009).

The pH depends strongly on the initial $CO_2$ concentration in solution and fluid/rock ratios (Gysi and Stefánsson, 2012b; Gysi, 2017). Experiments conducted over a broad range of initial $CO_2$ (<50 to ~300 mmol/kg) and fluid/rock ratios (100 to 550 g basalt/L solution) indicated variations from a $CO_2$-water-buffered system (depending on temperature) with pH values of ~3.5 to 4.5, to basalt and/or secondary-mineral buffered solutions with pH values ranging between ~5 to >8 (Gysi and Stefánsson, 2012b). The pH of pure water generally increased from ~ 7 to 9-9.5 after reaction with basalt, i.e., within the range of naturally equilibrated groundwaters with basalt, whereas initially acidic pH in basalt-$CO_2$-water interaction systems was buffered to a near-neutral region (Galeczka et al., 2014). This usually happened when the $CO_2$ supply was limited and/or the fluid-rock ratio was low (i.e., the pH is said to be rock buffered) (Gysi and Stefánsson, 2011; Gysi, 2017). In cases where $pCO_2$ was fixed (i.e., $CO_2$ supply was sufficiently high), the solution pH remained as low as ~3 (e.g., Adeoye et al., 2017).

The alkalinity in the starting solution may act as a pH buffer. Alkalinity is the "Acid Neutralizing Capacity" of a solution. Carbonate alkalinity can be defined as Alkalinity = $mHCO_3^- + 2mCO_3^{2-} + mOH^- - mH^+$ (Zhu and Anderson, 2003). Some experiments used DI water that has no initial alkalinity (e.g., Brady and Gislason, 1997; Clark et al., 2019), whereas others either added electrolyte background (e.g., $NaHCO_3$; Adeoye et al., 2017) or replicated the *in situ* groundwater/formation water that include some alkalinity (e.g., Gysi and Stefánsson, 2012 a,b,c).

pH is related to the results of basalt-$CO_2$-water interactions in three ways: (1) The dissolution rate of basaltic glass and Al-bearing minerals shows a strong pH dependence (**Appendix C**), with higher rates at low pH; (2) Secondary mineral types differ with pH. At ~4.5 to 6.5, the secondary minerals formed are Ca-Mg-Fe carbonates, smectites, and Al-Si minerals (i.e., allophane/kaolinite and amorphous Si) whereas at pH >6.5 the secondary minerals formed consist of zeolites and Mg-rich clays, thereby limiting the mobility of Ca and Mg while competing with carbonate mineralization (Gysi and Stefánsson, 2012b); and (3) The Fe oxidation rate is pH-dependent, especially at pH 3-7 (Morgan and Lahav, 2007).

### 2.3.2. $CO_2$ partial pressure

Carbon dioxide partial pressure ($pCO_2$) decreased with time in almost all batch-type $CO_2$-basalt-water experiments, whereas in some mixed flow or core flooding experiments, it could be fixed at a certain value (e.g., Luhmann et al., 2017a, b). This is because, in batch experiments, the $CO_2$ supply was usually limited to that incorporated at the beginning, whereas in mixed flow experiments, the fresh fluid with dissolved $CO_2$ could continuously replenish the reactor. $pCO_2$ had several effects on the kinetics of basalt-$CO_2$-water reactions: (1) higher $pCO_2$ lowered the pH while increasing primary mineral dissolution rates; (2) $HCO_3^-$ increased the dissolution rates of primary minerals (Daval et al., 2013); (3) higher $pCO_2$ increased the $HCO_3^-$ activity in the solution and thus could increase the rates and the amounts of carbonate mineral precipitation. Wolff-Boenisch et al. (2011) found that the rates of all reactions increased by 0.3–0.5 log units in the presence of a $pCO_2$ of 4 bars compared to those at atmospheric $CO_2$ partial pressure.

Whether or not certain carbonates precipitate is a complex function of many varibles, the precipitation of some carbonate minerals seems to require high $pCO_2$ levels. High $pCO_2$ is not a limitation for the precipitation of calcite and aragonite; their precipitation can occur under atmospheric $pCO_2$ (**Figure 2**). However, other carbonate minerals appear to require elevated $pCO_2$ to promote their precipitation. Siderite precipitation occurred at a slightly elevated $pCO_2$ (0.6 bar and above) (Clark et al., 2019). At least 2.75 bars of $pCO_2$ are required for the formation of ankerite, 5 bars for magnesite, dolomite, and huntite, and 100 bars for some rare carbonate minerals (e.g., kutnohorite, rhodochrosite) (**Figure 2**). The observation that different $pCO_2$ levels induce precipitation of different carbonate minerals should be tested and verified through geochemical modeling.

### 2.3.3. Supercritical $CO_2$ vs. $CO_2$-charged water

A supercritical $CO_2$ fluid phase was injected in basalt at the Wallula CCS site in contrast to the CarbFix CCS site where $CO_2$-charged water was injected. Schaef et al. (2011) conducted experiments on basalt focusing on the different reactivity between supercritical $CO_2$ and $CO_2$-charged water. In the high-pressure vessels they used, the water remained in the lower part and the headspace was charged with supercritical CO$_2$. The basalt-supercritical-CO$_2$ reaction was investigated by holding basalt chips above the water level (only H$_2$O vapor was present in the headspace). They found that the morphology and amount of secondary minerals formed under the two conditions were different. Discrete nodule-shaped carbonate minerals formed in CO$_2$-saturated water, whereas a surface coating (15-25 μm thick) of tiny clusters of aragonite needles formed in the humid supercritical CO$_2$ phase (Schaef et al., 2011). The carbonation rate in CO$_2$-charged water was also found to be higher than in the wet supercritical CO$_2$ phase (Schaef et al., 2011).

*2.3.4. Temperature*

Temperature has a significant role in the basalt-CO$_2$-water reactions. Only calcite and aragonite were observed in low $T$ experiments (<40 °C); other carbonate minerals were observed at elevated temperatures (**Figure 2**). Both thermodynamic and kinetic controls may contribute to carbonate stability. Most carbonate minerals (e.g., calcite, aragonite, magnesite, siderite, and dolomite) display a retrograde solubility (Lu et al., 2022a), i.e., carbonates are less soluble at elevated temperatures favoring their precipitation. Furthermore, the precipitation of Fe- and Mg-bearing carbonate minerals is kinetically inhibited at low temperatures (Golubev et al., 2009; Saldi et al., 2012). Fe-bearing carbonate precipitation was observed at $T \geq 40$ °C for Fe-dolomite, >50 °C for siderite, and >75 °C for ankerite (**Figure 3**). Magnesite and dolomite precipitation requires even higher temperatures (90 and 100 °C). Rarely observed carbonate minerals such as kutnohorite, rhodochrosite, and huntite, also require high temperatures (90 - 100 °C) for their formation.

The optimal temperature for mineral carbonation is below 100 °C because at $T > 100$ °C the tendency for Mg and Fe to be incorporated into clays is greatly enhanced, limiting the availability of the metal ions for the carbonate mineralization (Gysi and Stefánsson, 2012c). Gysi and Stefánsson (2012c) showed in batch-type experiments that at $T \geq 150$ °C almost all uptake of Mg and Fe was sequestrated into clays (smectite and/or chlorite), with the formation ~20% of secondary carbonates, primarily calcite. The formation of clay minerals (montmorillonite and nontronite) and zeolite (stilbite) that may remove some divalent cations can even happen at lower temperature (e.g., 60 °C; Phukan et al. 2021a). At temperatures > 90 °C, anhydrite may precipitate from seawater due to the presence of sulfate in these experiments (**Table 1**). Anhydrite precipitation removes part of $Ca^{2+}$ from the solution and decreases the carbonation potential of the system.

### *2.3.5. Redox potential*

Fe and Mn released during basalt dissolution will participate in the formation of various secondary minerals, depending on the prevailing oxidation state (**Figure 4**). Fe (II) usually dominates in basalts with only a small fraction of Fe(III) being present. Water-saturated basalts in the subsurface (isolated from the atmosphere) are redox buffered by the formation of secondary magnetite through the Schikorr reaction. The overall reaction is written as

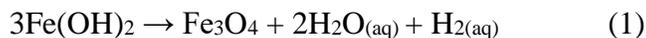
$$3Fe(OH)_2 \rightarrow Fe_3O_4 + 2H_2O_{(aq)} + H_{2(aq)} \qquad (1)$$

The generated hydrogen will be converted to methane by hydrogenotrophic methanogens at a temperature lower than 50 °C (Huser et al., 1982).

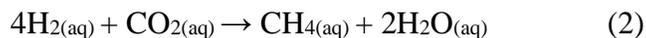
$$4H_{2(aq)} + CO_{2(aq)} \rightarrow CH_{4(aq)} + 2H_2O_{(aq)} \qquad (2)$$

Ingress of seawater containing sulfate leads to sulfate reduction mediated by bacteria with the production of $H_2S$ and iron sulfides. If fresh or seawater is used to dissolve $CO_2$ before injection, then dissolved oxygen at circa 10 ppm will be co-injected, and the extent of oxidation will be determined by how many cycles of injection take place.

Fe(III) is important in the formation of Fe-rich smectite (e.g., nontronite). In a reducing environment, Mn and Fe can participate in the formation of carbonate minerals (siderite, ankerite, and rhodochrosite), whereas in an oxidizing environment Mn and Fe oxides will form (**Figure 4**). At higher pH and elevated temperatures, the Fe(II) oxidation reaction rate is significantly accelerated. At pH >6.1, the mobility of Fe decreases upon oxidation of Fe(II) to Fe(III) and subsequent formation of ferrihydrite (Gysi and Stefánsson, 2012b). Whether $Fe^{2+}$ in basalt can be oxidized quickly is important because (1) it affects the carbonation potential of basalt (competition of Fe(II)-bearing carbonate minerals with Fe(III)-bearing oxyhydroxides and clay minerals), and (2) precipitation of Fe oxides can lead to co-precipitation of toxic metals (e.g., Martin et al., 2005; Lu et al., 2011). In some systems, the presence of anhydrite would sustain an elevated redox potential, inhibiting the formation of secondary Fe(II) carbonates until Fe(II) reduction of $SO_4^{2-}$ to $HS^-$ is complete. The redox effects in the basalt-$CO_2$-water interaction experiments are relatively understudied; greater effort should be dedicated to this topic given the importance of Fe for the precipitation of Fe-Mg carbonates at temperatures where Mg carbonates such as magnesite would be inhibited.

## 2.4. Influence of Geomechanical Properties

### 2.4.1. Reactions in fractures

Highly fractured and brecciated basalt is important in promoting basalt-$CO_2$-water interactions, as it usually has high porosity and permeability (McGrail et al., 2006). Carbonation reactions in basalt reservoirs may occur in fractures because the solution pH is relatively high near the dead-end fractures due to its isolation from the $CO_2$-acidified bulk solution, thereby facilitating potential carbonate mineral precipitation (Adeoye et al., 2017; Xiong et al., 2017a). Experimental studies focus on the interplay of primary mineral dissolution and secondary clay and carbonate mineral precipitation locations and volumes and flow and transport in fractures and pore networks (Adeoye et al., 2017; Xiong et al., 2017a; Menefee et al., 2018; Peuble et al., 2015; Phukan et al., 2021a).

Most experiments utilized artificial fractures in one-inch (2.54 cm) basalt cores to investigate the reactions in fractures. The cores were split into two half-cylinder sections and channels were etched on the surface of the half-core. Adeoye et al. (2017) artificially generated a 1 mm-wide × 95-105 μm depth meandering channel in core samples to mimic the fractures and focused on the role of fluid transport regimes on the dissolution and precipitation in fractures. Enhanced dissolution along fractures at large grains of reactive minerals (e.g., pyroxene and olivine) was found in the flow-through experiment, but no secondary mineral precipitation was detected up to 140 h at 45 and 100 °C. However, in a batch experiment conducted by the same authors, 50 μm size Mg-bearing siderite crystals formed along the entire fracture and clusters along the fracture of a serpentinized basalt reactant. Menefee et al. (2018) found that carbonate precipitation increased with diffusion distance into dead-end fractures, while noncarbonated secondary alteration products (e.g., clays, iron oxides) formed in more acidic advection-controlled flow paths. Carbonates likely first fill the asperities of the dead-end fractures, which could potentially allow for carbonation reactions to proceed to near completion without self-sealing in zones of restricted flow where the majority of $CO_2$ mineralization is expected to occur (Menefee et al., 2018). The implication is that carbonate precipitation is more likely to occur in the dead-end fractures, away from the wellbores, and less likely to occur near the wellbores or along interconnected fracture pathways due to the strong advective flow of $CO_2$-acidified solutions (Adeoye et al., 2017).

Peuble et al. (2015) proposed that the formation of low-density carbonate minerals at the expense of olivine may increase the solid volume, with corresponding reductions in porosity and permeability. Menefee et al. (2018) estimated that total fracture volume was reduced by 48% in an experiment at 150 °C, compared to 35% in a 100 °C experiment at 100 bars $pCO_2$ (1 ml/h flow rate, in 0.64 M $NaHCO_3$ and with a duration up to 12 days). Xiong et al. (2017a) observed that precipitation filled 15% of the exposed fracture volume at 150 °C compared to 5.4% at 100 °C in experiments where flood basalts were reacted with $CO_2$-acidified DI water at 100 bars $pCO_2$ after 28 days. Both carbonate and clay precipitation were enhanced at 150 °C in comparison to experiments conducted at 100 °C based on X-ray µCT segmentation data by Xiong et al. (2017b), in which estimated precipitates filled 5.4% and 15% (by volume) of the flood basalt fracture after 40 weeks of reaction at 100 °C and 150 °C, respectively.

Phukan et al. (2021b) conducted a diffusion-controlled experiment on a fractured basalt half-core plug (6.0 cm x 2.54 cm) at 60 °C and 80 bars to investigate the early stage (about 12 weeks) of geochemical reactions in fractures and the adjacent pore network. Four grooves with the size of 4 x 0.2 x 0.1-0.2 cm were milled into the surface to mimic the fracture system. They observed a 7.7% reduction in initial pore volume. Although there was an overall self-sealing of fractures and adjacent pore network, precipitation primarily occurred in the pores (15% decrease in the number of pores) and to a lesser extent in the throats (4%). In addition, a decline in the number of isolated pores was also noticed due to the primary mineral dissolution.

*2.4.2. Porosity and Permeability Changes in Core Samples*

Basalt-$CO_2$-water interactions induce primary mineral dissolution and secondary mineral precipitation. The former generates secondary porosity and increases permeability while the latter causes porosity and permeability reductions. Whether the interactions result in an overall porosity/permeability increase or decrease at the reservoir condition is important to the $CO_2$ injection operations.

Luhmann et al. (2017a,b) conducted whole-core flow-through experiments to assess the changes in porosity and permeability. Permeability was calculated using Darcy's law and flow rate changes and porosity changes were determined using small-angle and ultra-small-angle neutron scattering. They found that the permeability changes of core plugs were related to the flow rate of $CO_2$-rich brine: permeability decreasing slightly during the lower flow rate experiments (from ~1.5 x $10^{-16}$ to ~1 x $10^{-16}$ $m^2$ or from ~0.15 to ~0.1 md in 20-33 d) and increasing during the higher flow rate experiments (from 4.4 x $10^{-16}$ to 9.8 x $10^{-15}$ $m^2$ or from ~0.44 to ~9.8 md in 0.5 d). For porosity, they found that secondary minerals decrease the volume of large pores, but overall, the net porosity increases by 0.25-0.7%.

Kanakiya et al. (2017) employed a helium pycnometer, automated nitrogen gas porosimeter-permeameter, and nuclear magnetic resonance analysis to measure the changes in porosity, permeability, and pore geometry (size), respectively, for their batch experiments. Consistent with Luhmann et al. (2017a; b), they found that rock dissolution was more significant than secondary mineral precipitation after reaction in $CO_2$-charged water (about 0.45 mol/L $CO_2$), which resulted in an increase in porosity and a decrease in the rigidity of all samples. The basalt sample with the highest initial porosity and volcanic glass volume had the greatest potential for secondary mineral precipitation. At the same time, this sample exhibited the greatest increase in porosity (15.3%) and permeability (1.34 x $10^{-14}$ $m^2$ or 13.4 md), and a decrease in rock rigidity (Young's modulus -28.3%) post-reaction.

## 2.5. Insights from Geochemical Modeling

Some investigators of the basalt-$CO_2$-water experimental studies conducted geochemical modeling to assess the controlling reactions from solution chemistry and/or via comparison with the characterized precipitated solids. Different types of models have been used. Speciation-solubility modeling calculates the distribution of aqueous species and the mineral saturation indices based on Ion Association (IA) Theory according to mass balance and mass action equations (Zhu and Anderson, 2003). Reaction path models calculate a sequence of equilibrium states involving incremental or step-wise mass transfer between the phases within a system, or incremental addition or subtraction of a reactant from the system (Helgeson, 1968; Helgeson et al. 1969). Reactive transport model (RTM) couples mineral dissolution and precipitation with water flow and calculates the impact of mineral reactions on porosity and permeability evolution (Lu et al., 2023).

In batch-type experiments, speciation-solubility modeling was used to calculate *in situ* pH, aqueous speciation, and saturation indices (SI) of minerals. Some have used reaction path modeling to track the reaction progress and quantify the carbonation potential (Schaef et al., 2010; Gysi and Stefánsson, 2012c; Rosenbauer et al., 2012; Schaef et al., 2013; Schaef et al., 2014; Peuble et al., 2015; Kumar et al., 2017; Luhmann et al., 2017b; Xiong et al., 2017a; Voigt et al., 2021). However, in most studies, reaction path modeling was conducted for predictive purposes, or for comparing data trends, but with no attempt to match modeling results with experimental data. In only a few studies (e.g., Gysi and Stefansson, 2012a; Galeczka et al., 2014;

Voigt et al., 2021), attempts were made to calibrate the modeling results with experimental data, and approximate matches of the solution chemistry were achieved, whereas the secondary mineral sequences and their compositions could be reproduced fairly well via reaction path modeling (Gysi, 2017).

In core flooding column experiments, reactive transport modeling (RTM) has also been used, generally building on the reaction path models mentioned above, and enabling the interpretation of core flooding and column experiments to quantify the temporal and spatial evolution of mineralogy and porosity and to assess the carbonation potential along a 1-D flow path (Menefee et al., 2018; Clark et al., 2019; Kelland et al., 2020; Menefee and Ellis, 2021). However, conducting high-quality RTM is even more challenging than reaction path modeling, because it requires the consideration of the effects of multiphase, multi-component flow, and transport on reactions and mass transfer.

The uncertainties and limitations of the reaction path kinetic models have been discussed and summarized in many studies (e.g., Gysi and Stefansson, 2011; 2012a; 2012b; Galeczka et al., 2014; Gysi, 2017; Hellevang et al., 2017). These include: (1) Reaction kinetics such as dissolution stoichiometry, RSA and kinetics of the basaltic glass dissolution, and secondary mineral precipitation; (2) Redox chemistry, particularly Fe chemistry (Redox equilibrium may not be achieved among redox couples and redox state may be changed during the basalt-$CO_2$-water interactions, especially at higher pH and elevated temperatures, i.e., see Stumm and Morgan, 1996); (3) Thermodynamic properties and approach to equilibrium with certain complex minerals that are difficult to model (e.g., smectite conversion to chlorite, which is likely to be kinetically inhibited; the presence of metastable mineral phases, etc.), especially complex clay solid solutions and zeolites; (4) Calculations of *in situ* pH in experiments dependent on the uncertainties of DIC.

# 3. KNOWLEDGE GAPS AND RESEARCH NEEDS

The basalt-$CO_2$-water system is one of the most studied experimentally. Here, we summarize the current knowledge that can be extracted from previous experiments and identify research needs (**Table 3**) to facilitate future experiments to gain additional information and make further progress on this topic.

## 3.1. Secondary mineral characterization

Several previous experiments produced information on what secondary phases did or did not precipitate at a given $T$, pH, and $pCO_2$. However, most experiments did not have sufficient information regarding the quantity and paragenesis of secondary minerals. Characterization of secondary phases was challenging because the quantity of the precipitates was small and often below the detection limits of many analytical methods. Solid characterization was also expensive and time-consuming. In most studies reviewed in this work, a comprehensive characterization of the product minerals was not performed, perhaps owing to these limitations, or those imposed either by insufficient analytical capabilities or the lack of funding.

In most experiments, SEM and EDS analyses were conducted on the end products; in some studies, XRD analyses were performed on the reacted primary minerals and precipitated secondary minerals. However, in most of the low-temperature (≤40 °C) experiments, the secondary precipitates were not produced in sufficient volumes to be identifiable using XRD, whereas high-temperature (>75-150 °C) experiments allowed the study of faster basalt reaction progress and larger volumes of formed secondary minerals (e.g. Gysi and Stefánsson, 2012c). FTIR was used, e.g., Prasad et al. (2009), Schaef et al. (2014), Phukan et al. (2021a), which is a technique particularly useful for early, poorly crystalline (and hydrous) phases.

The analysis of reaction products in low-temperature experiments is challenging, even for SEM-EDS or FTIR analysis, because the secondary minerals formed are commonly small in size (i.e., <1-5 µm; Gysi and Stefánsson, 2012b). In scarcely any studies were the compositions of the small mineral intergrowths measured using EMPA and WDS-element mapping, which can help identify semi-quantitative compositional trends of secondary Ca-Mg-Fe smectites and carbonates that are difficult to distinguish at the micron scale (Gysi and Stefánsson, 2012b,c). For flow-through reactor type experiments, similar difficulties and challenges are encountered as in the batch-type experiments for the characterization of the solid reaction products, which are more difficult because of the low fluid/rock ratios and far-from-equilibrium design of these experiments.

### 3.2. Passivation layer

A Si–Al-rich alteration layer (the leached layer or passive layer) at temperatures < 150 °C has been observed on the surface in basalt glass dissolution experiments (Bourcier et al., 1989; Daux et al., 1997; Oelkers and Gislason, 2001; Gislason and Oelkers, 2003). Whether this layer exists widely for basalt-$CO_2$-water reactions for both basalt glass and crystalline minerals in basalts (e.g., olivine, pyroxene, and plagioclase) is unknown and constitutes another knowledge gap. Furthermore, how this layer controls the differential release of elements via diffusion is also not clear. An extensive radioactive waste literature deals with both natural and synthetic glass alteration, and some of the findings might be applied to the current issue. Sissmann et al. (2013)

found that coatings of secondary by-products on olivine decreased and/or controlled its dissolution rate and also affected the extent and rate of carbonation in their 90-170 °C experiments.

### 3.3. Knowledge of thermodynamics and kinetics of geochemical reactions

Many studies have already demonstrated that geochemical modeling predictions in multi-mineral systems using the off-the-shelf thermodynamic databases, e.g., *llnl.dat*, and disparate far-from-equilibrium rate parameters in single-mineral systems using kinetic databases (e.g., Palandri and Kharaka, 2004) cannot match the experimental data. Modifications of both thermodynamic and kinetic data were required (Gysi and Stefansson, 2011; 2012a; 2012b; Galeczka et al., 2014; Voigt et al., 2021).

#### *3.3.1. Thermodynamic properties*

To model the experimental system of basalt-$CO_2$-water interaction, Gysi and Stefánsson (2012a, b,c), Jones et al. (2012), Galeczka et al. (2014), Hellevang et al. (2017) and Luhmann et al. (2017b) updated the thermodynamic properties of selected aqueous species and minerals, building on the data files, e.g., *llnl.dat*, from the geochemical modeling program PHREEQC (Parkhurst and Appelo, 2013). Gysi and Stefánsson (2011) included thermodynamic data for clay minerals, carbonate solid solutions, and zeolites, as well as updated aqueous species for Al and Si based on experimental data available at that time.

Voigt et al. (2021) adjusted the log $K_s$ of the dissolution reactions of seven minerals (Mg-Mg-Saponite, Mg-Fe-saponite, Ca-Stilbite, anhydrite, aragonite, chalcedony, and allophane) and assigned 40-80% faster dissolution rates for basaltic glass to the first 50 days than for longer durations. With these adjustments along with the application of kinetics for secondary mineral precipitation, they were able to match approximately their modeling results with experimental data. Uncertainties in the thermodynamic properties of secondary minerals seem to be the most serious limitations in modeling when used in association with models to illustrate the effects of solution stoichiometry on the crystal growth (Zhang and Nancollas, 1998; Hellevang et al., 2014; Kang et al., 2022).

*Ad hoc* changes or additions to thermodynamic data in an extant database for a given modeling code must be internally consistent. Such is difficult to achieve in practice. Much to be preferred would be the construction of a unified database dedicated to the study of geochemical reactions in basalt. Some efforts are responsive to this need, e.g., *carbfix.dat* (Voigt et al., 2018), but there is still room for improvement.

*3.3.2. Kinetics*

Our current knowledge of geochemical kinetics is largely based on compilations of experimental data of single mineral dissolution rates far from equilibrium such as the kinetics parameter databases by the USGS (Palandri and Kharaka, 2004), CO2CRC (Black et al., 2015), French Geological Surveys (Marty et al., 2015), and French National Centre for Scientific Research (CNRS; Heřmanská et al., 2022; 2023). To model the kinetics of mineral dissolution and precipitation, a rate law is also required. A general form of rate law (Lasaga et al., 1994; Lasaga, 1998) has been proposed for mineral dissolution and precipitation

$$R = \frac{r}{s} = \pm k a_{H^+}^i (f \Delta G_r) \qquad (3)$$

where, $R$ is the dissolution/precipitation rate in mol m$^{-2}$ s$^{-1}$ (positive values indicate dissolution, and negative values for precipitation), $r$ is the dissolution/precipitation rate in mol s$^{-1}$ kgw$^{-1}$, $s$ is the specific reactive surface area per kg H$_2$O, $k$ is temperature dependent rate constant (mol m$^{-2}$ s$^{-1}$), $a_{H^+}^i$ is the activity of H$^+$ and $i$ is empirical reaction order accounting for catalysis by H$^+$ in solution. $f(\Delta G_r)$ is the rate dependence on the chemical driving force of the reaction, $\Delta G_r$. Possible ionic strength dependence of the rates and catalytic and inhibitory effects of aqueous species may also need to be considered (Schott et al., 2009).

The rate constants (usually at 25 °C) are extrapolated to other temperatures through Arrhenius-type equations (Palandri and Kharaka, 2004).

$$k = Ae^{-Ea/RT} \quad (4)$$

where $E$a is the apparent activation energy and $A$ the pre-exponential factor. A function of Gibbs free energy of the reaction, widely described as based on Transition State Theory (TST) (Aagaard and Helgeson, 1982; Lasaga, 1981), is typically used for modeling basalt-CO$_2$-water reaction systems

$$f(\Delta G) = 1 - \exp\left(\frac{\Delta G_r}{RT}\right) \quad (5)$$

Acid, neutral, basic promoted mechanisms are utilized in order to account for the significant effect of pH on the reaction rate for most minerals (Palandri and Kharaka, 2004).

Only a limited amount of near-equilibrium reaction rate and precipitation rate data is available (Schott et al., 2009; Zhu et al., 2021). However, the aqueous solutions are predominantly near-equilibrium with respect to the participating minerals several days to months to years after $CO_2$ injection. Generally, there is a lack of experimental data on the near-equilibrium, multi-mineral kinetics of such systems (Schott et al., 2009; Zhu et al., 2020; Zhu et al., 2021). Because of the lack of relevant data, most current geochemical models use far-from-equilibrium dissolution rate constants and assume their validity under near-equilibrium conditions for both dissolution and precipitation through *ad hoc* assumptions of rate laws.

For example, transition state theory (TST) rate law (Lasaga, 1981; Aagaard and Helgeson, 1982) is widely used for both mineral dissolution and precipitation. However, actual relationship between rates and Gibbs free energy of reaction deviates from TST rate law for near equilibrium dissolution of alkali-feldspars (Burch et al., 1993; Gautier et al., 1994; Hellmann and Tisserand, 2006; Zhu et al., 2010). Therefore, when TST is applied to far-from-equilibrium kinetic data and the data extrapolated to affinity = 0, this extrapolation often overpredict rates near equilibrium.

As for precipitation, Kumar et al. (2017) and Hellevang et al. (2017) indicate that the mineral reaction rate law based on TST overestimated carbonate mineral precipitation rates. The Burton-Cabrera-Frank (BCF) theory for crystal growth (Burton et al., 1951), has instead been used for carbonate precipitation at modest degrees of supersaturation, and was found to reproduce experimental data for magnesite precipitation (Saldi et al., 2009).

Reactive surface area (RSA) is a critical parameter for surface controlled reaction kinetics (Helgeson et al., 1984). The common practice in geochemistry is to use the BET surface area of a mineral as a proxy. Other methods (e.g., using geometric techniques) may result in orders of magnitude difference in estimating specific surface area (Black et al., 2015). However, there are several challenges in using BET surface area in the kinetic modeling. It is difficult to measure BET RSA for a mineral within a rock. The surface areas of secondary minerals are unknown. In an experiment, RSA may vary due to the growth or reduction of grain sizes and geometries. To account for the RSA changes during dissolution or precipitation, an empirical relationship of RSA ($S$) as a function of the initial total surface area ($S^o$) has often been used

$$S = S^o(m/m_o)^n$$

where $m$ and $m_o$ are current and initial amount of the mineral; $n$ is a coefficient that depends on the shape of the crystal and the relative rates of dissolution (or growth) on different surfaces. $n$ equals to 2/3 if the shape of the crystals remains unchanged and rates on all faces are equal. $n$ values of 0.5 indicate that dissolution or growth occur predominantly in two directions while $n$ values of 0 indicate one direction (e.g., Witkamp et al., 1990).

For secondary minerals, thermodynamic data and kinetic data of some amorphous secondary minerals $SiO_{2(a)}$, pyrolusite (amorphous $MnO_2$), $Fe(OH)_{3(a)}$, allophane, as well as the disordered dolomite-ankerite solid-solution series are lacking. Thermodynamic data of some low temperature clay minerals (e.g., smectite and chlorite) are questionable (Oelkers et al., 2009). Pyrite precipitation kinetics under the conditions of interest is also unknown. Many modeling studies need ankerite, but there is no experimental data for precipitation available. People have used dolomite precipitation rates as a substitute, but these dolomite precipitation rates themselves are not reliable (Arvidson and Mackenzie, 1999).

The uncertainties are propagated when a geochemical model is coupled with transport processes (diffusion, dispersion, and advection) in coupled RTMs (Steefel and MacQuarrie, 1996; Steefel et al., 2015; Li et al., 2017) and are applied to $CO_2$ storage in aquifers and soils. Dai et al. (2020) provides a review of these models. Although such studies provide useful insights (Liu et al., 2011; Zhang et al., 2015; Zhang et al., 2016; Zhang et al., 2021; Lu et al., 2022b; Shabani et al., 2022), it is also prudent, for the sake of contributing to the implementation of climate change mitigation strategies, to ask to what extent do the attendant uncertainties affect the magnitude and time required for sequestration of $CO_2$.

## 3.4. How silicate dissolution and carbonate and aluminosilicate precipitation reactions are coupled

As mentioned earlier, our current knowledge base of geochemical kinetics is largely largely based on single mineral dissolution rates far-from-equilibrium. Experimental, field, and modeling evidence shows that, in a multi-mineral system, the overall reaction rate in the system is determined by the coupling of dissolution and precipitation reactions of reactant and product minerals (Alekseyev et al., 1997; Alekseyev et al., 2004; Zhu et al., 2004; Zhu, 2005; Maher et al., 2009). Zhu et al. (2004) and Maher et al. (2009) hypothesized that the coupling of reactions brings the reacting solutions close to equilibrium with respect to feldspars, which decreases the feldspar dissolution rates via the chemical affinity term in the rate laws and contributes to the apparent discrepancy between laboratory and field rates. Because basalt-$CO_2$-water systems are so complex, it is still not clear how each primary mineral and secondary mineral is coupled and how these couplings affect each other. The cited investigators further speculated that the slow precipitation of (Ca, Fe, Mg) clay and carbonate minerals could decrease the dissolution rates of the basalt glass and (Ca, Mg, Fe) primary silicates over time. The overall reactivity of the system can also be reduced with the formation of leached layers or coating of precipitates on the reactants. In addition, Daval et al. (2010) showed that the reduction of labradorite and diopside dissolution rates at conditions of "near-equilibrium" could decrease by six fold the reactivity of the systems and carbon mineralization.

Saudi Aramco: Public

**3.5. CO$_2$-water-basalt reaction paragenesis**

Paragenesis in geology means the sequential order of mineral precipitation (Klein et al., 1993), which is controlled by the relative effective rates of dissolution and precipitation (Lasaga, 1998; Zhu, 2009; Zhu and Lu, 2009). Knowledge of paragenetic sequences has two practical impacts for accurately predicting CO$_2$ mineralization. First, different primary minerals have different carbonation potentials (Gadikota et al., 2020). Second, whether and how much precipitating non-carbonate minerals will compete for the divalent cations involved in carbonate precipitation (Gysi and Stefánsson, 2012a; Hellevang et al., 2017; Pollyea and Rimstidt, 2017) could affect the porosity and permeability of fractured basalts (Navarre-Sitchler et al., 2009; Noiriel et al., 2016; Pollyea and Rimstidt, 2017).

Even for basaltic glass (instead of composite basalt, which has both glass component and olivine, pyroxene, and plagioclase), secondary mineral types and paragenetic sequences can be very complex and these types and sequences vary with temperature (Gysi et al., 2012a; 2012b, 2012c). Fluid chemistry data should be integrated with meticulous mineralogical characterizations to obtain the knowledge of which minerals are stable at which pH and $p$CO$_2$. It is still a challenge to identify which minerals dissolved first, which minerals also dissolved, and how much each of these minerals dissolved during the course of the experiments. Another problem in some experiments is to effectively tune the geochemical models to be consistent with paragenetic information gained from these experiments.

**3.6. Calibration of reaction path models**

Although many of the aforementioned 49 basalt-CO$_2$-water interaction studies have used reaction path models to interpret experimental data, only a small fraction of these studies showed a good match of modeling results against experimental data, e.g., see Voigt et al. (2021). Model calibration remains an issue because (1) the basalt-$CO_2$-water system is so complex as it forms a geochemical reaction network of multiple primary and secondary minerals (Zhu, 2009); (2) the use of major ion chemistry data in systems of many solid phases poses a undetermined inverse mass balance model problem. Many assumptions have to be made and variables adjusted; (3) quantification of secondary mineral abundances and their sequence and timing is a major challenge for petrographic characterization. Future experiments need to better constrain experimental datasets, e.g., through use of multiple isotopic tracer experimental data to calibrate reaction path models.

### 3.7. Geochemical models under a wide range of CCUS conditions

$CO_2$ storage conditions vary widely in terms of temperature, $pCO_2$, basalt mineralogical and chemical compositions, and solution chemistry. Reaction rates and mineral parageneses differ under these conditions. It is impractical to conduct experiments to cover all conditions of $CO_2$ removal and storage. Geochemical modeling can help bridge the gaps in discrete experimental data by conducting a large number (e.g., thousands) of geochemical simulations covering a wide range of storage conditions. Recently, Ely (2020) simulated ~7.7 million cases of basalt-seawater interactions from a grid of water-rock ratios, temperature, and basalt compositions. The vast amount of data generated from modeling was used to compare observed fluid compositions and laboratory experimental data. Leong et al. (2021) simulated the potential of $H_2$ gas generation from the serpentinization of ultramafic and mafic rocks by reacting water with 9,414 recorded rock compositions.

## 4. DISCUSSION

The science associated with CCS in basalt formations has much wider application in areas such as soil amendment with comminuted basalt (Schuiling and Krijgsman, 2006; Köhler et al., 2010; Power et al., 2010; Hartmann et al., 2013; Köhler et al., 2013; Power et al., 2013; Moosdorf et al., 2014; Renforth et al., 2015; Taylor et al., 2016; Kantola et al., 2017; Meysman and Montserrat, 2017; Montserrat et al., 2017; Renforth and Henderson, 2017; Taylor et al., 2017; Beerling et al., 2018; Rigopoulos et al., 2018; Beerling et al., 2020; Swoboda et al., 2022), and dispersal of comminuted basalt over marine beaches or in oceans (National Academies of Sciences and Medicine, 2021) to capture atmospheric $CO_2$ directly. Furthermore, research concerning the natural atmospheric alteration of mafic and ultramafic mine tailings (National Academies of Sciences Engineering Medicine, 2019), the deep-sea natural alteration of basalts by magmatic $CO_2$, or seawater bicarbonate at spreading plate centers (Snæbjörnsdóttir and Gislason, 2016), can also contribute to an understanding of $CO_2$ mineralization over a wider range of ambient basalt formation temperatures, including those relating to $CO_2$ injection into sub-seafloor basaltic aquifers (Figure 1).

In order to provide some perspective as to how basaltic rocks would be affected by the range of conditions determined by both natural and enhanced $CO_2$ mineralization rates of basaltic rocks, we have, in the following sections, assessed the kinetics of primary rock and mineral dissolution and secondary mineral precipitation, using the limited data currently available. These data consist primarily of far-from-equilibrium mineral dissolution rates. Both near-to-equilibrium dissolution and precipitation rates can be estimated by assuming that these data are consistent with the application of transition state theory over the range of applicable temperatures. Given that such assumptions are in question and could lead to a substantial overestimation of reaction rates, it must be recognized that the following is a preliminary, albeit valuable qualitative analysis, which provides further insight into the future scope of scientific inquiry.

## 4.1. Far-from-equilibrium reaction rates of primary and secondary basaltic minerals

In this study, we selected the primary and secondary minerals in glassy and crystalline basalt, which are of interest for CCS. The primary minerals include forsterite, fayalite, diopside, enstatite, microcline, labradorite, andesine, apatite, ilmenite and phlogopite, and secondary minerals include calcite, dolomite, magnesite, siderite, chlorite (clinoclore 7A/14A), smectite, chalcedony, albite, hematite, magnetite, pyrite, clinoptilolite-Na (zeolite) and anhydrite. Far-from-equilibrium rates of primary mineral dissolution were calculated over the temperature (25-300 °C) and pH (3-10) ranges of interest to CCS in basalt. The pH ranges for each secondary mineral were individually determined based on their occurrence in the experiments (**Figures 2 and 3**). The calculated pH-dependent far-from-equilibrium reaction rates are presented in **Appendix C**. Based on these reaction rates, the ranges of the rates at different temperatures can be summarized in bar diagrams. **Figures 5 and 6** show the ranges of primary and secondary mineral far-from-equilibrium reaction rates at 25-300 °C, respectively.

## 4.2. Basalt reactivity index

Mineral reaction rates (*r*) can be expressed as Eq. 4. We consider the reactivity as the product of rate constant with RSA considering the effects of pH dependence and temperature and without considering the $f(\Delta G_r)$ term,

$$Reactivity = k \, exp\left[\frac{-E_a}{R}\left(\frac{1}{T} - \frac{1}{298.15}\right)\right] a_{H^+}^i \qquad (5)$$

Reactivity index (RI) can be used to evaluate the reactivity of basalt, which is the logarithm sum of the reactivity of each component mineral in the rock multiplied by its mineral abundance.

$$RI = log\left(\sum_{i=1}^{n} Reactivity_i * f_i\right) \quad (6)$$

where $i$ is each mineral component in a basalt. $f$ denotes the fraction (0-1) of the mineral component. In calculating RI, we assume that surface area is proportional to the mineral mass fraction and that the mineral dissolution rates are normalized to the mole of a mineral. The Brunauer–Emmett–Teller (BET) surface area (Braunauer et al., 1938) can be used as a proxy for RSA when it is available; otherwise, geometric surface areas can be used instead. For example, the geometric surface area is 0.0542 $m^2/g$ for the basalt with a particle size of 53-150 microns (Kelland et al., 2020). The effect of temperature and pH should be considered when calculating RI.

The application scenarios of basalt-$CO_2$-water interactions includes soils with comminuted basalt (pH 5-8 and average earth surface temperature of 15 °C; Kelland et al., 2020), Sub-seafloor basaltic aquifers (0.01−65 °C and pH 3-4.5; Tutolo et al., 2021), and subsurface basaltic formations (pH 7-9, lower temperature between 20-65 °C and upper temperature at 250 °C; Aradóttir et al., 2011; McGrail et al., 2014; Matter et al., 2016).

We used the basalt samples from Lewis et al. (2021) as an example for RI calculation (**Table 1**); 15 °C and 6.5 are selected for the respective temperature and pH regimes for soil amendment, 35 °C and 3.5 respectively for CCS in sub-seafloor basaltic aquifers, 50 °C and 8 respectively for CCS in low temperature subsurface basaltic formations, and 250 °C and 8 respectively for CCS in high temperature subsurface basaltic formations.

**Table 1 Calculated reactivity index (log mol/s) for basalt samples from Lewis et al. (2021).**

|  | Blue Ridge | Cragmill | Hillhouse | Oregon | Tawau | Tichum |
|---|---|---|---|---|---|---|
| **Soil amendment (pH 6.5; *T* 15)** | -9.75 | -5.38 | -6.14 | -3.69 | -4.67 | -4.56 |
| **Sub-seafloor (pH 3.5; *T* 35)** | -8.16 | -4.92 | -5.15 | -3.23 | -4.19 | -4.08 |
| **Subsurface basaltic formations low (pH 8; *T* 50)** | -8.77 | -5.35 | -6.06 | -3.87 | -4.60 | -4.74 |
| **Subsurface basaltic formation high (pH 8; *T* 250)** | -4.88 | -1.65 | -2.06 | -3.11 | -0.85 | -0.06 |

Three groups can be separated for the basalt reactivity for each utilization based on the RI values (**Figure 7**). One basalt rock may be in different groups if the application scenarios are different. Oregon, Tichum, and Tawau belong to the high reactivity group, Cragmill and Hillhouse are in the medium reactivity group, and Blue Ridge is in the low reactivity group for the soil amendment, CCS in the sub-seafloor basaltic aquifer, and CCS in the low-temperature subsurface basaltic formation applications (the RI value for Cragmill basalt for the sub-seafloor basaltic aquifer is close to 5). However, for CCS in the high-temperature subsurface basaltic formations, Tichum, Tawau, and Cragmill fall in the high reactivity group, Hillhouse and Oregon are in the medium reactivity group, and Blue Ridge are in the low reactivity group. Although the RI value of Cragmill is lower than that of Oregon at low temperatures (e.g., 50 °C), it is higher at high temperatures (250 °C) because Cragmill contains a higher abundance of minerals of high activation energy than that of Oregon.

Another method is the direct measurement of rock reactivity. In most experiments, basalt dissolution rates were estimated by measuring the steady-state silica release rate normalized to the unit surface area of the rock, although others utilized dissolution rates with respect to other elements (e.g., Al, Ca, Mg, Fe, and Na; Schaef and McGrail, 2009). However, interpreting basalt dissolution is challenging, because basalt is complex in both mineralogical and chemical compositions, and measuring mass changes or the variation of fluid concentrations does not tell us what components in basalt are actually dissolving and what are not. Future benefits might be gained from (1) better-constrained solution chemistry (e.g., multiple isotope tracer) (e.g., Chen et al., 2023) and mineralogical characterization data and (2) associated geochemical modeling.

39Saudi Aramco: Public

The dissolution rates of naturally altered basalt are generally one to three orders of magnitude lower than their unaltered counterparts. Elevated temperatures (e.g., > 100 °C) may be required to compensate for their lower reactivity during subsurface carbonation (Delerce et al., 2023).

## 5. PROPOSED STEPS FORWARD

### 5.1. Experiments for near-equilibrium dissolution and precipitation

To build our knowledge base of geochemical kinetics, we need to progress in complexity from the study of single mineral dissolution far-from-equilibrium to the coupled dissolution - precipitation reactions near-equilibrium, and ultimately to to multi-reactant-multi-product (whole rock) systems, such steps being independent of the added complexity of transport processes (**Figure 8**). The end stage0 in the context of the current review would be to measure basalt and single mineral dissolution rates at the near-equilibrium conditions and understand how (Ca,Mg,Fe) silicate dissolution and (Ca,Mg,Fe) carbonate precipitation reactions are coupled.

### 5.2. Solution stoichiometry for carbonate precipitation

Some basalt-$CO_2$-water experiments found carbonate precipitation whereas others did not. Current practice uses an affinity-based rate law, which does not account for lattice ion ratios in the aqueous solutions. For modeling carbonate mineral precipitation, ionic ratios of [Me]/[$CO_3$] should be provided (where Me is a divalent metal cation). The effect of ionic ratios on insoluble AB (or di-ionic; e.g., halite and calcite) crystal precipitation rates was recently demonstrated by Hellevang et al. (2014) and Kang et al. (2022). At the same SI or chemical affinity, mineral precipitation rate constants may vary one order of magnitude or more with different A/B ratios (Hellevang et al., 2014; Kang et al., 2022). The control of precipitation rates by both affinity and ionic ratios is explained with Zhang and Nancollas's (1998) (ZN98) process-based AB crystal growth model and is also found for carbonate minerals (see Kang et al., 2022 for a review).

## 5.3. Multi-isotope doping to unravel the multi-sources, multi-sinks problem in a multi-mineral system

A promising approach was a study by Seimbille et al. (1998) who used multiple isotope tracers ($^{84}$Sr, $^{39}$K) for granite-water experiments, and were therefore able to identify which minerals dissolved and the extent of dissolution of each mineral in a given system. They also derived rates of multiple reactions simultaneously. Therefore, the multi-sources, multi-sinks problem in a multi-mineral system was overcome with the innovative application of multiple isotope tracers.

Reaction path modeling leads to greater complexity than is found in batch type reaction path modeling. Currently, there are still many challenges associated with RTM (see **Table 3**). On the positive side, after we have incorporated the findings of batch system chemical and isotopic data in our geochemical models, we will have greater confidence that our RTM models can be applied in the field. This is the proper progression of model complexity (**Figure 8**). Some parts of the rock can be inaccessible to the aqueous reactant, e.g., at mineral grain interiors, and is defined by the total vs. RSA. Therefore, experiments should use the whole rock instead of a single mineral, e.g., use a core plug for core flooding, with experimental conditions as close to *in* *situ* conditions as possible. The experimental duration should be sufficiently long (e.g., >1 yr) to maximize to the extent feasible confidence in predictive modeling of the full carbonation potential of the rock.

### 5.4. Better characterization of secondary precipitates

Considering the small quantities of secondary minerals, XRD and TEM analysis can be conducted to identify the secondary minerals and quantify their abundances. Ultrasonication can be used to free the secondary minerals from the bulk material, concentrated by filtering, and subsequently mounted onto a zero-background quartz plate for XRD characterization or onto a Cu-TEM-grid for TEM analysis (e.g., Lu et al., 2013). Parallel experiments terminated at different times should be performed to acquire time-series secondary mineral information.

### 5.5. Natural analogue studies to calibrate the experimentally derived parameters

One limitation of experimental studies is their timescale. Parameters obtained from short-term experiments (hours to months) are difficult to extrapolate to the field scale operations, but they can be calibrated with natural analogue studies with some limitations such as differences in site-specific conditions. To resolve the lab-field discrepancies in the mineral dissolution and precipitation rates, experimental studies are required at near-equilibrium conditions. Reaction-induced RSA changes at the thousands to millions of years' time scale is also a challenge to be investigated. The model of a natural analogue permits direct comparison of model prediction and field observations of secondary minerals and their abundance, as well as carbonation potential of the basalt rocks. Natural analogue studies need to be conducted to establish their value in making longer-term predictions.

**ACKNOWLEDGMENTS**

CZ's research is partially supported by U.S. National Science Foundation grant EAR-2221907. The authors would like to apologize to friends and colleagues for any omissions of important papers in such a wide-ranging review.

**Author contributions**

P.L. and C.Z. conceived the study and wrote the original draft. P.L. analyzed most of the literature. G.Z. contributed to the summary of the secondary mineral occurrences and mineral reaction rates. J.A. reviewed, wrote, edited drafts of the manuscript extensively.  A.G. contributed to the revision of the manuscript.

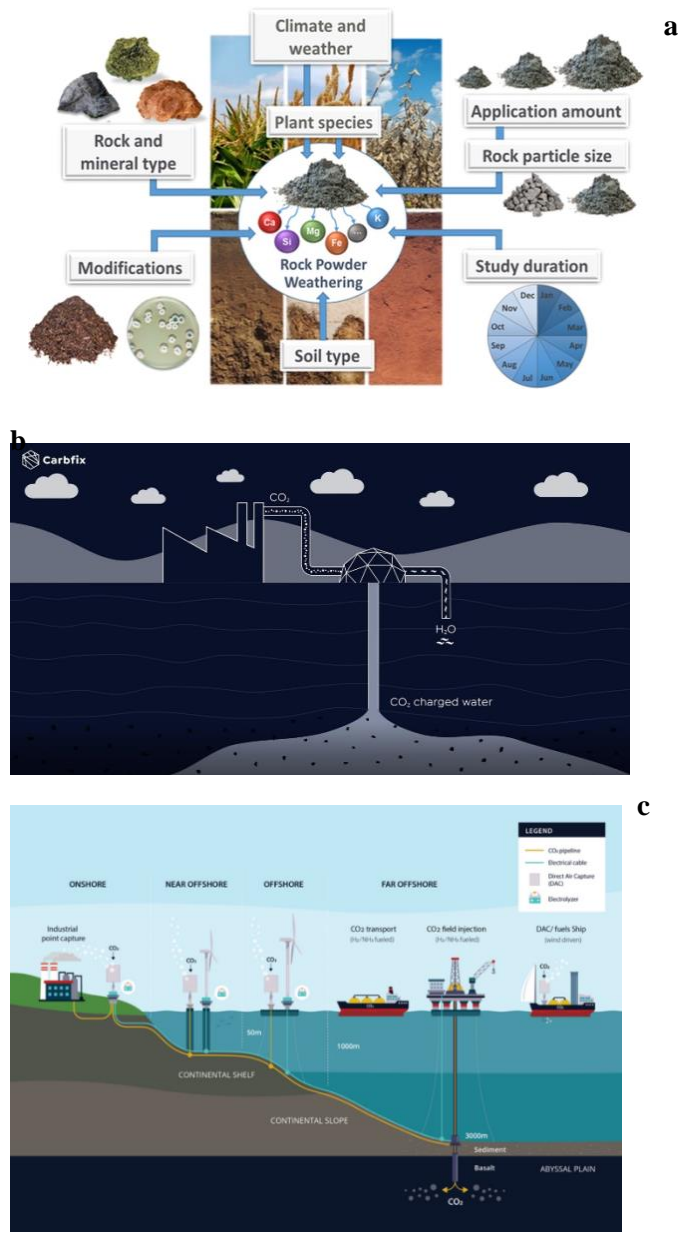

**Figure 1.** Schematic diagrams of some example applications of CCS in basalt. (a) Soil amendment with comminuted basalt in croplands. From Swoboda et al. (2022) with the permission from Elsevier. (b) CCS in subsurface basaltic formation (Carbfix project as an example). (c) CCS in offshore oceanic basalt (e.g., Cascadia Basin, Offshore Washington State and British Columbia).

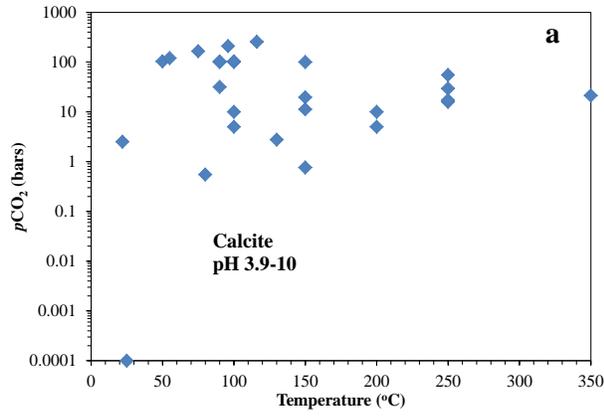
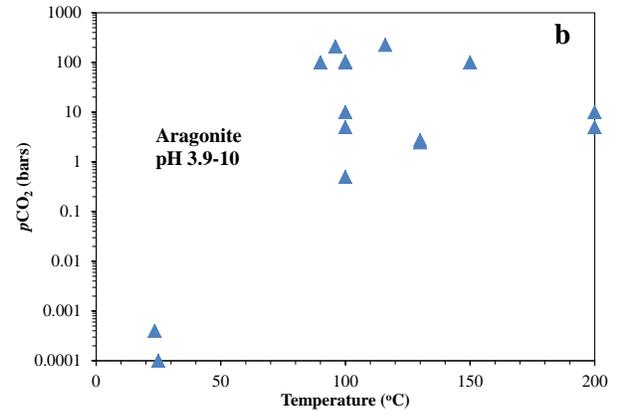
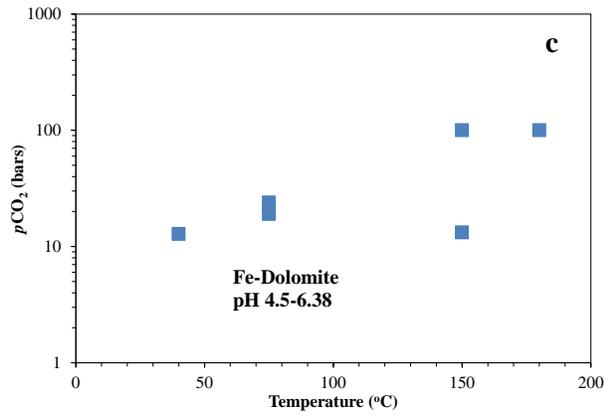
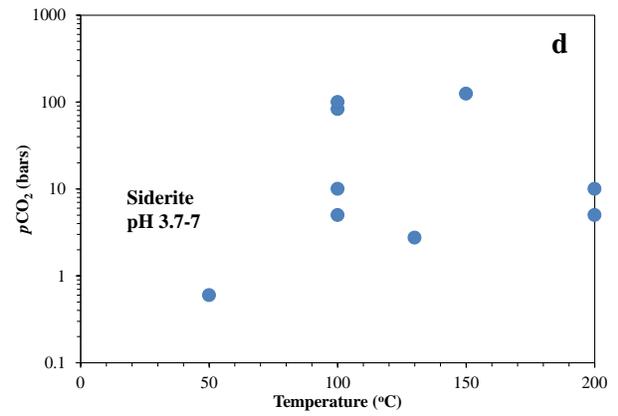
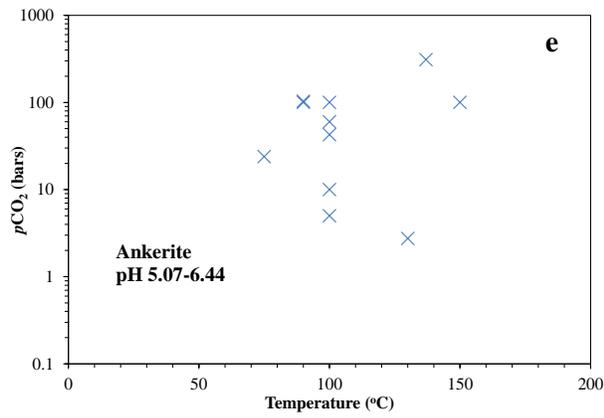
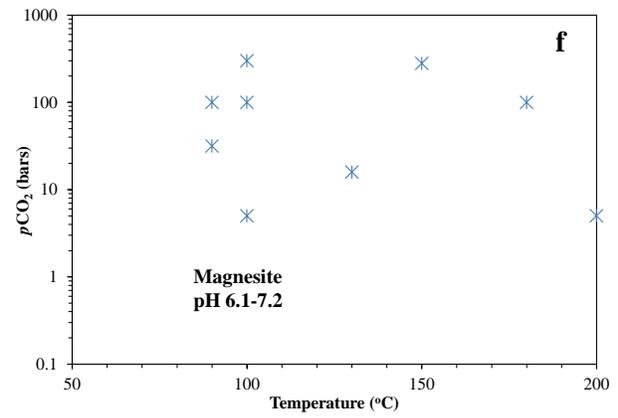

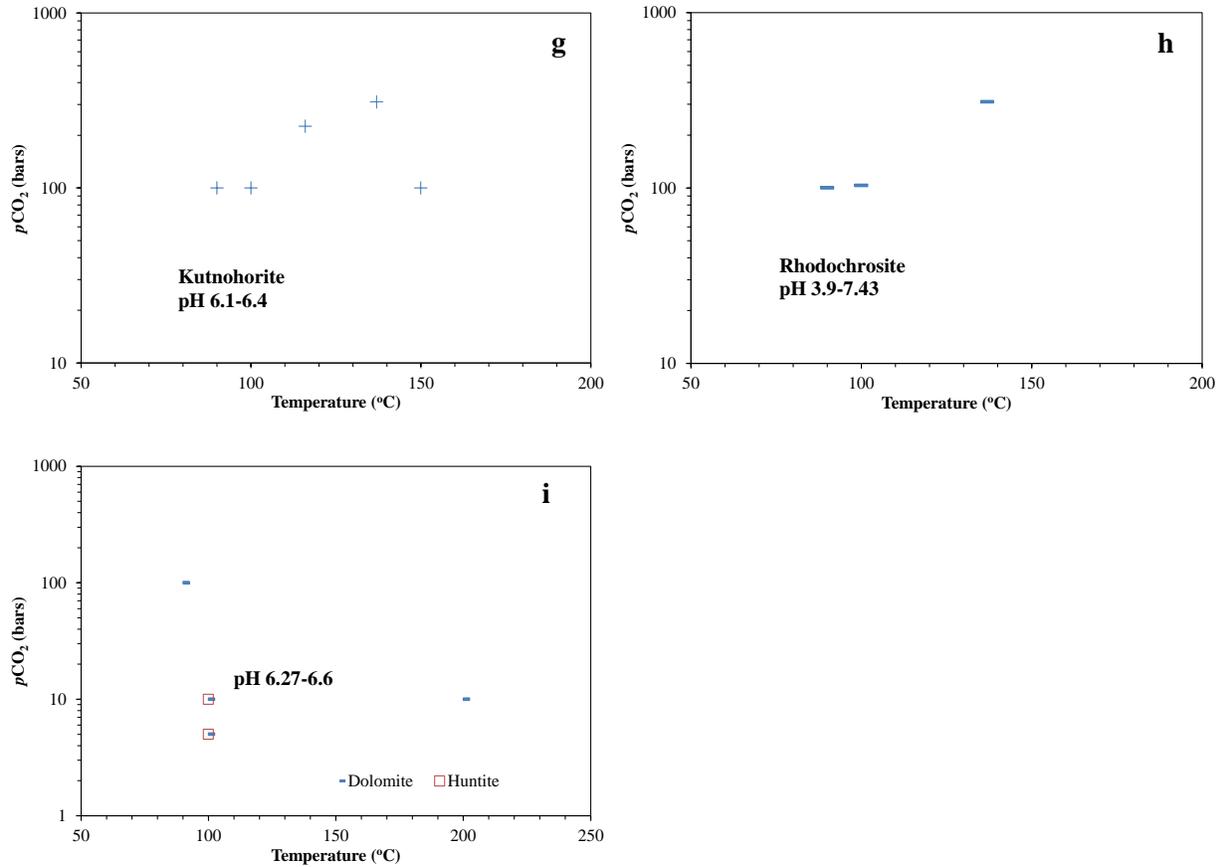

**Figure 2**. Observed secondary carbonate minerals in the experiments. (a) Calcite; (b) Aragonite; (c) Fe-dolomite; (d) Siderite; (e) Ankerite; (f) Magnesite; (g) Kutnohorite ($CaMn(CO_3)_2$); (h) Rhodochrosite ($MnCO_3$); (i) Dolomite and Huntite ($Mg_3Ca(CO_3)_4$).

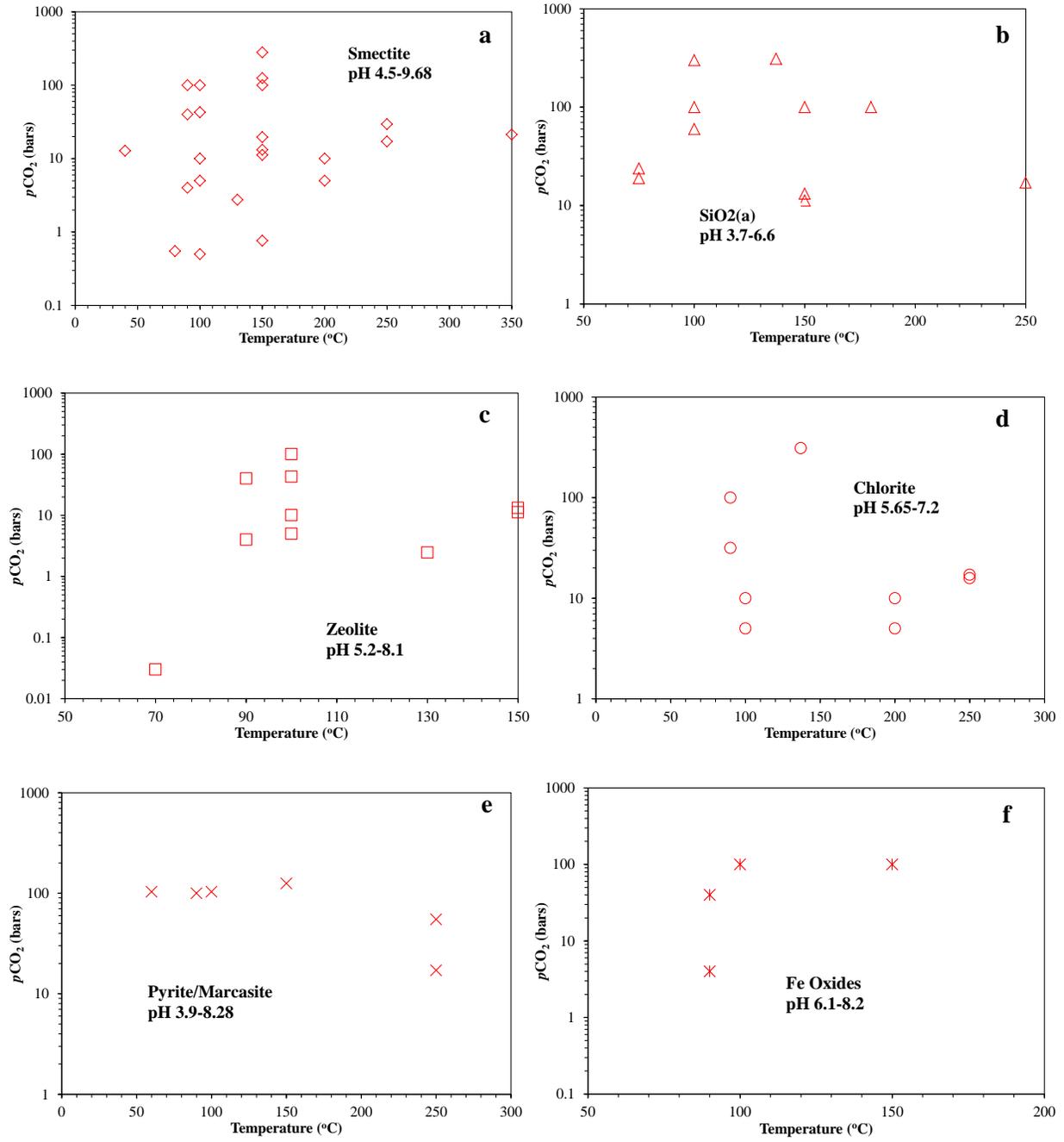

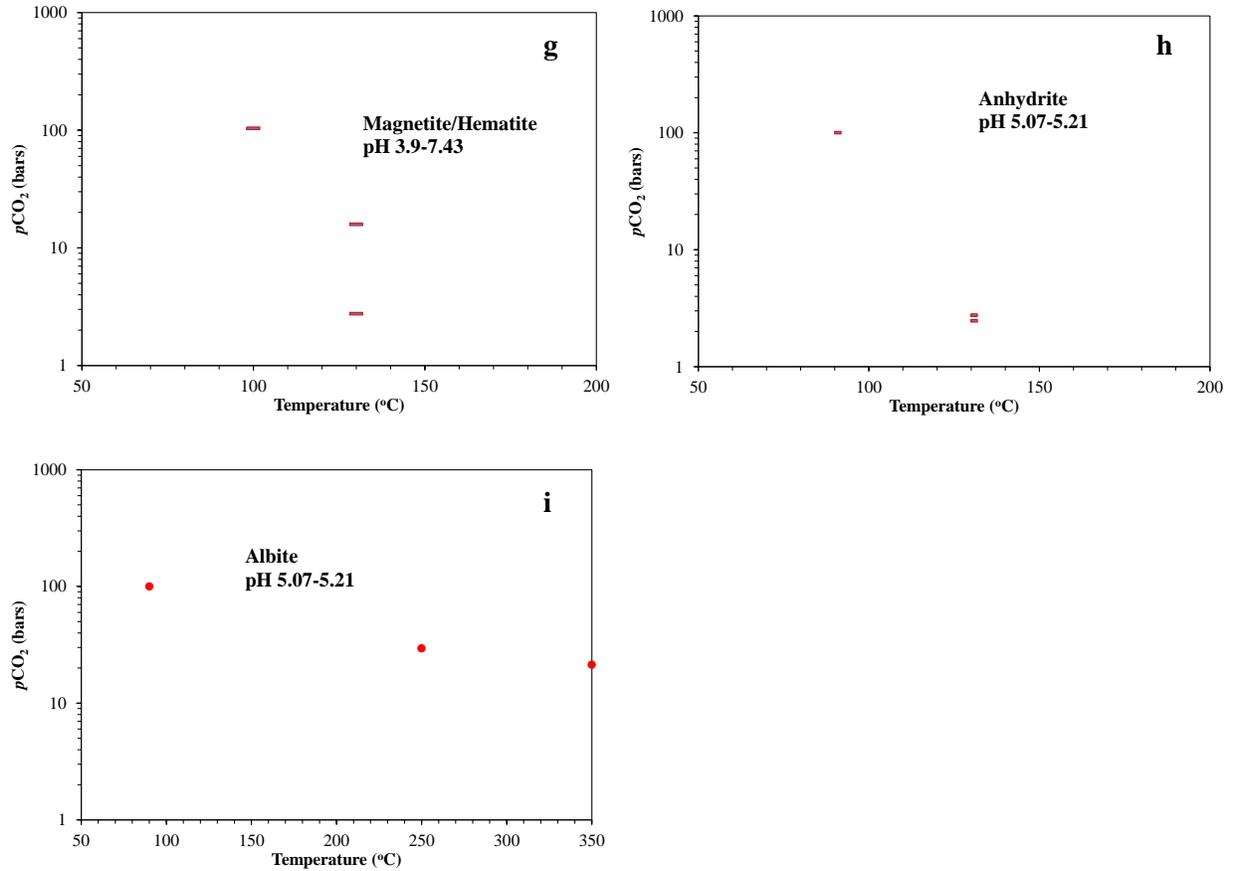

**Figure 3**. Observed secondary clay, sulfide, and oxide minerals in the experiments. (a) Smectite; (b) SiO$_2$(a); (c) Zeolite; (d) Chlorite; (e) Pyrite/Marcasite; (f) Fe oxides; (g) Magnetite/Hematite; (h) Anhydrite; (i) Albite.

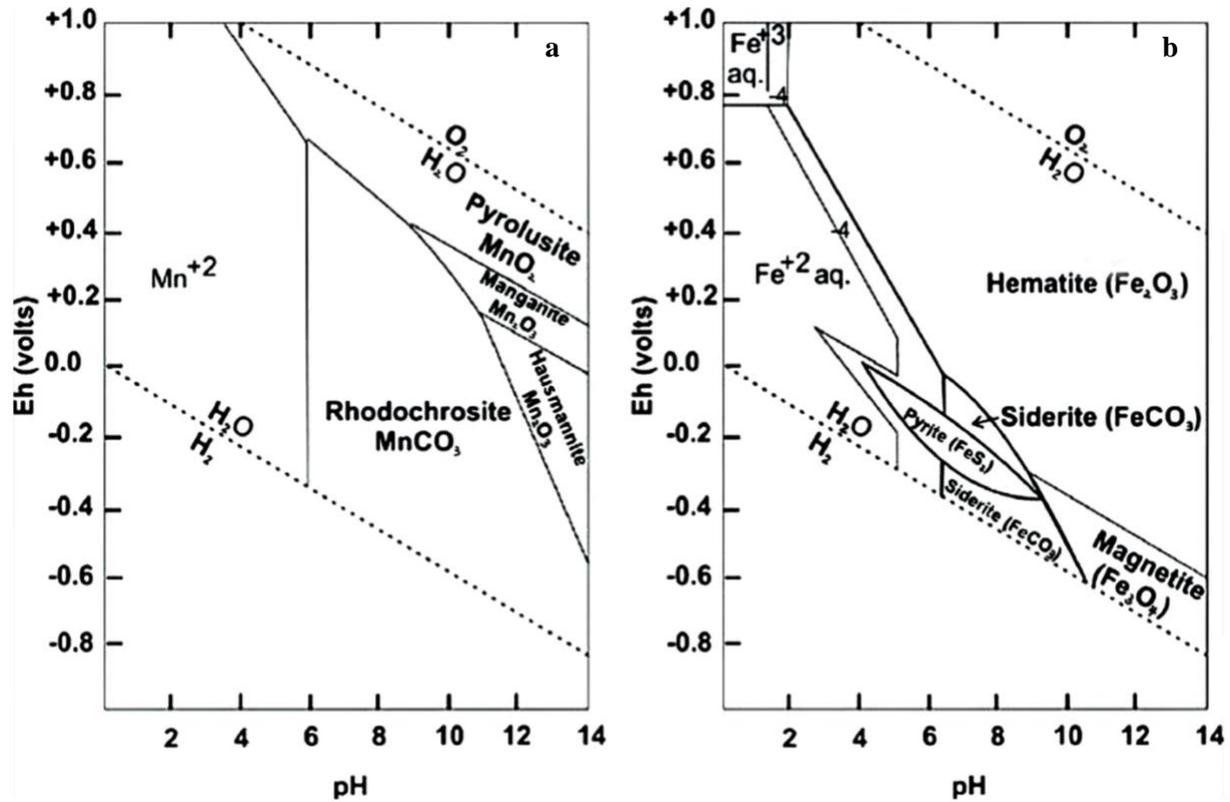

**Figure 4**. Eh-pH diagram showing the stability fields of the common manganese and iron minerals in water at 25 °C and 1 atm total pressure (diagrams after Christ, 1965; Brookins, 1987; Krauskopf and Bird, 1995; Takeno, 2005). (a) The total concentration of dissolved carbon is 1 $m$ and the total dissolved sulfur and manganese are $10^{-6}$ $m$, respectively (b) The total concentration of dissolved carbon is 1 $m$ and the total dissolved sulfur and iron are $10^{-6}$ $m$, respectively.

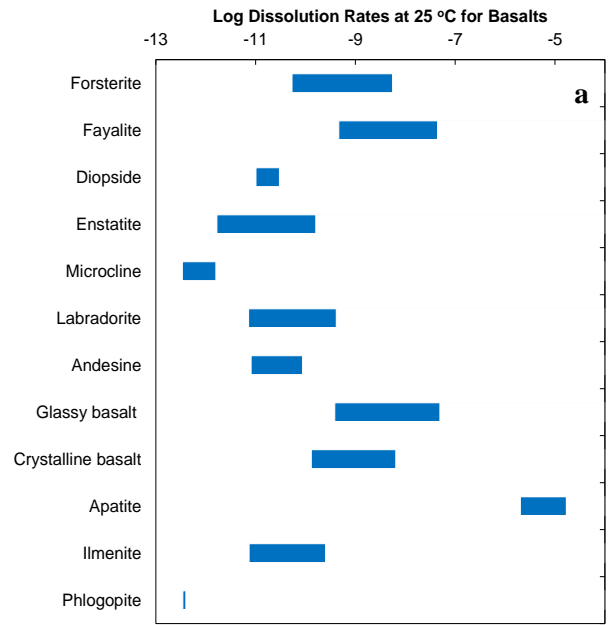
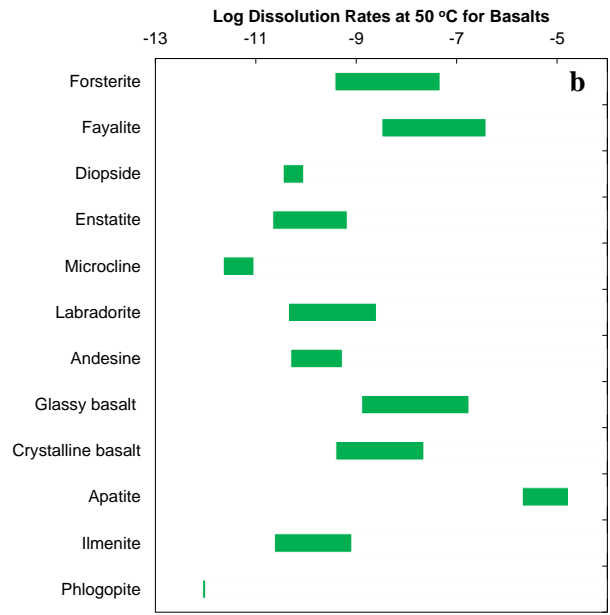
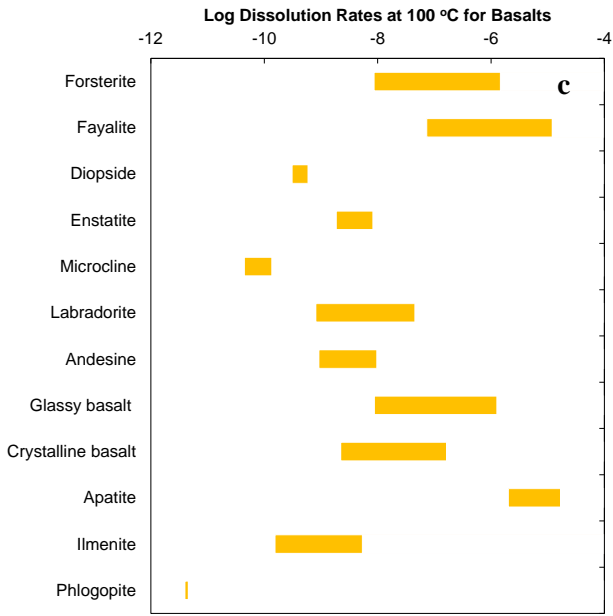
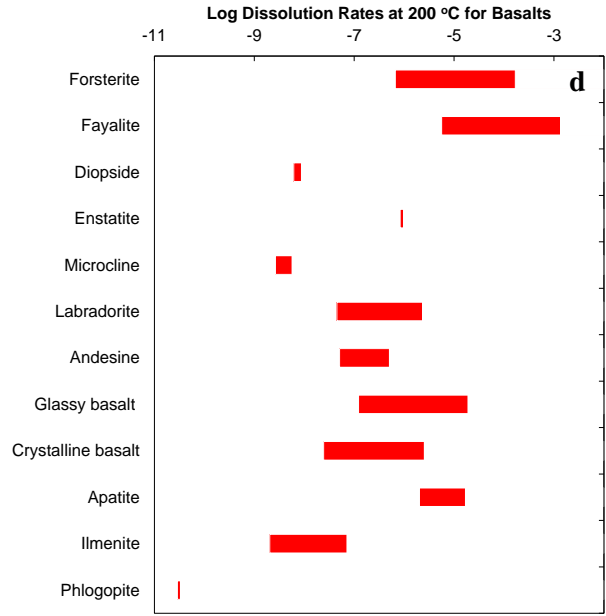

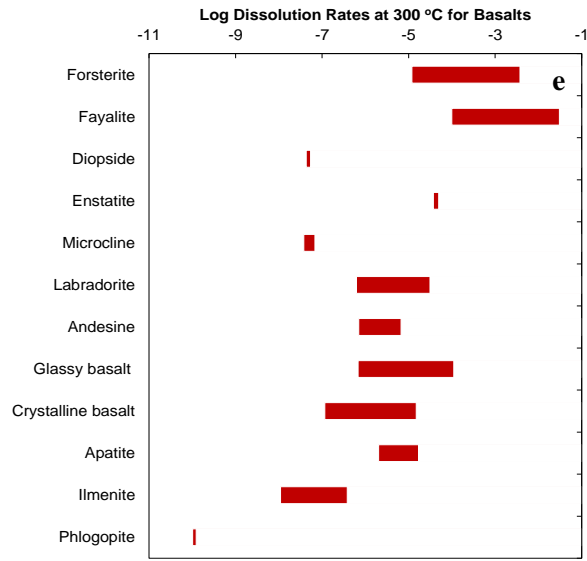

**Figure 5**. Range of primary mineral far-from-equilibrium dissolution rates for basalts in terms of the logarithm of mol/m$^2$/s. (a) T = 25 °C; (b) T = 50 °C; (c) T = 100 °C; (d) T = 200 °C; (e) T = 300 °C. The pH range is 3-9.5. Mineral reaction rates are calculated based on: forsterite, fayalite, diopside, enstatite, microcline, labradorite, and andesine from Heřmanská et al. (2022); glassy basalt and crystalline basalt from Pollyea and Rimstidt (2017); apatite (hydroxyapatite), ilmenite, and phlogopite from Palandri and Kharaka (2004).

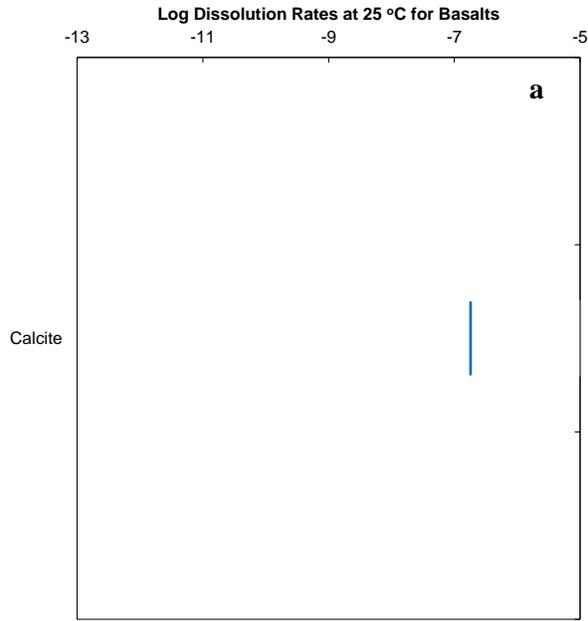
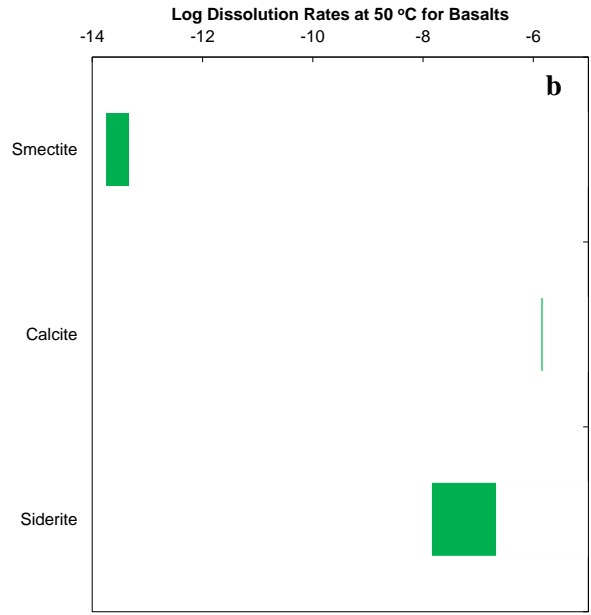
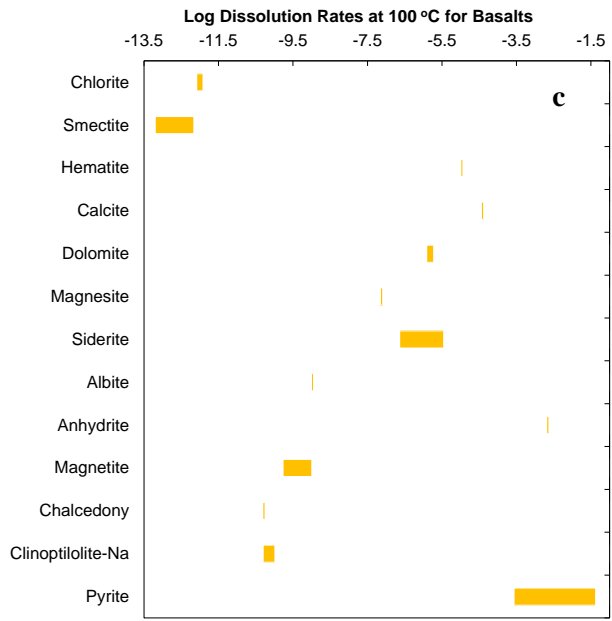
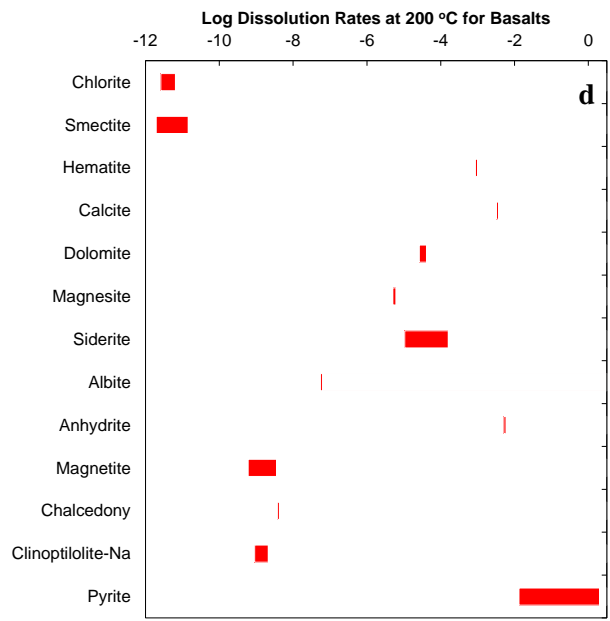

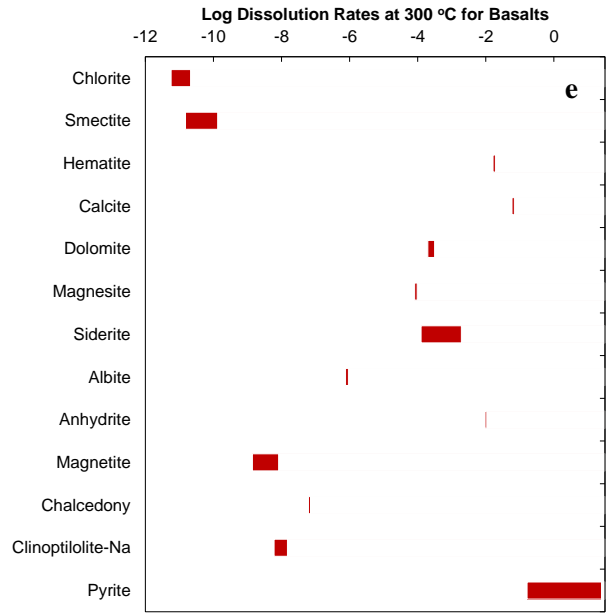

**Figure 6**. Range of secondary mineral far-from-eqbilirium reaction rates for basalts in the logarithm of mol/m$^2$/s. (a) T = 25 °C; (b) T = 50 °C; (c) T = 100 °C; (d) T = 200 °C; (e) T = 300 °C. The pH ranges are respectively 4.7-5.7 for chlorite and smectite, 3.9-7.47 for hematite and magnetite, 4.5-6.4 for calcite, 4.5-6.7 for dolomite and siderite, 4.8-6.4 for magnesite, 5.3-6.7 for albite, 3.4-4.4 for anhydrite, 5-7.5 for chalcedony and 5.2-6.7 for pyrite. Calculated mineral reaction rates are based on: chlorite (clinoclore 7A/14A), smectite, and Clinoptilolite-Na (zeolite) from Heřmanská et al. (2023); albite from Heřmanská et al. (2022); hematite, magnesite, anhydrite, magnetite, chalcedony, and pyrite from Palandri and Kharaka (2004); calcite, dolomite, and siderite from Marty et al. (2015).

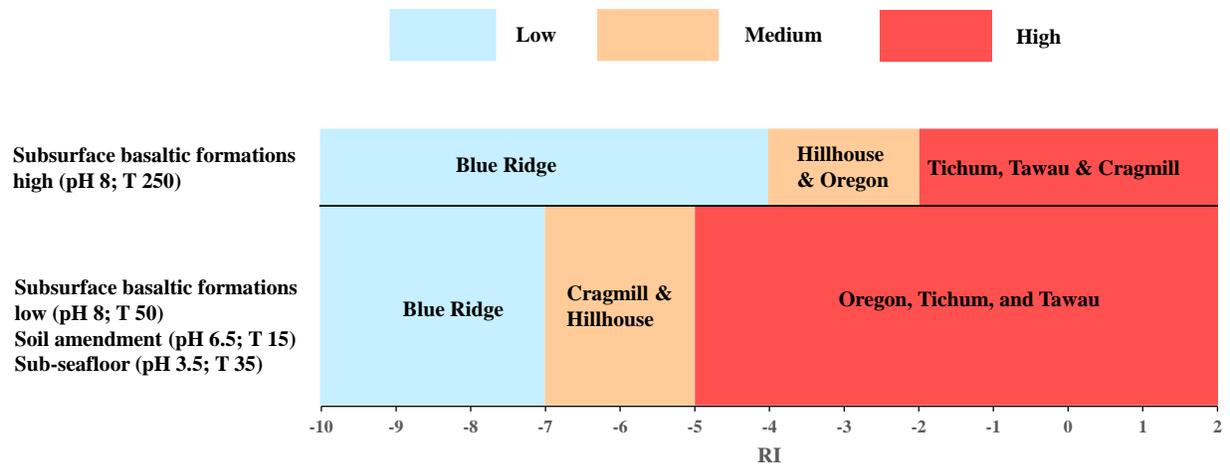

**Figure 7**. Basalt reactivity groups for each utilization scenario. The basalt samples from Lewis et al. (2021) as an example for RI calculation.

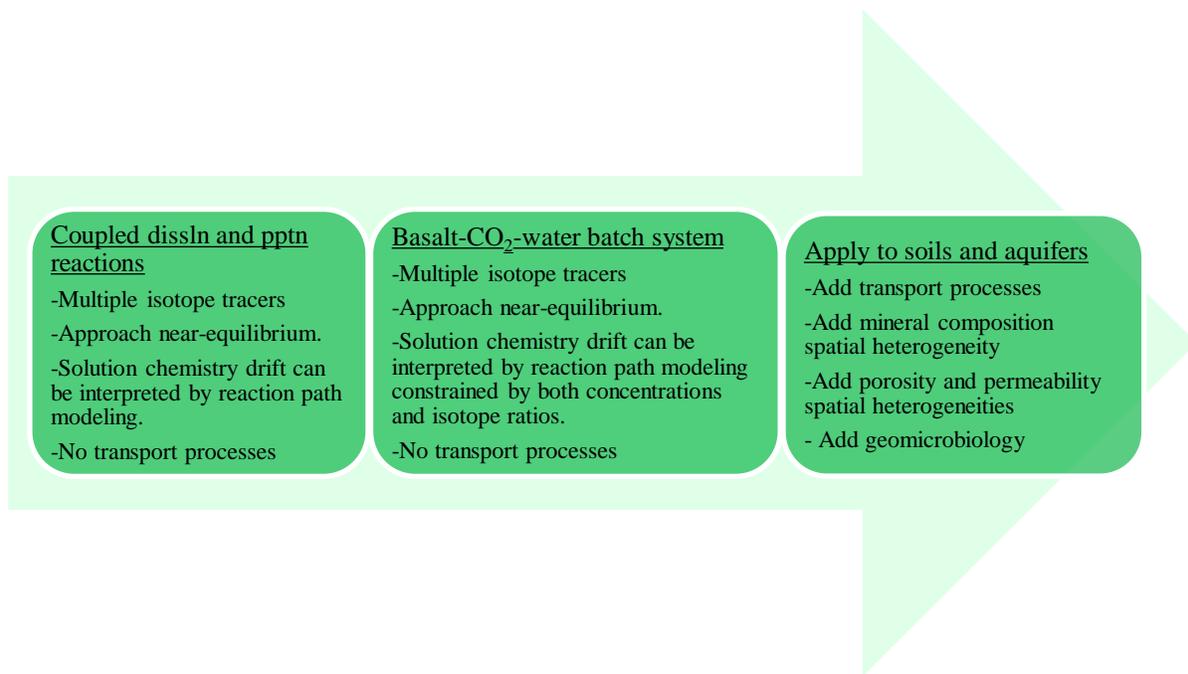

**Figure 8.** Progression of geochemical modeling complexity.

**Table 3. Current knowledge and knowledge gaps for basalt-CO$_2$-water systems**

|  | **Existing knowledge** | **Knowledge gap** |
|---|---|---|
| Solution chemistry | pH and major ion concentration changes | Higher resolution and better constrained solution chemical data (e.g., trace elements and multi-isotope doping) |
| Secondary mineralogy | Qualitative secondary mineral information | Quantitative secondary mineral information; paragenetic sequence and its effective modeling |
| Basalt reactivity | Using the abundance of fast-reacting minerals as a proxy; calculating basalt reactivity index; measuring steady-state release rate of Si or other elements | Knowledge of actual dissolution components in different basalts under different conditions of $T$ and $p$CO$_2$ |
| Passivation layer | Observed occasionally | To what extent it forms and how it controls the differential release of elements via diffusion |
| Thermodynamics | Off-the-shelf database and *ad hoc* data modification | Unified and dedicated database; Thermodynamic data of amorphous, disordered minerals, and low temperature clay minerals |
| Primary mineral kinetics | Single mineral, far-from-equilibrium dissolution rates and activation energies | Multi-minerals, near-equilibrium dissolution rates. |
| Secondary mineral kinetics | Usage of far-from-equilibrium dissolution rates as a proxy | Multi-minerals, near-equilibrium precipitation kinetics; amorphous minerals, pyrite, ankerite |
| Rate laws | TST for both mineral dissolution and precipitation | Appropriate and customized rate laws for both mineral dissolution and precipitation |
| RSA | BET or geometric surface area as a proxy. Using shrinking sphere model to account for the RSA changes during dissolution | A large uncertainty exists in estimating RSA for precipitating secondary phases. Lack of knowledge of accounting RSA changes during precipitation |
| Coupled dissolution and precipitation reactions | Far-from-equilibrium single mineral dissolution rates | How each primary mineral and secondary mineral are coupled and how these couplings affect each other |
| Geochemical model calibration | Matching approximately the evolution of major ionic composition and dissolved inorganic carbon | Improved calibration with higher resolution and better constrained solution data and quantification of secondary mineral abundances and their occurring sequence and timing |
| Model applicability | Models for certain specific conditions of experiments | Models for a wide range of CCS conditions |

Online Proceedings Library (OPL) Volume 176: Symposium U – Scientific Basis for Nuclear Waste Management XIII , 209.

Brady P. V. and Gíslason S. R. (1997) Seafloor weathering controls on atmospheric $CO_2$ and global climate. Geochim. Cosmochim. Acta 61, 965-973.

Brantley S. L. (2008) Kinetics of mineral dissolution. In Kinetics of Water-Rock Interaction (eds. S. L. Brantley, J. D. Kubicki and A. F. White). Springer, New York. pp. 151-210.

Braunauer S., Emmett P. H. and Teller E. (1938) Adsorption of gases in mutlimolecular layers. J. Am. Chem. Soc. 60, 309–319.

Brookins D. G. (1987) Platinoid element Eh-pH diagrams (25° C, 1 bar) in the systems M-O-H-S with geochemical applications. Chemical geology 64, 17-24.

Burch T. E., Nagy K. L. and Lasaga A. C. (1993) Free energy dependence of albite dissolution kinetics at 80 °C and pH 8.8. Chem. Geol. 105, 137–162.

Burton W.-K., Cabrera N. and Frank F. (1951) The growth of crystals and the equilibrium structure of their surfaces. Philosophical Transactions of the Royal Society of London. Series A, Mathematical and Physical Sciences 243, 299-358.

Chen, M., Lu, P., Pan, R, Song, Y. and Zhu, C. (2023) Geochemical modeling to aid experimental design for multiple isotope tracer studies of coupled dissolution and precipitation reaction kinetics. Acta Geochimica, submitted.

Christ C. L. (1965) Solutions, minerals, and equilibria. Harper & Row.

Clark D. E., Galeczka I. M., Dideriksen K., Voigt M. J., Wolff-Boenisch D. and Gislason S. R. (2019) Experimental observations of $CO_2$-water-basaltic glass interaction in a large column reactor experiment at 50 C. International Journal of Greenhouse Gas Control 89, 9-19.

Coogan L. A. and Gillis K. M. (2013) Evidence that low-temperature oceanic hydrothermal systems play an important role in the silicate-carbonate weathering cycle and long-term climate regulation. Geochemistry, Geophysics, Geosystems 14, 1771-1786.

Dai Z., Xu L., Xiao T., McPherson B., Zhang X., Zheng L., Dong S., Yang Z., Soltanian M. R., Yang C., Ampomah W., Jia W., Yin S., Xu T., Bacon D. and Viswanathan H. (2020) Reactive chemical transport simulations of geologic carbon sequestration: Methods and applications. Earth-Science Reviews 208, 103265.

Daux V., Guy C., Advocat T., Crovisier J.-L. and Stille P. (1997) Kinetic aspects of basaltic glass dissolution at 90 C: role of aqueous silicon and aluminium. Chemical Geology 142, 109-126.

Daval D., Hellmann R., Martinez I., Gangloff S. and Guyot F. (2013) Lizardite serpentine dissolution kinetics as a function of pH and temperature, including effects of elevated $pCO_2$. Chemical Geology 351, 245-256.

Daval, D., Hellmann, R., Corvisier, J., Tisserand, D., Martinez, I., & Guyot, F. (2010). Dissolution kinetics of diopside as a function of solution saturation state: Macroscopic measurements and

# Appendix A. Summary of experiments

| Publication | Experimental conditions | Material | Objective | Results & Comments |
|---|---|---|---|---|
| Adam et al. (2013) | Batch experiment; 100 °C, 8.3 mPa, 3.5 $M$ dissolved $CO_2$. 60mL of the basalt-equilibrated water and 14.6 g of $CO_2$. Duration: up to 210 d | Three ferrobasaltic samples from the Snake River Plain, Idaho, U.S. Sample B1 B2 B3 Plagioclase % 62.7 67.3 66.3 Pyroxene % 19.3 15.1 13 Olivine % 13.8 12.2 15.5 Gypsum % 1.7 .7 .6 Ilmenite % 2.5 4.7 4.6 | To quantify the changes in seismic velocity due to rock alteration; the feasibility of seismic monitoring for $CO_2$ sequestration in basalts. | Carbonates (Fe-Mg carbonates) precipitated in the pore space, but in particular within fractures, reducing the rock permeability and stiffening the rock. Porosity decreased by 2-3%. |
| Adeoye et al. (2017) | Flow-through and static batch; 45 °C and | Columbia River Flood basalt; Serpentinized basalt from Colorado Flow-through: | Role of fluid transport regimes on the dissolution | Net dissolution in flow-through experiments, whereas carbonate mineral precipitation along basalt fractures after 6 wks reaction in batch experiments. Calculated pH kept around 3.2-3.8 up to 140 h |
| | 100 °C, 100 bars $pCO_2$ (fixed); **Injection of $CO_2$-charged water for flow-through experiment**; Solution: DI water & 1.2 m$M$ NaHCO$_3$, 13.8 m$M$ NaCl; Rock-to-water ratio: 1:3 Duration: up to 42 d. | 2.54cm (diameter) x 42-45 mm (thick) core, etched with 1 mm-wide x 95-105 μm depth meandering pattern<br><br>Batch: Core with a groove with 11 mm width x 95-105 μm depth<br><br>Flood basalt: Olivine 9%, Pyroxene (diopside 22%, orthopyroxene 1%), Plagioclase 31% (Labradorite), Ilmenite 3%, and Si- and K-rich matrix 33%;<br><br>Serpentinized basalt (coarse grained 100-150 μm): Olivine 1%, Pyroxene (diopside 21%, orthopyroxene 1%), Serpentine 14%, Plagioclase 28% (labradorite An 0.55), [apatite, chromite, glass] 3% and Si- and K-rich matrix 32%. | and precipitation in fractures; critical factors promoting $CO_2$ mineralization; Used batch and flow-through to compare the diffusive and advective transport regimes. | Flow-through: Dissolution ↑ when T↑ & Salinity ↑. Enhanced dissolution along fractures at large grains of pyroxene and olivine.<br><br>Batch: 50 μm size secondary mineral by SEM; Mg-bearing siderite along the entire fracture in fresh basalt, but form clusters along the fracture of serpentinized basalt; Small amount of $SiO_2(a)$.<br><br>**Pyroxene was the most reactive (even than olivine)**<br><br>Diffusive mass transport conditions and reactive mineral distribution are the key factors.<br><br>**Problems:** Used mass balance calculation to estimate net changes in minerals and fracture volume (not comprehensive and accurate).<br><br>Thermodynamic database for calculation SUPCRT92 with DPRONS92; obsolete and needs improvement. |
| Brady and Gíslason (1997) | Flow-through; 25-50 °C; Seawater $pCO_2$: $10^{-3.5}$ and $10^{-2.3}$. Duration: 2-3 d. | Unaltered pillow basalt from the Southern Rift Valley of the Eastern Galapagos Rift.<br><br>Min. comp.: Plagioclase and Clinopyroxene.<br><br>Grain size: 75–125 μm<br><br>BET: 0.199 m$^2$/g. | To measure reaction rates and derive activation energy. | The activation energy for initial basalt weathering in seawater is 41-65 kJ/mol. Basalt log dissolution rate: At Log $pCO_2$ -3.5, 25 °C: -14.35 mols/cm$^2$/s 37 °C: -14.05 50 °C: -13.79 Increasing the $pCO_2$ in equilibrium with seawater increased basalt dissolution rates slightly At Log $pCO_2$ -2.3 25 °C: -14.07 mols/cm$^2$/s |
| Clark et al. (2019) | Flow-through; 50 °C and 80 bar; DI water, pH 4, ~20 mmol/kg DIC ($pCO_2 = 0.6$ bar). Duration: 108 d. | Wolff-Boenisch, SW Iceland. $Si_{1.000}Ti_{0.024}Al_{0.358}Fe_{0.188}Mg_{0.28}1Ca_{0.264}Na_{0.079}K_{0.008}O_{3.370}$ (Oelkers and Gislason, 2001);<br><br>Grain size: 45–100 μm; BET (N$_2$) 0.124 cm$^2$/g, Geometric SA 0.0286 m$^2$/g. | To verify the carbon mineralization sequence by Snæbjörnsdóttir et al., (2018). pH ~5 siderite first, pH increases, mixtures of Fe-, Mg-, and Ca-carbonates; finally calcite | Initial pore water pH 9.8. Within 18 h (1.5 Pore Volume), pH dropped to 6-6.5 and DIC increased to 20 m$M$ with the injection of $CO_2$-charged water. After 12 h, Ca-Mg-Fe-carbonates supersaturated.<br><br>From 4-52 d, pH ~5.6, DIC fluctuated. Concentrations of major elements stabilized in 6 days.<br><br>After 56 d, injecting fresh water (no $CO_2$), pH peak at 8.7 after 61 d (5 PV freshwater flooding). Carbonate mineral is undersaturated. At 61 d, injection of $CO_2$-charged water; pH restabilized to ~5.7 at 64 d (6 PV of $CO_2$-charged water).<br><br>Carbonate minerals (siderite) are rare if at all, not detected by EDXS and XPS.<br><br>1D RTM using PHREEQC with *carbfix.dat* database.<br><br>The sequence of carbonate minerals with increasing pH is worth |
|  |  |  |  | noting. |
|  |  |  |  | Reducing the geometric SA to 1/10 has a good agreement with major ion concentrations in modeling. |
|  |  |  |  | A "**sweet spot**" **pH 5.2–6.5** favors the saturation of carbonates rather than zeolites and clays (Snæbjörnsdóttir et al., 2018). |
|  |  |  |  | **Smectite is one of the most abundant secondary minerals in basaltic rocks at<100 °C** and had been identified in all wells drilled at the CarbFix site in Hellisheiði, SW-Iceland (e.g. Schiffman and Fridleifsson, 1991). |
|  |  |  |  | **Problems:** Did not use XRD analysis for post-experiment products. Secondary mineral indication only by SI calculation. |
|  |  |  |  | Modeling is not comprehensive. No analysis of rates and fate of $CO_2$. Assuming secondary mineral at equilibrium. |
|  |  |  |  | Generally, **the authors failed the objective of verifying the sequence of carbonate minerals. Still a knowledge gap.** |
| Delerce et al. (2023) | Flow through, pH 3, 120 °C. | Eight altered basalts from Iceland, one sample from the Deccan traps in India, and two non-basaltic rock samples from a geothermal field in Western Turkey. Grain size: 40–200 µm; BET specific surface area 0.29-4.98 $m^2/g$, Geometric SA 0.029 $m^2/g$. | To assess the relative reactivity of altered basalts to determine the efficiency of carbonating these rocks. | Eight altered basalts from Iceland, one sample from the Deccan traps in India, and two non-basaltic rock samples from a geothermal field in Western Turkey. 40-200 micron grains.<br><br>The steady-state element release rates of altered basalts are one to three orders of magnitude slower than corresponding release rates of basaltic glass and fresh crystalline basalt. Efforts to carbonate subsurface altered basalts may be best targeted at systems having temperatures in excess of 100 °C to compensate for the lower reactivity of these rocks.<br><br>XRD results indicate the total dissolution of calcite, the significant dissolution of zeolites, epidote, and chlorite, and partial dissolution of plagioclase, pyroxenes, magnetite, and titanite from the basaltic samples. SEM, XRD: Boehmite and iron oxyhydroxide as secondary minerals.<br>Zeolites and epidote can release Ca for mineral carbonation, comparable to plagioclase. |
| Galeczka et al. (2014) | Flow-through; 22 and 50 °C, $10^{-5.7}$ to 22 bar $pCO_2$; DI water; Duration: up to 104 d. | Stapafell basaltic glass, SW Iceland. $Si_{1.0}Ti_{0.024}Al_{0.355}Fe_{0.207}Mg_{0.276}Ca_{0.265}Na_{0.073}K_{0.007}O_{3.381}$; Grain size: 45–100 µm; BET specific surface area ($N_2$) 2.2 $m^2/g$, Geometric SA 0.029 $m^2/g$. | To improve the understanding of $CO_2$-charged fluid–basalt interaction. | **Pure water**: pH from 6.7 to 9-9.5. Fe consumed by secondary ppt. Al concentration remains high, contrast to natural system.<br>**$CO_2$ charged water**: pH increases from 3.4-4.5.<br>**No carbonates ppt observed**, but once this fluid exited the reactor, carbonates precipitated. 'flash scaling' (Wolff-Boenisch et al., 2016).<br>Basaltic glass dissolution in the $CO_2$-charged water was closer to stoichiometry than in pure water.<br><br>Mn, Cr, Al, and As exceeded the allowable drinking water limits; Aqueous $Fe^{2+}/Fe^{3+}$ ratio increased along the flow path.<br><br>1D RTM using PHREEQC with *phreeqc.dat* database; updated with elected aqueous speciation and mineral solubility constants from Gysi and Stefansson (2011).<br><br>**Problems:**<br>Modeling is not comprehensive. No analysis of rates and fate of $CO_2$. No mineral SI calculation. Assuming secondary mineral at equilibrium. |
| Gysi and Stefánsson (2012a) | Batch experiment 75, 150, and 250 °C, ~80 to 270 mmol/kg $CO_2$ = $pCO_2$ ~10 to 25 bar. | Natural basaltic glass from Stapafell ($K_{0.008}Na_{0.08}Ca_{0.27}Mg_{0.26}Mn_{0.003}S_{0.002}Ti_{0.02}Fe^{2+}_{0.169}Fe^{3+}_{0.012}Al_{0.35}Si_{1.00}O_{3.327}$) Grain size: 45–125 µm. | To improve the understanding of the integrated effects of temperature, acid supply and | Mineral carbonation was most favorable at 75 °C. At 75 °C, the pH was buffered at ~4.5 with Ca, Mg, Fe and Si being incorporated into Ca–Mg–Fe carbonates and chalcedony. At 150 and 250 °C, the pH increased from ~5.5 to >6 with Ca being incorporated into calcite and Mg, Fe, Al and Si into smectite and/or chlorite depending on temperature.<br>Secondary minerals: 75 °C Ca–Mg–Fe carbonate solid solutions and $SiO_2(a)$.<br>150 °C, smectites, calcite, $SiO_2(a)$, and zeolites; |
| | ~80 g basaltic glass and ~400 ml degassed Vellankatla natural spring water<br><br>Stirred at 100 rpm<br><br>Duration: 100 d | | reaction progress on the fluid composition, secondary mineralogy, and mass of $CO_2$ mineralization | 250 °C, chlorite, and calcite.<br><br>Fe(II) predominate at 75 °C and Fe(III) at 150 °C and 250 °C. Overall redox equilibrium is not attained at <100 °C and even at much higher temperatures<br><br>PHREEQC, llnl.dat with updated selected aqueous speciation and all mineral solubility constants relevant to $CO_2$-water-basalt system (carbonates, zeolites, phyllosilicates, Al-Si and Si-minerals), see Gysi and Stefansson (2011).<br><br>The favorability of mineral carbonation at temperatures below 100 °C was associated with the tendency for Mg and Fe to be incorporated into clays at temperatures greater than 100 °C, limiting the availability of the metal ions for carbonate mineralization.<br><br>Compared equilibrium vs. kinetics for secondary minerals. Partial equilibrium is fair for T > 150 °C, but for T = 75 °C, smectite ppt is retarded in the kinetic model. |
| Gysi and Stefánsson (2012b) | Batch experiment at 40 °C, $CO_2$ concentrations ranging from 24 to 305 mmol/kg.<br><br>Vellankatla natural spring water<br><br>Duration: up to 260 d | Natural basaltic glass from Stapafell ($K_{0.008}Na_{0.08}Ca_{0.27}Mg_{0.26}Mn_{0.003}S_{0.002}Ti_{0.02}Fe^{2+}_{0.169}Fe^{3+}_{0.012}Al_{0.35}Si_{1.00}O_{3.327}$); Olivine (2–5%) and Plagioclase (1–2%). Accessory phases (<1%) were composed of Clinopyroxene and Titanomagnetite.<br><br>Grain size: 45–125 μm size;<br><br>$SA_{BET}$: 2.3 m$^2$/g<br>$SA_{GEO}$: 0.025 m$^2$/g | To gain insight into the alteration mineralogy and water chemistry associated with low-temperature $CO_2$-water-basalt interaction as a function of reaction progress. | 40 °C, secondary minerals: poorly crystalline Ca–Mg–Fe carbonates, Fe-hydroxides and/or oxyhydroxides, and Ca–Mg–Fe clays.<br><br>The dissolution of basaltic glass in $CO_2$-rich waters was observed to be incongruent. Initially, the pH increased rapidly from <4.5 to ~4.5–6, Mg and Ca were observed to be mobile and dissolved stoichiometrically relative to Na, whereas Si, Al, and Fe were immobile. The pH increased to >6.5 ($CO_2$ mineralization) and the mobility of most elements consequently decreased including Mg and Ca.<br><br>PHREEQC, phreeqc.dat updated with updated selected aqueous speciation and all mineral solubility constants relevant to $CO_2$-water-basalt system (carbonates, zeolites, phyllosilicates, Al-Si and Si-minerals), see Gysi and Stefansson (2011). |
| Gysi and Stefánsson (2012c) | Batch experiment<br><br>75, 150, and 250 °C, ~80 to 270 mmol/kg $CO_2$ = $pCO_2$ ~10 to 25 bar.<br><br>~80 g basaltic glass and ~400 ml degassed Vellankatla natural spring water<br><br>Stirred at 100 rpm<br><br>Duration: 124 d | Natural basaltic glass from Stapafell ($K_{0.008}Na_{0.08}Ca_{0.27}Mg_{0.26}Mn_{0.003}S_{0.002}Ti_{0.02}Fe^{2+}_{0.169}Fe^{3+}_{0.012}Al_{0.35}Si_{1.00}O_{3.327}$) and natural spring water from Vellankatla, SW Iceland;<br><br>Grain size: 45–125 μm.<br><br>$A_{BET}$: 2.3 m$^2$/g<br>$A_{GEO}$: 0.025 m$^2$/g | Secondary mineral evolution and mineralization mechanisms upon basalt alteration from low (< 150 °C) to high temperatures (≥150 °C). | Secondary minerals:<br>T < 100 °C, Ankerite, Fe-Dolomite, $SiO_2$(a).<br>T ≥150 °C, mixed Ca–Mg–Fe smectites and chlorite (~80%), Calcite (~20%), $SiO_2$(a), and zeolites.<br><br>T ≥150 °C, Competing for reactions between carbonates and clays for major divalent cations (Ca, Mg and Fe), limiting carbonate potential.<br><br>Dolomite and magnesite were not observed→kinetic inhibition.<br><br>Thermodynamics and kinetics of Fe-dolomite ss.<br><br>Leached layer issue. |
| Hellevang et al. (2017) | Batch experiment<br><br>80, 100, 150 °C, pH 5.7-9.86.<br><br>Duration: up to 52 d. | Basaltic glass from Stapafell, Reykjanes Peninsula, Iceland.<br><br>Grain size: 63–125 μm size. | To better understand the reactions competing for the divalent metal cations and carbonatization | No mineral carbonate formation at pH < 8 at 80 °C and pH <7 at 100 and 150 °C.<br>At higher pHs various forms of Ca-carbonates formed together with the smectite (nontronite).<br>No Mg-Fe-carbonates.<br><br>PHREEQC, phreeqc.dat updated with mineral solubility data derived by Gysi and Stefansson (2011a).<br>TST over prediction. |
| | Na$_2$CO$_3$ 10 m$M$, HCl 0-18 m$M$, MgCl$_2$ 0-100 m$M$, CaCl$_2$ 0-100 m$M$. | | on potential of basalt | |
| Jones et al. (2012) | Batch experiment<br><br>Seawater, pH 8.1,<br>5 and 21 °C.<br><br>Atmospheric $p$CO$_2$ and $p$O$_2$.<br><br>250 g sediments + 900 mL seawater.<br><br>Duration: 270 d | Riverine bedload material (from sandbank) and estuarine sediment (estuarine mouth) from western Iceland.<br><br>Hvıta River (BET 6.358 m2/g, ~200 um)<br>Microcrystalline 55.07%<br>Volcanic glass 18.94%<br>Bytownite 10.57%<br>Olivine (Fo$_{0.8}$Fa$_{0.2}$) 7.93%,<br>Fe–Ti oxides 4.41%,<br>Anorthoclase (Na$_{0.7}$K$_{0.2}$Ca$_{0.1}$Al$_{1.1}$Si$_{2.9}$O$_8$) 2.64%<br>Quartz 0.44%<br><br>Borgarfjörður Estuary (BET 7.357 m$^2$/g, ~75 μm)<br>Microcrystalline 50.25%<br>Volcanic glass 18.23%<br>Bytownite 10.84%<br>Olivine 7.39%<br>Fe–Ti oxides 4.93%<br>Anorthoclase 2.46%<br>Quartz 1.48%<br>Calcite 4.43% | To assess the role of riverine particulate material dissolution in seawater with closed-system experiments | Riverine transported basaltic particulate material can significantly alter the composition of seawater.<br>Seawater in contact with bedload material shows considerable enrichment of Ca, Mg, Mn, and Ni, whereas Li and K concentrations decrease. Moreover, the 87Sr/86Sr of seawater decreases with time with little change in Sr concentrations.<br>3% of the Sr contained in the original riverine bedload was released during 9 months of reaction.<br><br>**We can study artificial processes (dump basaltic particles into rivers or estuarine) to accelerate this natural process.**<br><br>Model with PHREEQC, phreeqc.dat with updated selected aqueous speciation and mineral solubility constants.<br><br>Problems:<br>**No agitation or stirring.**<br>Did not characterize post-experiment secondary mineral products. |
| Kanakiya et al. (2017) | Batch<br>100 °C and 5.5 MPa, CO$_2$ 4.5 MPa (approximately 0.45 $M$ CO$_2$).<br><br>DI water equilibrated with crushed rock samples<br><br>Duration: 140 d | Auckland Volcanic Field young basalts. Fine-grained alkali basalt (E1F and P4F) and basanite (P2F, P3, and P5). 2.5 cm cores. | How whole core basalts with similar geochemistry but different porosity, permeability, pore geometry and volcanic glass content alter due to CO$_2$-water-rock reactions | Ankerite and smectite, zeolite<br><br>The basalt with the highest initial porosity and volcanic glass volume shows the most secondary mineral precipitation. At the same time, this sample exhibits the greatest increase in porosity and permeability and a decrease in rock rigidity post-reaction.<br>A correlation between volcanic glass volume and rock porosity increases due to rock-fluid reactions.<br><br>Ankerite solid solution, compositional bands. |
| Kelland et al. (2020) | Column reactor,<br>Room temperature (25 °C?),<br>$p$CO$_2$ 10$^{-3.38}$ bar.<br><br>Basalt leachate | Mixed basalt powders and soil.<br><br>Pulverized basalt from the Cascade Mountain Range, Oregon similar to the Columbia River Basalt.<br>Glass 24.9%, | To investigate ERW carbon removal and co-benefits with the C4 crop Sorghum bicolor grown at | Significantly increased the yield (21 ± 9.4%, SE) of the important C4 cereal Sorghum bicolor under controlled environmental conditions, without accumulation of potentially toxic trace elements in the seeds.<br><br>Shoot silicon concentrations also increased significantly (26 ± 5.4%, SE).<br><br>Elemental budgets indicate substantial release of base cations important for inorganic carbon removal and their accumulation |

| | | Plagioclase (labradorite) 35.1%, Alkali-feldspar 23%, Diopside (Mn) 6.2%, Diopside 4.4%, Apatite 2.8%, Montmorillonite 1.2%, Olivine (forsterite) 1%, Quartz < 9,5%, Calcite 0.34%. | bench-scale in mildly acidic clay-loam agricultural soil amended with basaltic rock dust. | mainly in the soil exchangeable pools. $CO_2$ sequestration rates of 2–4 t $CO_2$/ha, 1–5 years after a single application of basaltic rock dust, including via newly formed soil carbonate minerals whose long-term fate requires assessment through field trials. This represents an approximately four-fold increase in carbon capture compared to control plant-soil systems without basalt. |
|---|---|---|---|---|
| | Duration: 120 d | Mildly acidic soil (pH = 6.6; measurement protocol in Section 2.5) with a clay-loam texture (31.8%, 35.4% and 32.8% by mass of clay [<2 μm], silt [2–60 μm] and sand [60–2,000 μm], respectively) collected from an agricultural field at Leicestershire, United Kingdom | | |
| Kumar et al. (2017) | Batch reaction. 100 °C, 5 bar, and 10 bar $pCO_2$. 10 g basalt + 100 ml DI water. Initial pH 6.94-7.49; Final pH 6.27-6.44. Duration: 4.2 d. | Deccan basalt. Clinopyroxene (augite to diopsidic augite), Plagioclase (andesine to labradorite), Olivine (rarely found), Magnetite, and marginal amounts of interstitial glass. Grain size: 90-106 μm. $A_{geo}$ 0.025 $m^2$/g and $A_{BET}$ 2.3 $m^2$/g. | To understand the $CO_2$ pressure, temperature, and time conditions of mineral carbonation reactions | Aragonite, Calcite, Dolomite, Siderite, Ankerite, Huntite; Chlorite, Smectite (Saponite), Zeolite (Chabazite) by XRD, SEM-EDS, and Raman. Reaction path modeling using EQ3/6. TST works well to predict the dissolution rates of most minerals but overpredicts the growth of non-hydrous Mg carbonates at low temperatures. The formation of secondary carbonates predominated over that of silicates in short-term experiments; however, with increasing reaction time, the carbonates no longer persisted in the system, as they were dissolved and replaced by silicates. However, the degree of carbonation increased with the increasing $pCO_2$ and pH of the solution. |
| Luhmann et al. (2017a) | Flow through, flow rate: 0.01-0.1 mL/min; 150 °C, 150 bars, $pCO_2$ 125 bars (0.65 $M$); **1$M$ NaCl solution.** Initial pH 3.3; Duration: up to 33 d. | Basalt core plugs from the Eastern Snake River Plain, Idaho. 48.5 wt% Plagioclase, 13.8 wt% Olivine, 12.5 wt% Diopside, 12.4 wt% Hypersthene, 5.43 wt% Ilmenite, 4.71 wt% Orthoclase, 1.74 wt% Apatite, 1.13 wt% Magnetite, and 0.09 wt% Zircon. | To assess changes in porosity, permeability, and surface area caused by $CO_2$-rich fluid-rock interaction | Permeability decreases slightly during the lower flow rate experiments and increases during the higher flow rate experiments. At the higher flow rate, core permeability increases by more than one order of magnitude in one experiment and less than a factor of two in the other due to differences in preexisting flow path structure. Secondary minerals reduce large pores, but overall there is a net porosity increase. (U)SANS data suggest an overall preservation of the pore structure with no change in mineral surface roughness from the reaction. |
| Luhmann et al. (2017b) | Flow through, flow rate: 0.01-0.1 ml/min; 150 °C, 150 bars, $pCO_2$ 125 bars (0.65 M); **1$M$ NaCl solution.** | Basalt core plugs from the Eastern Snake River Plain, Idaho. 48.5 wt% plagioclase, 13.8 wt% olivine, 12.5 wt% diopside, 12.4 wt% hypersthene, 5.43 wt% ilmenite, | To understand basalt alteration due to reaction with $CO_2$-rich brine and to explore feedbacks | Permeability increased for both experiments at the higher flow rate but decreased for the lower flow rate experiments. Fluid became enriched in Si, Mg, and Fe; Relatively high mobility of the alkali metals; up to 29% and 99% of the K and Cs Secondary minerals: Smectite? Pyrite? Siderite? SI calculations and reaction path modeling using GWB, database consistent with SUPCRT92 with updates on Al-bearing minerals and species, quartz, $H_4SiO_4$(aq), plagioclase, dawsonite, and analcime. |
| | Initial pH 3.3; Duration: up to 33 d. | 4.71 wt% orthoclase, 1.74 wt% apatite, 1.13 wt% magnetite, and 0.09 wt% zircon. | between chemical reactions and permeability changes | Problems: Secondary minerals post-experiment products are not well characterized. |
| Marieni et al. (2018) | Flow-through. 250 °C, 55 bars $CO_2$ 15.4 m$M$, $H_2S$ 2.15 m$M$; Initial pH 7.12, final pH 8.28 Duration: 7.7 d. | Basaltic glass (BG) (Stapafell Mountain, SW Iceland); Grain size: 45–125 μm. $A_{geo}$ 0.025 m$^2$/g and $A_{BET}$ was 2.3 m$^2$/g | To quantify $CO_2$ and $H_2S$ mineralization rates in natural systems | Secondary minerals: Calcite and Pyrite. The measured mineralization rates indicate that ∼0.2–0.5 t of CO2, and ∼0.03–0.05 t of H2S can be sequestrated annually per cubic meter of rock, depending on reservoir lithology and surface area |
| Marieni et al. (2021) | Batch reactor, 40 °C, 1.1-1.76 bar $p$CO$_2$, Initial pH 4.81. Artificial seawater, salinity 35. Duration: up to 20 d. | Mid-ocean ridge basalts from the Juan de Fuca and Mid-Atlantic Ridges. 38 vol% labradoritic Plagioclase, 23 vol% augitic Clinopyroxene, 14 vol% mesostasis, 15 vol% secondary minerals, with minor Olivine, Magnetite, and glass. The secondary mineralogy is composed of Mg-saponite, Celadonite, Iron oxyhydroxides, and ~1 vol% Calcite. The pillow basalt CD80WP132 from the Mid-Atlantic Ridge (MAR). A vesicular cryptocrystalline olivine-bearing basalt consisting of 31 vol% labradoritic Plagioclase, 20 vol% glass, 17 vol% Augitic clinopyroxene, 11 vol% mesostasis, 6 vol% Olivine, <1 vol% secondary minerals, and 15 vol% unfilled vesicles. 63–125 μm size JdF mix 1: $A_{BET}$ 5.724 $A_{GEO}$ 0.0406 JdF mix 2: $A_{BET}$ 5.724 $A_{GEO}$ 0.0489 JdF mix 2bis: $A_{GEO}$ 0.0279 JdF mix 3: $A_{BET}$ 0.876 $A_{GEO}$ 0.0781 MAR: $A_{BET}$ 2.214 | To study the low-temperature dissolution potential of crystalline submarine basalts (from Juan de Fuca Plate and Mid-Atlantic Ridges) | Significant contributions of plagioclase dissolution in all the rocks. $CO_2$-rich saline solutions react with mafic rocks at higher rates than fresh water with low $p$CO$_2$, at the same pH. At low pH and high $p$CO$_2$, basalt dissolution rates may be nearly as high as those for olivine or peridotite (Matter and Kelemen, 2009; Wolff-Boenisch et al., 2011). |
| | | $A_{GEO}$ 0.0188 m$^2$/g | | |
| Matter et al. (2007) | Single-well push-pull test; 15 °C; CO$_2$-saturated water (pH 3.5); Injection: 1400 L in 3 h. Solution (328 mg/L NaCl, 10 mg/L Kbr), $p$CO$_2$ 8 bars; 7-day incubation; Pull: 12 l/min for 20 h.<br><br>Batch experiment: 20 °C, 1 bar. | Dolerite rocks in the Palisades Sill, Newark Basin. Quartz, Plagioclase, Alkali-feldspar, Hypersthene, Diopside, Magnetite, Ilmenite. (Olivine-free).<br><br>Lab exp: 0.2 mm grains; Geometrical SA = 0.218 m$^2$/g. | To investigate the extent of *in situ* water-rock reactions in basaltic rocks after injection of CO$_2$ in a natural environment to understand dissolution rates of Ca, Mg silicate bearing rocks | Carbonic acid was neutralized within hours of injection. Calculated cation release rates decrease with increasing pH. Large differences between release rates obtained from the field and laboratory experiments may be mainly due to uncertainties in the estimation of the reactive surface area in the field experiment and in hydrological and geological factors.<br><br>The field and laboratory experiments agree that Ca is released faster than Mg. However, the difference between the Ca and Mg release rates is less than half an order of magnitude for the laboratory experiment, whereas a difference of 1.5 orders of magnitude for the field experiment.<br><br>SI calculations using PHREEQC llnl.dat database.<br><br>**Problem:** Lab experiment condition does not match the field. (20 °C in lab vs. 15 °C in the field, 4e-4 bar $p$CO$_2$ in lab vs 8 bars in the field, pH 2-4 in lab vs. 6-7 in the field, reactive surface area: lab using crushed grains). |
| McGrail et al. (2006) | Batch experiment<br><br>90 °C, 103.4 bars. Supercritical CO$_2$, almost no water Duration: 224 d | Columbia River Basalt. Crystalline components: Plagioclase and pyroxenes, minor amounts of quartz and olivine. Secondary mineral phases are carbonates, clays, and zeolites. | To better understand the chemical reactivity of basalts with CO$_2$-water. | Coated with calcite. > 365 d calcite transition to ankerite (Ca(Fe,Mg)(CO$_3$)$_2$).<br><br>An atlas of the major LIPs in the United States and India is developed. pH calculation using EQ3NR.<br><br>**Problem:** No grain size information. |
| McGrail et al. (2009) | Batch experiment<br><br>50 °C, 103.4 bars. Supercritical CO$_2$, almost no water Duration: 95 d | Columbia River Basalt and Newark Basin basalt. | Water reactivity with metal and oxide surfaces. | Calcite? Ankerite?<br><br>**Problem:** No grain size information. |
| Menefee and Ellis (2021) | Core flooding Experiments 100 °C, 200 bars, Q: 1 mL/hr. CO$_2$ saturated DI water and 0.64 $M$ NaHCO$_3$.<br><br>Duration: 12 d and 8 d. pH 4.2 | Columbia River flood basalt from Pullman, WA, was obtained from Wards Scientific and subcored into 1 in. (2.5 cm) diameter by 1.5 in. (4.3 cm) samples. | To explore CO$_2$ mineralization patterns in diffusion-limited zones of reactive minerals where most carbonation is expected to occur | While volume expansion associated with carbonation reactions has been held critical to optimizing carbonation efficiency through the continual renewal of the reactive surface area, no evidence of reaction-induced fracturing was observed. Carbonation (58%). Calcite, aragonite, Mg-calcite, magnesite, chrysotile.<br><br>Silicate mineral carbonation is significantly enhanced at elevated [NaHCO$_3$]. 1D RTM with CrunchTope. |
| Menefee et al. (2018) | Core flooding Experiments 100 °C-150 °C, 100 bars $p$CO$_2$, 200 bars total P. flow rate: 1ml/h NaHCO$_3$ 6.3 - 640 m$M$. initial pH 4.2-6.4. | Serpentinized basalt samples from Valmont Butte, Colorado were supplied by Ward's Science. Cores measuring 2.54 cm (1 in.) in diameter and 3.8 cm (1.5 in). Heavily serpentinized (14% Serpentine and | To investigate transport limitations, reservoir temperature, and brine chemistry impact carbonation reactions | Increasing [NaHCO$_3$] to 640 mM dramatically enhanced carbonation in diffusion-limited zones, but an associated increase in clays filling advection-controlled flow paths could ultimately obstruct flow and limit sequestration capacity under such conditions. Carbonates tend to precipitate in diffusion-limited zones and localize on reactive minerals supplying critical divalent cations (i.e., olivine, pyroxene). Carbonate and clay precipitation were further enhanced at 150 °C, reducing the pre-reaction fracture volume by 48% compared to 35% at 100 °C. Low NaHCO$_3$: Fe oxides, SiO$_2$(a), Smectite? Calcite, Aragonite; High NaHCO$_3$, Ankerite, Kutnohorite, Smectite? |
| | Duration: 10-12 d. | 1% unaltered Olivine) and are rich in Plagioclase (28% labradorite), Pyroxene (21% diopside), and a potassium-rich matrix (32%). Mineral grains are coarse, and accessory minerals (e.g., Apatite, Chromite). | | Higher temperatures also produced more carbonate-driven fracture bridging.<br><br>RTM using CrunchTope.<br><br>Reducing and high-pH reservoirs are necessary to realize the full mineral trapping potential of host rocks.<br><br>Problems:<br>The sample choice (Serpentinized basalt) is not very good.<br>Using kaolinite as a secondary mineral. |
| Peuble et al. (2015) | Core flooding Experiments<br><br>T = 180 °C; $pCO_2$ 100 bars, total P 120 bars. $NaHCO_3$ 0.57 M. In situ pH 6.6.<br><br>1 mL/h for experiment H1 and 0.1 mL/h for H2.<br><br>Duration: 92.39 d for H1 and 55.8 d for $H_2$. | 6.35 x 13 mm cores. Sintered olivine grains. Beach sand sampled on the southern shore of Hawaii, Olivine, Basaltic glass, Augite, Plagioclase and Chromite.<br><br>$Mg_{1.73}Fe_{0.25}Ca_{0.01}Ni_{0.01}SiO_4$; $Fo87.5±2.5$<br><br>200–315μm grain size.<br>Initial porosity: 6.15% for H1 and 6.02% for H2. Permeability: 330-1.02 (initial to final) for H1 and 41-1.01 for H2. | To investigate the interplay between hydrodynamic and reactive processes during $CO_2$-mineralization in olivine-rich basaltic layers | Higher flow rates improve carbonation efficiency, as permeability reductions blocked transport in low-flow regimes<br><br>The measured bulk carbonation rates ranged from 4 to 7 x $10^{-8}$ $s^{-1}$. The carbonation rate per unit mass of rock was enhanced at higher flow rates, although more CO2 was carbonated per unit fluid volume at lower flow rates.<br><br>Secondary minerals: Fe-dolomite and magnesite.<br><br>Reaction path modeling using PHREEQC llnl.dat database. |
| Phukan et al. (2021a) | Batch Experiment. 60 °C, 80 bars, Basalt-equlibriated water Duration 44 d. | Flood basalt from the Bambstone Bluestone quarry, Australia. 15 basalt wafers (3.8 x 1.2 x 0.08 cm). Plagioclase (23.2% andesine), Alkali feldspar (21.3% anorthoclase), Olivine (16.6% forsterite) and Pyroxene (21.2% diopside). Ilmenite and Apatite. | To understand the nature of mineral precipitation due to $CO_2$-charged water reacting with fracture surface at reservoir condition. | Montmorillonite, nontronite, stilbite and carbpnate minerals (calcite, magnesite, and siderite) detected by FTIR. SEM-EDS analysis shows precipitates with elongated nanowire-like structure, showing Mg, O peaks and flaky material with Mg, C, O peaks. Kaolinite, Fe-carbonate mineral, Na-rich smectite.<br>Solution chemistry evlovution suggests the first 6 days dominance of mineral dissolution, and the remaining 38 days dominance of secondary mineral precipitation.<br>Speciation-solubility modeling using Geochemist's Workbench with thermo.com V8. R6+ database, updated with stochiometry of primary mineral compositions and solid solutions of ankerite (ideal mixing; Ca-Mg-Fe carbonate mineral).<br>Problem: the calculated pH (from 4.15 at 0 d to 3.15 at 44 d) is questionable and so does the mineral SI calculations. |
| Phukan et al. (2021b) | Core flooding Experiments 60 °C, 80 bars,<br><br>Duration 84 d. | Flood basalt from the Bambstone Bluestone quarry, Australia. A core plug of 2.56 cm x 6 cm. Plagioclase (23.2% andesine), Alkali feldspar (21.3% anorthoclase), Olivine (16.6% forsterite) and Pyroxene (21.2% diopside). Ilmenite and Apatite. | to better understand the influence of the early stage of geochemical reactions on preferential fluid flow pathways (faults and fractures) and their adjacent pore network | Micro-CT analysis of the core. Observed 7.7% reduction in initial pore volume. Although there was an overall self-sealing of fractures and adhacenet pore network, precipitation primarily occured in the pores (15% decrease in the number of pores) and minorly in the throats (4%). In addition, a decline in the number of isolated pores was also noticed due to the primary mineral dissolution. |
| Prasad et al. | Batch experiment. | Picritic basalts as interlayered flow | To probe the CCS | $SiO_2(a)$, ankerite by Raman and IR. |

| Reference | Conditions | Rock | Aim | Findings |
|---|---|---|---|---|
| (2009) | 100 °C, pCO2 60 bar, DI water. Duration 150 d. | units within the sequence of tholeiite basalts in the Igatpuri Formation. Deccan Basalt Medium-grained with porphyritic Olivine and Clinopyroxene set in a ground mass of fine-grained Clinopyroxene ± Olivine, ± Plagioclase and Titanomagnetite. Chips of ~2–3 mm size | potential of picritic basalts from the Deccan Basalt province | Problem: Did not provide additional information regarding the composition of the carbonates. Better to use μXRD and SEM-EDXS. |
| Rani et al. (2013) | Batch reactor, 100 and 200 °C, $pCO_2$ 5-10 bars, total pressure: 10-20 bars. 10 mg of basalt in 100 ml of DI water, stirring rate of 100 rpm Duration: 50-80 h. | Deccan flood basalts $A_{geo}$ 0.025 m$^2$/g, $S_{BET}$ 2.3 m$^2$/g. $SiO_2$= 48.64, $TiO_2$= 2.83, $Al_2O_3$=14.88, CaO= 10.02, $Fe_2O_3$= 14.85, $K_2O$=14.85, MgO=5.77, $Na_2O$=2.77 and $P_2O_5$= 10.19 | To study the reaction products of Deccan flood basalt with $CO_2$ and water. | 100 °C, calcite, aragonite, siderite and magnesite, and smectite, chlorite, and smectite/chlorite mixed layer clays. 200 °C, mostly calcite. |
| Rigopoulos et al. (2018) | Batch reactors, 23.5 ± 1.5 °C, P: 1 atm. rock/fluid ratio: 2 g/L, **Artificial seawater** to exclude any potential biological processes; continuously shaken using orbital shakers, 200 rpm; Duration: 2 mo | Rocks from the Troodos ophiolite complex, Cyprus; two peridotites were collected from the Troodos mantle section: one dunite and one harzburgite, both of which are partially serpentinized. Additionally, an olivine basalt was collected from the "Upper" Pillow Lava unit of the Troodos ophiolite. Pulverized to obtain the 104–150 μm. BET: SM15 (Un-milled Dunite), 0.4 m$^2$/g; BM38 (Milled Dunite), 4 h, 35.7 m$^2$/g BM46 (Milled Dunite), 20 h, 64.6 m$^2$/g SM17 (Un-milled Harzburgite), 0.5 m$^2$/g BM72 (Milled Harzburgite), 16 h, 53.1 m$^2$/g | To assess the potential drawdown of $CO_2$ directly from the atmosphere by the enhanced weathering of peridotites and basalts in seawater | Ball-milled dunite and harzburgite changed dramatically the chemical composition of the seawater within a few hours **precipitation of aragonite (up to 10% wt), a** small amount of sepiolite [$Mg_4Si_6O_{15}(OH)_2 \cdot 6H_2O$], and minor Fe-hydroxide ppt. Pulverized but unmilled rocks, and the ball-milled basalt, did not yield any significant changes in seawater composition. Ball milling can substantially enhance the weathering rate of peridotites in marine environments. To mimic the coastal ocean, where the rock material would be in suspension due to the action of waves and currents. |

| | | SM1 (Un-milled Basalt), 2 m²/g<br>BM7 (Milled Basalt), 4 h, 58.9 m²/g | | |
|---|---|---|---|---|
| Rosenbauer et al. (2012) | Reaction cell; 50 to 200 °C at 300 bar. 1 $M$ $CO_2$ or $scCO_2$, 0.5-1 $M$ NaCl; Initial water/rock ratios were 10:1–2:1; Initial water/liquid $CO_2$ volume ratios were ~ 2:1 to 20:1 Up to 182 d. | A mid-ocean ridge basalt (MORB) from the Juan de Fuca Ridge and a tholeiite from Mt. Lassen and an olivine-rich tholeiite from Hualalai Volcano.<br><br>Grain size: 75–150 μm. | To probe the controls on the extent and rates of reaction between $CO_2$ and basaltic rock. | Secondary mineral: ferroan magnesite and $SiO_2$(a). Calcite was not observed.<br>The amount of uptake at 100 °C, 300 bars ranged from 8% by weight for a typical tholeiite to 26% for a picrate.<br>Overall reaction extent is controlled by bulk basalt Mg content.<br><br>8 to 26% of the available $CO_2$ was mineralized within ~4 months at 100 °C, while lower carbon uptakes were observed at 50 and 200 °C, respectively.<br><br>Reaction path modeling with CHILLER.<br>Used $N_2$ to remove dissolved $O_2$<br><br>\| \| JDF (%) \| Lassen \| Picri \|<br>\|---\|---\|---\|---\|<br>\| Orthoclase \| 2.1 \| 5.8 \| 1.4 \|<br>\| Albite \| 24.4 \| 23.7 \| 8.4 \|<br>\| Anorthite \| 33.9 \| 32.0 \| 11.0 \|<br>\| Diopside \| 14.8 \| 16.3 \| 9.2 \|<br>\| Hypersthene \| 6.6 \| \| 13.8 \|<br>\| Olivine \| 13.7 \| 17.9 \| 50.0 \|<br>\| Magnetite \| 2.1 \| 2.1 \| 3.2 \|<br>\| Ilmenite \| 1.9 \| 1.9 \| 1.7 \|<br>\| Apatite \| 0.4 \| 0.4 \| 0.2 \|<br>\| Total \| 100.2 \| 100.1 \| 98.9 \| |
| Roy et al. (2016) | Batch reaction, Room temperature and low-pressure $CO_2$ charged water. DI water.<br><br>Up to 90 d. | Three types of basalts from Deccan Volcanic Province. Massive basalts (C1), small Plagioclase (C2), large plagioclase (C3). Cores with a 54.7 mm diameter. | To assess the strength properties after $CO_2$-water-basalt interactions. | Precipitation of new carbonate minerals (calcite) and Fe-rich smectite in the cracks and vesicles of the host rock.<br>Loss of strength and material integrity.<br><br>**Problem:**<br>Experimental conditions are poorly described. Results descriptions are not specific. |
| Schaef and McGrail (2009) | Flow-through exp. 25–90 °C, pH 3-7<br><br>Duration: 130 d.<br><br>Solution: 0.05 $m$ KHphth + $HNO_3$ or LiOH to buffer pH. | Columbia River Basalt. Flow top and flow interior samples from the Grand Ronde Basalt Formation. Plagioclase (An40–An60) (25–48 wt%), augite ($Wo_{40}En_{40}Fs_{20}$–$Wo_{30}En_{30}Fs_{40}$) (20–32 wt%), and pigeonite ($Wo_{13}En_{50}Fs_{37}$-$Wo_9En_{64}Fs_{29}$) (0–10 wt%), with trace amounts of orthopyroxene, titaniferous magnetite grains, and ilmenite. 20–60 wt% glass.<br>2.00 to 0.42-mm diameter size.<br>$N_2$ BET: 19.8 ± 0.097 (Top) and 7.75 ± 0.037 m²/g (interior).<br>$A_{geo}$ = 0.00215 ± 0.0004 m²/g | Dissolution kinetics of the Columbia River Basalt (CRB) was measured for a range of temperatures (25–90 °C) under mildly acidic to neutral pH conditions | Activation energy, $E_a$, has been estimated at 32.0 ± 2.4 kJ/mol.<br><br>No precipitation was detected.<br><br>$k_p = 3.93 \times 10^4 \pm 3.09 \times 10^4$ g m$^{-2}$d$^{-1}$<br>$E_a = 32.0 \pm 2.3$ kJ mol$^{-1}$<br>$\eta = -0.15 \pm 0.01$<br>$R^2 = 0.95$<br><br>Few measurements of dissolution rates of whole basalt rock samples exist under mildly acidic conditions.<br>$S_{BET}$ is more than 1000 times higher than the calculated $A_{geo}$. Brantley and Mellott (2000) point out that surface area measured by gas adsorption may not be appropriate for extrapolation of interfacial controlled dissolution of many silicates if internal surfaces are present that are substantially less reactive than the external surfaces. So, use middle of $S_{BET}$ and $A_{geo}$.<br>pH calculation using EQ3NR. |
| Schaef et | Batch | Five basalts: | To evaluate | Basalt from the Newark Basin in the U.S. is by far the most reactive |
| al. (2009) | reaction, 60 °C or 100 °C, 103.4 bar $pCO_2$, $scCO_2$.<br><br>$H_2S$: ~15,000 ppmv.<br><br>Duration, no $H_2S$ up to 1351 d, with $H_2S$ up to 181 d. | Columbia River (CR) and the Central Atlantic Mafic Provence (CAMP), Newark Basin (NB) basalt, Deccan basalt (DECCAN), KAROO basalt (KAROO).<br><br>2.0–0.42 mm sized | $CO_2$-water-basalt interactions through long-term, static, high-pressure $scCO_2$ experiments. | of any basalt tested to date. Carbonate reaction products for the Newark Basin basalt were globular in form and contained significantly more Fe the secondary carbonates that precipitated on the other basalt samples. Calcite grains with classic "dogtooth spar" morphology and trace cation substitution (Mg and Mn) were observed in post-reacted samples associated with the Columbia River basalts. The Karoo basalt appeared the least reactive.<br><br>No $H_2S$, calcite and Rhodocrosite;<br>With $H_2S$, Calcite, pyrite and marcasite. |
| Schaef et al. (2011) | Batch reaction,<br><br>7.6 to 31.0 MPa, 34 °C to 136°C,<br>Duration: 30-180 d.<br><br>$CO_2$-saturated water and water-rich $scCO_2$<br><br>DI water. | Well cuttings from the borehole Columbia River Basalt and core samples from the Central Atlantic Mafic Province | Experiments cover a wide range of pressures and temperatures to illustrate the impacts of depth on basalt reactivity and carbonation | Under shallow conditions, tiny clusters of aragonite needles began forming in the wet $scCO_2$ fluid, whereas in the $CO_2$ saturated water, calcite and Fe, Mg, and Mn-substituted calcite developed thin radiating coatings. Conditions corresponding to deeper depths showed increasing carbonate precipitation.<br><br>25.5 M$Pa$, 116 °C, 30 days coated in tiny nodules of precipitate (~100 μm in diameter) ankerite, $[Ca(Fe,Mg)(CO_3)_2]$. 180 d, iron staining along with minerals structures similar to rhodochrosite and kutnohorite, chlorite, cristobalite heulandite.<br><br>**Morphology of secondary minerals different: discrete nodules in $CO_2$-saturated water, but surface coating (15-25 μm thick) in wet $scCO_2$ phase.** |
| Schaef et al. (2010) | Batch reaction,<br><br>100 °C, 103.4 bar $pCO_2$, $scCO_2$.<br>$CO_2$ 0.8 $M$.<br><br>$H_2S$: ~15,000 ppmv.<br><br>25 g of crushed basalt was added to each reactor along with 12 mL DI water<br><br>Duration: $scCO_2$ up to 1387 days and aqueous dissolved $scCO_2$–$H_2S$ for 181 d. | Five basalts: Columbia River (CR) and the Central Atlantic Mafic Provence (CAMP), Newark Basin (NB) basalt, Deccan basalt (DECCAN), KAROO basalt (KAROO).<br><br>Plagioclase is andesine-labradorite; Pyroxene has chemistry similar to Augite (Wo31En40Fs29).<br><br>2.0–0.42 mm sized | To evaluate the geochemical reactions at $CO_2$–$H_2S$–$H_2O$-basalt mixtures under supercritical $CO_2$ ($scCO_2$) conditions | CR, NB, and Deccan basalt samples yielded solution pH values between 3.9 and 4.3. CAMP and Karoo basalt–$CO_2$–$H_2O$ mixtures measured more neutral at 6.45 and 7.43.<br><br>$CO_2$-$H_2O$, 1387 d: Calcite, aragonite rhodochrosite, kutnohorite, magnesite, siderite.<br><br>$H_2S$-$CO_2$-$H_2O$, 181 d: calcite? Kutnohorite? Rhodochrosite? pyrite, marcasite.<br><br>Basalt from the Newark Basin in the United States was by far the most reactive; the Karoo basalt from South Africa appeared the least reactive.<br><br>No convincing correlations could be found relating the reactivity of the basalts to differences in their bulk composition, mineralogy, or glassy mesostasis quantity or composition.<br><br>Addition of $H_2S$ to the $scCO_2$–$H_2O$ system enhanced the carbonation of some basalts (Karoo, CRB) but inhibited carbonate formation with NB basalt due to the precipitation of pyrite that appeared to armor the basalt grains.<br><br>| Sample ID | Interstitial glass (Mesostasis) | Plagioclase feldspar | Pyroxene | Opaques (magnetite |
|---|---|---|---|---|
| | | | (wt%) | |
| CR | 45.3 | 35.3 | 18.3 | 1.0 |
| CAMP | 30.7 | 43.3 | 26.0 | <1.0 |
| NB | 27.0 | 46.0 | 26.0 | 1.0 |
| Deccan | 20.7 | 48.7 | 29.3 | 1.3 |
| Karoo | 23.3 | 45.8 | 30.8 | <1.0 |<br><br>The pyrite coatings appeared to armor the surface and halted or at least delayed carbonate formation<br><br>Reaction path modeling using EQ3/6.<br><br>Ca-carbonates generally formed first regardless of starting mineralogy. Calcite grains with classic "dogtooth spar" morphology and trace cation substitution (Mg and Mn). |

| Schaef et al. (2013) | Batch reaction, 90 °C, 100 bar, 1% H$_2$S  28 g of crushed basalt was added to each reactor along with 12.5 mL DI water.  Duration: 3.5 yr. | Five basalts: Columbia River (CR) and the Central Atlantic Mafic Province (CAMP), Newark Basin (NB) basalt, Deccan basalt (DECCAN), KAROO basalt (KAROO). 2.0–0.42 mm sized | Long-term experiment of basalt-CO$_2$-water reaction with 1% H2S. | 181 -538 d: Calcite, aragonite rhodochrosite, kutnohorite, ankerite, magnesite, pyrite, marcasite. 1252 d: pyrite, marcasite, calcite, aragonite, magnesite, rhodochrosite, kutnohorite, ankerite, + dolomite, anhydrite/gypsum, smectite/chlorite/albite.  Pyrite appeared to precede carbonation.  Reaction path modeling using EQ3/6.  Long-term: diagenetic features: albitization, anhydrite, dolomite formation. |
|---|---|---|---|---|
| Schaef et al. (2014) | Batch reaction, 90 °C 100 bar. scCO$_2$ with 1 wt % sulfur dioxide (SO$_2$) and 1 wt % oxygen (O$_2$) Duration: 98 d. | Five basalts: Columbia River (CR) and the Central Atlantic Mafic Province (CAMP), Newark Basin (NB) basalt, Deccan basalt (DECCAN), KAROO basalt (KAROO). 2.0–0.42 mm sized | To characterize the products on CO$_2$ containing minor concentrations of SO$_2$ and O$_2$ | Gypsum (CaSO$_4$) was a common precipitate, jarosite−alunite group mineral.  EQ3/6 reaction patch modeling. |
| Shibuya et al. (2013) | Batch reaction, 250 °C and 300 °C, 500 bars. NaCl 0.4 M NaHCO$_3$ 0.4 M.  Final pH 6.6 (250 °C) and 7.2 (350 °C). Duration: 94 d (250 °C) and 58 d (350 °C). | Synthetic basalt, produced from 12 reagents (SiO$_2$, TiO$_2$, Al$_2$O$_3$, Fe$_2$O$_3$, MnO, MgO, CaCO$_3$, Na$_2$CO$_3$, K$_2$CO$_3$, P$_2$O$_5$, NiO, and Cr$_2$O$_3$) to replicate the Archean mid-ocean ridge basalts.  < 90 µm | To simulate the reactions between basalt and CO$_2$-rich seawater at 250 °C and 350 °C, 500 bars, | The reaction proceeding at 250 °C yielded calcite, smectite, analcime (zeolite), and albite as alteration minerals, while the 350 °C reactions produced all of these minerals plus diopside.  Calcite destabilizes as temperature increases in the H$_2$O–CO$_2$–basalt system and crustal basalts can absorb almost all CO2 in the fluid as calcite, at least at temperatures and initial CO$_2$ concentrations below 250 °C and 400 mmol/kg, respectively.  **Albitization was found.** |
| Shrivastava et al. (2016) | Batch experiment, 100 °C and 200 °C, 5 and 10 bars $p$CO$_2$ Final pH 7.42.  Duration: 3.3 d | Deccan flood basalt. Chlorophaeite 4.75%, magnetite 9.55%, Clinopyroxene (augite) 44.75%, Plagioclase (labradorite) 40.74%, glass 0.25%. The geometric surface area of the basaltic glass was 250 cm$^2$/g and the measured BET the surface area was 23000 cm$^2$/g | To evaluate the effects of temperature and extent of reaction during basalt carbonate. | Secondary minerals: Calcite, aragonite, siderite, dolomite, magnesite. |
| Sissmann et al. (2014) | Batch experiments, 150 °C and $p$CO$_2$ 280 bar, DI water. | Midocean ridge basalt (MORB) sample from the Stapafell Mountain, Iceland. 44 wt % Plagioclase, 38 wt % | To unravel the differences in olivine reactivity in these | 56 wt % of the initial MgO had reacted with CO$_2$ to form Fe-bearing magnesite, (Mg$_{0.8}$Fe$_{0.2}$)CO$_3$, along with minor calcium carbonates. The overall carbonation kinetics of the basalt was enhanced by a factor of ca. 40 than single olivine.  Olivine ((Mg,Fe)$_2$SiO$_4$) and pyroxene ((Mg,Ca,Fe)$_2$Si$_2$O$_6$) dissolve |
| | Duration: 45 d | Clinopyroxene (augitic pyroxene), 12 wt % Olivine, and 6 wt % Iron oxide (titanomagnetite), no glass.<br><br>33−80 μm, BET: 0.40 m$_2$/g | two series of experiments, the nature, texture, and chemical composition of interfacial secondary phases produced during the course of the dissolution | preferentially under acidic conditions, favoring the release of Mg and Fe into solution, while plagioclase (CaAlSi$_3$O$_8$) dissolution is favored above circumneutral pH.<br><br>Smectite?<br>iron-rich (Mg-free) clay mineral 47.2% Si, 43.6% Al, 8.4% Fe, and 0.8% Ca (Allophane?)<br><br>Reaction path modeling using CHESS with EQ3/6 database. Differences in the chemical and textural properties of the second silica layer that covers reacted olivine grains in both types of sample.<br>Laboratory data obtained on olivine separates might yield a conservative estimate of the true carbonation potential of olivine-bearing basaltic rocks.<br><br>Large amounts of Fe(II) may have contributed to lowering the nucleation barrier of magnesite, whose nucleation and growth is a much slower process in Fe-free systems<br><br>**Allophane precipitation?** |
| Stockmann et al. (2011) | Mixed-flow, 25 °C and 70 °C<br><br>NaHCO$_3$-NH$_4$Cl-CaCl$_2$-Na$_2$CO$_3$-NaCl solution<br><br>pH 3.81-10.19<br>Duration: up to 140 d | Basaltic glass from the Stapafell Mountain in SW Iceland. Si$_{1.000}$Al$_{0.365}$Fe$_{0.191}$Mn$_{0.003}$Mg$_{0.294}$Ca$_{0.263}$Na$_{0.081}$K$_{0.008}$Ti$_{0.025}$P$_{0.004}$O$_{3.405}$.<br>45–125 μm.<br>BET: 0.5878 m$^2$/g<br>Ageo: 0.0251 m$^2$/g | To elucidate the effect of secondary mineral precipitation on the dissolution rate of primary phases | CaCO$_3$ grew as discrete crystals; no pervasive CaCO$_3$ layers were observed.<br>Basaltic glass dissolution rates are unaffected by the precipitation of secondary CaCO$_3$ precipitation.<br>Calcite, aragonite, zeolite? |
| Voigt et al. (2021) | Batch experiment. 130 °C, $pCO_2$ 2.5 and 16 bars, Seawater and synthetic Mg-free seawater. Initial pH 4.35-5.18, final pH 5.03 -5.87;<br>Duration: 2-228 d | Basaltic glass from the Stapafell Mountain in SW Iceland. 45–125 μm. SiTi$_{0.02}$Fe$^{2+}_{0.17}$Fe$^{3+}_{0.02}$Mg$_{0.28}$Ca$_{0.26}$Na$_{0.08}$K$_{0.008}$O$_{3.45}$<br>BET: 0.124 m$^2$/g.<br>>90 wt.% glass; <10 wt.% forsterite | To explore the potential of injecting CO$_2$-charged seawater into basalt formations for accelerated mineral carbon storage. | A rapid initial increase in alkalinity and pH in all experiments. Subsequently, the pH is buffered by a balance between basalt dissolution and secondary mineral precipitation as well as by the presence of bicarbonate.<br>2.5 bar $pCO_2$, calcite and aragonite are the first carbonate minerals to form, later followed by only aragonite (±siderite and ankerite); 16 bar $pCO_2$, magnesite was the only carbonate mineral observed to form. 20% of the initial CO$_2$ in the reactors was mineralized within five months.<br><br>Mg-free water: aragonite, anhydrite, zeolite<br>2.5 bar $pCO_2$: aragonite, siderite, calcite, ankerite, smectite, anhydrite, hematite;<br>16 bar $pCO_2$: magnesite, hematite.<br><br>Reaction path modeling using PHREEQC with carbfix.dat database.<br><br>The magnitude of mineralization observed in the CO$_2$-charged seawater experiments performed in this study is comparable to that observed in experiments conducted with freshwater. This suggests that subsurface carbon storage using seawater, similar to existing injections in freshwater systems. |
| Wolff-Boenisch et al. (2011) | Mixed flow<br>25 °C, pH 3.6, 4 bars $pCO_2$, artificial seawater with varying ionic strength, F-concentration and dissolved organic carbon. | Basaltic glass and crystalline basalt from the Stapafell Mountain in SW Iceland. The peridotite is the 'green' variety from the Gusdal locality in the Almklovdalen peridotite complex within the Western Gneiss Region, southern Norway. | To determine far-from-equilibrium dissolution rates of basaltic glass, crystalline basalt and dunitic peridotite | Glassy and crystalline basalts exhibit similar $rSi$ in solutions of varying ionic strength and cation concentrations.<br>Rates of all solids are found to increase by 0.3–0.5 log units in the presence of a $pCO_2$ of 4 bars compared to $P_{atm}$.<br>Glass r increased due to increased F- and crystalline enhanced due to the addition of organic ligands.<br>Peridotite no ligand-promoting effect.<br>Basalt no more than 0.6 log units lower than peridotite.<br>A first approximation, stoichiometric dissolution can be assumed for the glass and peridotite rates reported in this study.<br>Divalent metal cations tend to be released approximately 2–3 times faster than Si from the crystalline basalt.<br>CO$_2$-charged seawater injected into basalt might be nearly as |
| | Stirred at 300 rpm<br><br>Duration: 430 h | G, 100% glass, $Si_{1.000}Al_{0.365}Mg_{0.294}Ca_{0.263}Na_{0.081}K_{0.008}Ti_{0.025}P_{0.004}Mn_{0.003}Fe_{0.191}O_{3.403}$, A(BET) 0.5878 $m^2/g$, Ageo 0.0251 $m^2/g$<br><br>SS, 100% glass, $Si_{1.000}Al_{0.414}Mg_{0.189}Ca_{0.228}Na_{0.150}K_{0.018}Ti_{0.039}P_{0.006}Mn_{0.003}Fe(II)_{0.174}Fe(III)_{0.042}O_{3.457}$, A(BET): 0.1945 $m^2/g$, Ageo 0.0255 $m^2/g$<br><br>HEI, 100% glass, $Si_{1.000}Al_{0.377}Mg_{0.075}Ca_{0.147}Na_{0.224}K_{0.040}Ti_{0.033}P_{0.013}Mn_{0.004}Fe(II)_{0.129}Fe(III)_{0.056}O_{3.238}$, A(BET): 0.071 $m^2/g$, Ageo 0.0272 $m^2/g$<br><br>X: 41% Labradorite (An65), 34% Augite, 16% Olivine (Fo85), 5% Iron oxides, 4% Glass, $Si_{1.000}Al_{0.329}Mg_{0.310}Ca_{0.273}Na_{0.061}K_{0.007}Ti_{0.025}P_{0.003}Mn_{0.003}Fe(II)_{0.174}Fe(III)_{0.019}O_{3.374}$, A(BET): 0.703 $m^2/g$, Ageo 0.0255 $m^2/g$<br><br>P: 90–95% Olivine (Fo92), 5% Mg–clinochlore (Mg/(Mg + Fe) = 95–97 mol%), A(BET): 0.3286 $m^2/g$, Ageo 0.0232 $m^2/g$, 45–125 μm size | | efficient as injection into peridotite.<br><br>DFOB was used in this study as a proxy for siderophores present in seawater. Marine siderophores are known to be major iron chelators (Macrellis et al., 2001; Yoshida et al., 2002) which may serve a pivotal role in the supply of dissolved iron in many parts of the oceans.<br><br>An added benefit of the addition to the fluid phase of ligands that can complex aqueous aluminum is that they could reduce the potential for the precipitation of secondary aluminosilicate phases that consume those divalent cations that could otherwise be used for carbonate precipitation.<br><br>The sulfate present in seawater may prove problematic as anhydrite precipitation. Anhydrite solubility is retrograde and this mineral may, at least at higher temperatures, scavenge considerable amounts of Ca from solution and clog pore space. |
| Wolff-Boenisch and Galeczka (2018) | Mixed flow, 90 °C, 6 bars, 0.04-0.44 $M$ $NH_4HCO_3$ and seawater; Initial pH 7-7.7<br>Duration up to 68 d. | Basaltic glass and crystalline basalt from the Stapafell Mountain in SW Iceland.<br>45–125 μm size | To investigate carbonate vs. silicate formation in water basalt-$CO_2$ systems relevant to post-$CO_2$ injection conditions, when the pH is high (neutral to slightly alkaline) and the divalent cation load low. | 0.04 $M$ $NH_4HCO_3$: Smectite, Zeolite, Fe oxides(a);<br>0.4 $M$ $NH_4HCO_3$: $NH_4$-Dawsonite and Smectite, Zeolite, Fe oxides(a);<br>0.315 $M$ $NH_4HCO_3$ and seawater: $NH_4HCO_3$, Chlorite.<br><br>If the pore water is deficient in divalent cations, smectites and/or zeolites will dominate the secondary mineralogy of the pore space. $A_{BET}$ presents maximum values. Even if based on Ageo, the specific surface area may well overestimate the available reactive surface area as pore connectedness and preferential flow have not been considered (Aradóttir et al., 2012; Sonnenthal et al., 2005; Steefel et al., 2015).<br><br>Wolff-Boenisch et al. (2016) discussed the potential of carbonate precipitation ('flash scaling') that is impossible to control after degassing a pressurized batch experiment and that may have caused some of the carbonate precipitation described in the literature.<br><br>This is the first time that dawsonite formation is reported under experimental conditions in a basaltic matrix, which is an artifact of the experimental conditions. |

| Reference | Experiment | Rock/Sample | Objective | Results |
|---|---|---|---|---|
| | | | | **The advantage of using seawater is that it contains considerable concentrations of divalent Ca and Mg. Furthermore, seawater sulphate also increases basaltic glass dissolution in the acid medium by a factor of around two-three (Flaathen et al., 2010).** |
| Xiong et al. (2017a) | Batch reaction. 100 °C, 100 bars $pCO_2$, 3.6 g with 1.3 g of DI water; Packed bed porosities were 0.48−0.50 pH 5.7 Duration: 28 d. | Packed beds of basalt powder. Columbia River flood basalt (Pullman, Washington), and serpentinized basalt (Valmont Butte, Colorado) rocks from Ward's Science. 53–106 μm size BET: flood basalt: 6.4 m$^2$/g; serpentinized basalt: 0.9 m$^2$/g; basalt glass: 0.2 m$^2$/g. | To determine the influence of mafic rock and mineral composition and particle size on the location and extent of carbonate mineral formation | Precipitation filled 5.4% of the exposed fracture volume at 100 °C compared to 15% at 150 °C over 4 weeks. Mg- and Ca-bearing siderite, lizardite and zeolite formed. The average inorganic carbon amount in the entire packed beds was 1.6 mg/g, 0.5 mg/g, and 0.05 mg/g for flood basalt, serpentinized basalt, and basaltic glass, respectively. **Reactions in basalt reservoirs may primarily occur in poorly mixed fractures and pores. Diffusion-limited zones can achieve extents of supersaturation with respect to potential carbonate precipitates much higher than in the bulk solution.** GWB for reaction path modeling. Siderite is expected to be the first carbonate mineral to form in this study. The solubility products (Ksp) of calcite, magnesite and siderite, are $10^{-9.22}$, $10^{-9.41}$, $10^{-11.45}$ at 100 °C (calculated from SUPCRT 92). With the same amount of dissolved $CO_2$ in the system, the concentration of $Fe^{2+}$ required to reach supersaturation for siderite is 2 orders of magnitude lower than the concentrations of $Mg^{2+}$ or $Ca^{2+}$ to reach supersaturation for magnesite or calcite. However, siderite has much slower precipitation kinetics than calcite. |
| Xiong et al. (2017b) | Batch experiment, 100 °C or 150 °C, 100 bars $pCO_2$, Initial pH 3.2, final pH 4-5, Duration: up to 280 d. | Columbia River flood basalt (Pullman, Washington), and serpentinized basalt (Valmont Butte, Colorado). 2.54 cm x 4.3 cm plugs. | To determine when, where, and what types of carbonate minerals form in fractures of basalt and to explore the influence of carbonation on transport and reactions in the fracture. | Mg and Ca-bearing siderite formed in both basalts reacted at 100 °C and Mg-Fe-Ca carbonate minerals formed in the flood basalt reacted at 150 °C. A small amount of amorphous silica also formed. Precipitates filled 5.4% and 15% (by volume) of the flood basalt fracture after 40 weeks of reaction at 100 °C and 150 °C, respectively. 1D RTM with CrunchTope. |
| Xiong et al. (2018) | Batch experiment. 100 °C and 100 bars Final pH 5.28 DI water. Duration 42-280 d | Columbia River basalt cores with engineered dead-end 100 μm fractures. Plagioclase 58%, Pyroxene 14%, Ilmenite 3%, and Glass 25%. Cores 40 mm x 25.4 mm. | To quantify the rate of mineral trapping of $CO_2$ in Grand Ronde basalt, to determine the location of carbonate minerals in the pore and to identify the carbonate formation type. | Carbon mineral trapping rate of 1.24 ± 0.52 kg of $CO_2$/m$^3$ of basalt per year was estimated. Available pore space within the Grand Ronde basalt formation would completely carbonate in ∼40 years, resulting in solid mineral trapping of ∼47 kg of $CO_2$/m$^3$ of basalt. Aragonite and calcite, $SiO_2$. **Carbonate formation primarily occurs in pores and fractures where transport is controlled by diffusion.** |

# Appendix B. Occurrence of secondary minerals in basalt-water-$CO_2$ experiments

| Ref. | Temp (°C) | $pCO_2$ (bars)* | pH (final) | Solution | Basalt | Duration (days) | Carbonates | Non-carbonates | Analysis | Type |
|---|---|---|---|---|---|---|---|---|---|---|
| 1 | 100 | 83 | ~7 | Basalt-equilibrated water; 3.5 M $CO_2$ | Ferrobasaltic from the Snake River Plain | 210 | Fe-Mg carbonates (Mg-bearing siderite?) | | XRD, thin section petrography | Batch |
| 2 | 100 | 100 | 3.7 | DI water | Columbia River Flood basalt; Serpentinized basalt from Valmont Butte, Colorado | 42 | Mg-bearing siderite | $SiO_2(a)$ | Raman spectra | Flow-through and batch |
| 3 | 50 | 0.6 | 6 | DI water | Stapafell basaltic glass, SW Iceland | 108 | Siderite | | SEM-EDS, XPS | Flow-through |
| 4 | 75 | 23.9 | 4.83 | Vellankatla natural spring water | Stapafell basaltic glass, SW Iceland | 124 | Fe-Dolomite | $SiO_2(a)$ | SEM-EDS | Batch |
| | 150 | 11.2-19.6 | 5.66-6.38 | | | 26-123 | Fe-Dolomite | $SiO_2(a)$, Smectite, Zeolite | | |
| | 250 | 17.1 | 5.65 | | | 49 | Calcite | Chlorite | | |
| 5 | 75 | 23.9 | 4.83 | Vellankatla natural spring water | Stapafell basaltic glass, SW Iceland | 124 | Ankerite, Fe-Dolomite | $SiO_2(a)$ | SEM-EDS, EPMA, XRD | Batch |
| | 150 | 11.2 | 6.38 | | | 123 | Calcite | $SiO_2(a)$, Smectite, Zeolite | | |
| | 150 | 19.6 | 5.66 | | | 26 | Calcite | Smectite | | |
| | 250 | 17.1 | 5.65 | | | 49 | Calcite | $SiO_2(a)$, Smectite, Chlorite. Pyrite | | |
| 6 | 40 | 1.0-12.8 | 4.03-8.06 | Vellankatla natural spring water | Stapafell basaltic glass, SW Iceland | 103 | Fe-Dolomite | Smectite | SEM-EDS, EPMA | Batch |
| 7 | 80 | 0.55 | 9.36-9.95 | 0.01 M $Na_2CO_3$ | Stapafell basaltic glass, SW Iceland | 52 | Calcite | Smectite | SEM-EDS, XRD | Batch |
| | 100 | 0.5 | 7.89-9.68 | 0.01 M $Na_2CO_3$ | | 52 | Aragonite | Smectite | | |
| | 150 | 0.76 | 7.8 | 0.01 M $Na_2CO_3$ + 0-0.1 M $CaCl_2$ + 0-0.1 M $MgCl_2$ | | 52 | Calcite | Smectite | | |
| 8 | 100 | 42.7 | ? | Basalt-equilibrated water; 0.45 M $CO_2$ | Auckland Volcanic Field young basalts | 140 | Ankerite | Smectite, Zeolite | EPMA | Batch |
| 9 | 100 | 5 | 6.44 | DI water | Deccan basalt | 4.2 | Aragonite, Calcite, Dolomite, Siderite, Ankerite, Huntite | Chlorite, Smectite (Saponite), Zeolite (Chabazite) | XRD, SEM-EDS, Raman | Batch |
| | 100 | 10 | 6.27 | | | | | | | |
| 10, 11 | 150 | 125 | 4.6 | 1M NaCl | Whole rock basalt from Snake River Plain, Idaho | 33 | Siderite? | Smectite? Pyrite? | SEM-EDS | Flow-through |
| 12 | 250 | 55 | 8.28 | ~ 15.5 mmol DIC and ~ 2 mmol $H_2S$ | Stapafell basaltic glass, SW Iceland | 7.7 | Calcite | Pyrite | SEM-EDS | Flow-through |
| 13 | 90 | | ? | DI water | | 224 | Calcite | | | Batch |
| | | 103.4 | | | Columbia River Basalt | > 365 | Ankerite | | XRD, SEM-EDS, | |
| 14 | 50 | 103.4 | ? | | Columbia River Basalt and Newark Basin basalt. | 95 | Calcite? | | SEM-EDS | Batch |
| 15 | 100 | 100 | 6.1 | 0.64 $M$ NaHCO$_3$ | Columbia River Basalt | 14 | Calcite, Aragonite, Mg-Calcite, Magnesite | | Raman spectra | Core flooding |
| 16 | 100 | 100 | 6.1 | 6.3 m$M$ - 640 m$M$ NaHCO$_3$ | Serpentinized basalt from Valmont Butte, Colorado | 10-12 | Calcite, Aragonite, Kutnohorite, Ankerite | Smectite? Fe oxides, SiO$_2$(a) | Raman spectra, SEM-EDS | Core flooding |
| | 150 | | 6.4 | 640 m$M$ NaHCO$_3$ | | 12 | | | | |
| 17 | 180 | 100 | 6.6 | 0.57 $M$ NaHCO$_3$ | Beach sand from Hawaii | 92 | Dolomite, Magnesite, Fe-Ca carbonates | SiO$_2$(a) | SEM-EDS, TEM, Raman spectra | Core flooding |
| 18 | 100 | 60 | ? | DI water | Deccan flood basalts | 150 | Ankerite | SiO$_2$(a) | FTIR, optical microscope | Batch |
| 19 | 100 | 5 | ? | DI water | Deccan flood basalts | 3.5 | Calcite, Aragonite, Siderite, Magnesite | Smectite | SEM-EDS, XRD | Batch |
| 19 | 200 | 5 | ? | | | | Calcite, Aragonite, Siderite, Magnesite | Smectite, Chlorite | | |
| 19 | 200 | 10 | ? | | | | Calcite | Smectite, Chlorite | | |
| 20 | 23.5 | 4e-4 | 8.2 | Synthetic seawater | Olivine basalt from the Troodos mantle section | 60 | Aragonite | Fe oxides | SEM-EDS, XRD, HR-TEM | Batch |
| 21 | 100 | 300 | ? | 0.5 $M$ NaCl | A mid-ocean ridge basalt from the Juan de Fuca Ridge and a tholeiite from Mt. Lassen and an olivine-rich tholeiite from Houalalai Volcano. | 60 | Fe-Magnesite | SiO$_2$(a) | SEM-EDS, XRD | Batch |
| 22 | 22 | 2.5 | ? | DI water | Basalts from Deccan Volcanic Province | 90 | Calcite | | SEM-EDS, XRD | Batch |
| 23 | 100 | 103.4 | ? | DI water | Columbia River (CR) and the Central Atlantic Mafic Provence (CAMP), Newark Basin (NB) basalt, Deccan basalt (DECCAN), KAROO basalt (KAROO) | 854 | Calcite, Rhodochrosite | | XRD, SEM-EDS | Flow-through |
| | 60 | | ? | DI water with 1.5% H$_2$S(g | | 180 | Calcite | Pyrite, Marcasite | | |
| 24 | 100 | 103.4 | 3.9-7.43 | DI water | Columbia River (CR) and the Central Atlantic Mafic Provence (CAMP), Newark Basin (NB) basalt, Deccan basalt (DECCAN), KAROO basalt (KAROO) | 1387 | Calcite, Aragonite, Rhodochrosite | Magnetite | XRD, SEM-EDS | Batch |
| | | 103.4 | 3.9-7.43 | DI water with 1.5% $H_2S(g)$ | Columbia River (CR) and the Central Atlantic Mafic Provence (CAMP), Newark Basin (NB) basalt, Deccan basalt (DECCAN), KAROO basalt (KAROO) | 181 | Calcite, Aragonite, Rhodochrosite | Pyrite, Marcasite. | | |
| 25 | 55 | 120 | ? | DI water | Columbia River Basalt and Central Atlantic Magmatic Province | 180 | Calcite | | XRD, SEM-EDS | Batch |
| | 75 | 165 | ? | | | | Calcite | | | |
| | 96 | 210 | ? | | | | Calcite, Aragonite | | | |
| | 116 | 255 | ? | | | | Calcite, Aragonite, Kutnohorite | | | |
| | 137 | 310 | ? | | | | Rhodochrosite, Kutnohorite, Ankerite. | Chlorite, $SiO_2(a)$ | | |
| 26 | 90 | 100 | ? | DI water 1.5% $H_2S(g)$ | Columbia River (CR) and the Central Atlantic Mafic Provence (CAMP), Newark Basin (NB) basalt, Deccan basalt (DECCAN), KAROO basalt (KAROO). | 1252 | Calcite, Aragonite, Kutnohorite, Ankerite, Dolomite, Magnesite, Rhodochrosite, | Pyrite, Marcasite, Smectite, Chlorite, Albite, Anhydrite | XRD, SEM-EDS | Batch |
| 27 | 90 | 100 | ? | DI water with 1% $O_2(g)$, 1% $SO_2(g)$, | Columbia River (CR) and the Central Atlantic Mafic Provence (CAMP), Newark Basin (NB) basalt, Deccan basalt (DECCAN), KAROO basalt (KAROO). | 98 | | Gypsum, Natrojarosite, Hexahydrate, jarosite. | XRD, FTIR, SEM-EDS | Batch |
| 28 | 250 | 29.5 | 6.6 | Seawater | Synthetic basalt similar to Archean mid-ocean ridge basalt | 94 | Calcite | Smectite, Zeolite (Analcime), Albite | XRD | Batch |
| | 350 | 21.3 | 7.2 | | Synthetic basalt similar to Archean mid-ocean ridge basalt | 58 | | | | |
| 29 | 100 | 10 | ? | DI Water | Deccan Traps basalt | 3.3 | Calcite, aragonite, siderite and dolomite | | XRD, SEM-EDS | Batch |
| | 200 | 10 | ? | | | | Calcite, aragonite, siderite and dolomite | | | |
| | 100 | 5 | ? | | | | Calcite, aragonite, siderite and magnesite | | | |
| 30 | 150 | 280 | ? | DI Water | Stapafell crystal basaltic, SW Iceland | 45 | Magnesite | Smectite; Allophane?? | XRD, SEM-EDS, HR-TEM | Batch |
| 31 | 25 | 1e-4 | 10 | 0.01 $M$ $NaHCO_3$ + 0.01 $M$ $Na_2CO_3$ | Stapafell basaltic glass, SW Iceland | 140 | Calcite, Aragonite | | XRD, SEM-EDS | Flow-through |
| | 25 | 1e-4 | 6.8-8.6 | 0.035 $M$ NaHCO$_3$ + 0.01 $M$ CaCl$_2$ | | 69 | Calcite, Aragonite | | | |
| | 70 | 0.03 | 8.1 | 0.035 $M$ NaHCO$_3$ | | 26 | | Zeolite | | |
| 32 | 130 | 2.75 | 5.07 | Seawater | Stapafell basaltic glass, SW Iceland | 150 | Calcite, Aragonite, Siderite, Ankerite | Smectite, Anhydrite, Hematite | XRD, SEM-EDS | Batch |
| | | 15.81 | 4.35 | Seawater | | 140 | Magnesite | Hematite | | |
| | | 2.46 | 5.21 | **Mg-free** Seawater | | 150 | Aragonite | Zeolite, Anhydrite | | |
| 33 | 90 | 4 | ~7 | 0.04 $M$ NH$_4$HCO$_3$ | Stapafell crystal basaltic, SW Iceland | 49 | | Smectite, Zeolite, Fe oxides(a) | SEM-EDS, XRD | Flow-through |
| | | 40 | ~7.7 | 0.4 $M$ NH$_4$HCO$_3$ | | 68 | NH$_4$-Dawsonite | Smectite, Zeolite, Fe oxides(a) | | |
| | | 31.5 | ~7.2 | Seawater | | 4.6 | Calcite, Magnesite | Chlorite | | |
| 34 | 100 | 100 | 5.7 | DI water | Columbia River Basalt and serpentinized basalt (Valmont Butte, Colorado) | 28 | Siderite | Lizardite, Zeolite | Raman spectra, SEM-EDS | Batch |
| 35 | 100 | 100 | 4-5 | DI water | Columbia River Basalt and serpentinized basalt (Valmont Butte, Colorado) | 280 | Siderite | SiO$_2$(a) | Raman spectra, SEM-EDS | Batch |
| | 150 | | 4.8 | | | | Fe-dolomite | SiO$_2$(a) | | |
| 36 | 100 | 100 | 5.28 | DI water | Columbia River Basalt | 280 | Calcite, Aragonite | SiO$_2$(a) | XRD, Raman | Batch |

* When not available, $p$CO$_2$ was calculated at experimental conditions using PHREEQC 3.0 with *diagenesis.dat* database. Ankerite-dolomite solid solution is named "Fe-dolomite". SiO$_2$(a) = amorphous silica. References: 1. Adam et al. (2013); 2. Adeoye et al. (2017); 3. Clark et al. (2019); 4. Gysi and Stefánsson (2012a); 5. Gysi and Stefánsson (2012c); 6. Gysi and Stefánsson (2012b); 7. Hellevang et al. (2017); 8. Kanakiya et al. (2017); 9. Kumar et al. (2017); 10. Luhmann et al. (2017b); 11. Luhmann et al. (2017a); 12. Marieni et al. (2018). 13. McGrail et al. (2006b); 14. McGrail et al. (2009); 15. Menefee and Ellis (2021); 16. Menefee et al. (2018); 17. Peuble et al. (2015); 18. Prasad et al. (2009); 19. Rani et al. (2013); 20. Rigopoulos et al. (2018); 21. Rosenbauer et al. (2012); 22. Roy et al. (2016); 23. Schaef et al. (2009)24. Schaef et al. (2010); 25. Schaef et al. (2011); 26. Schaef et al. (2013); 27. Schaef et al. (2014); 28. Shibuya et al. (2013); 29. Shrivastava et al. (2016); 30. Sissmann et al. (2014); 31. Stockmann et al. (2011); 32. Voigt et al. (2021); 33. Wolff-Boenisch and Galeczka (2018). 34. Xiong et al. (2017a); 35. Xiong et al. (2017b); 36. Xiong et al. (2018).

**Appendix C.**

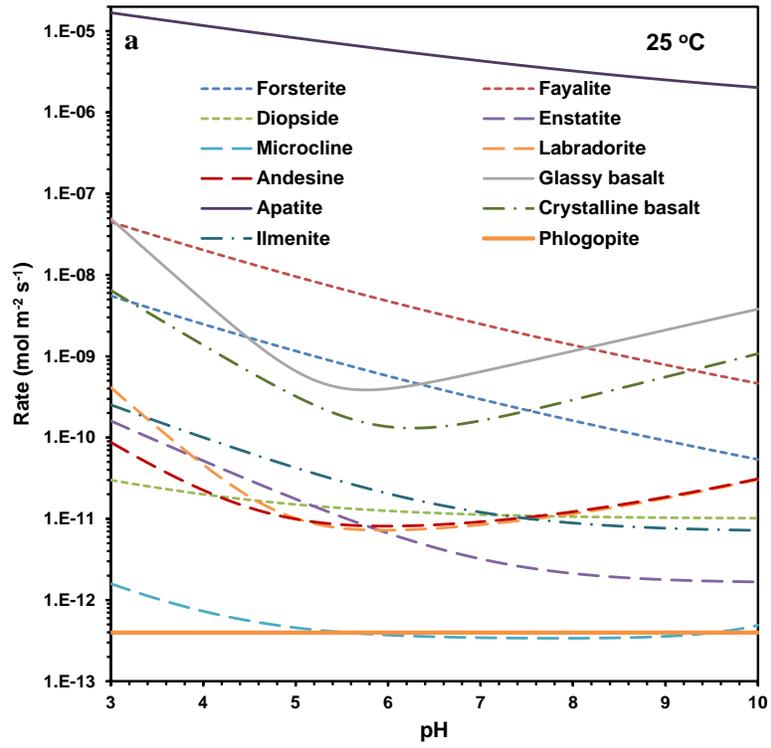

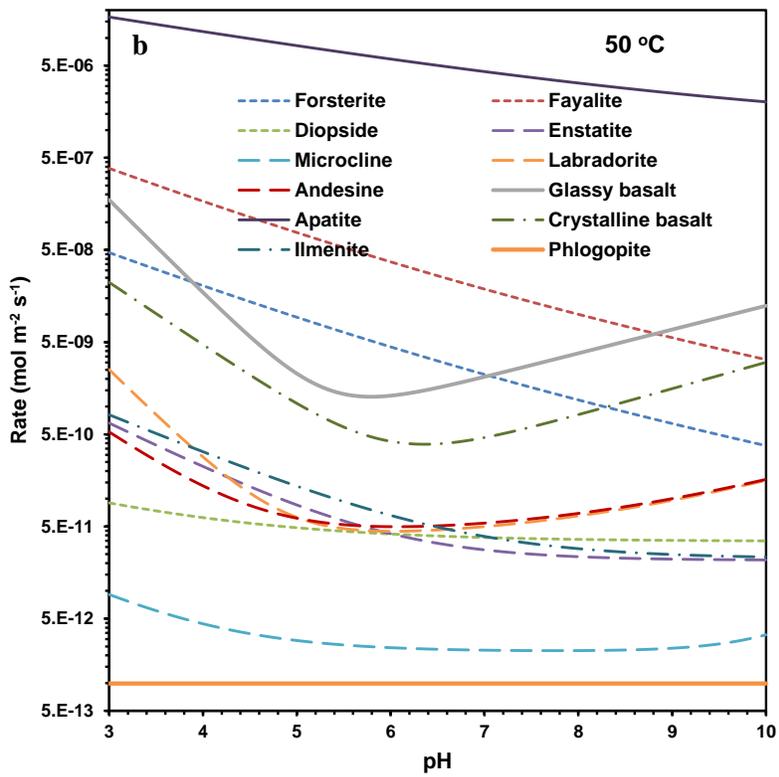

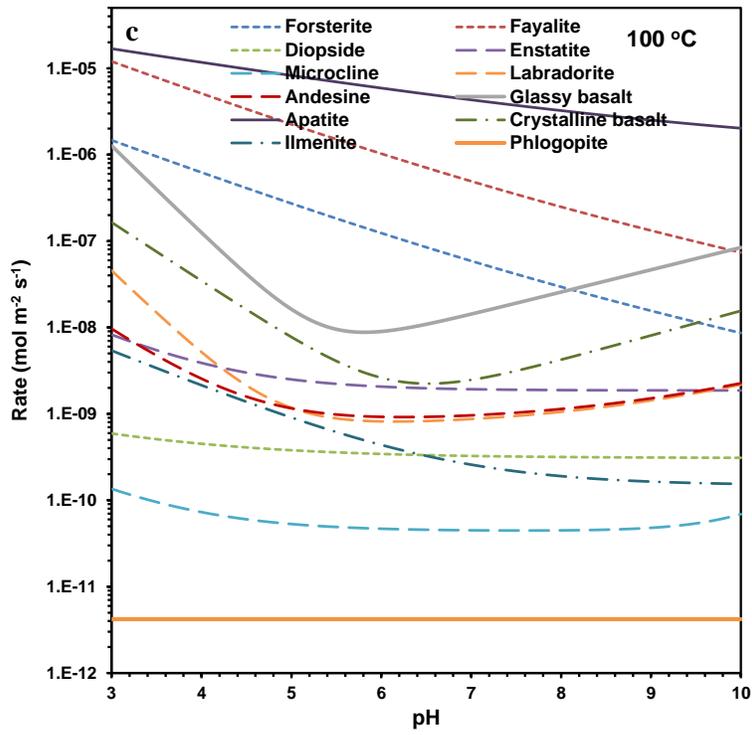

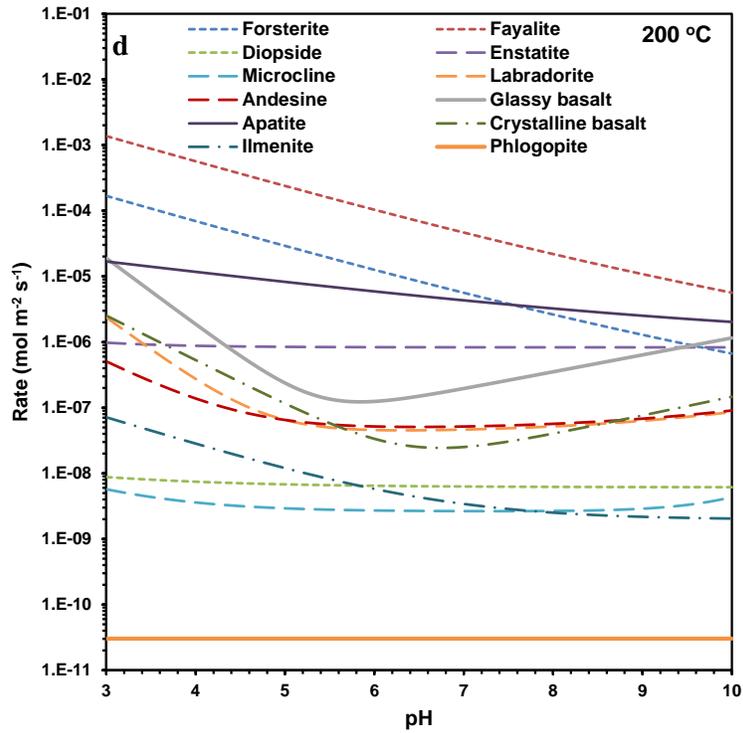

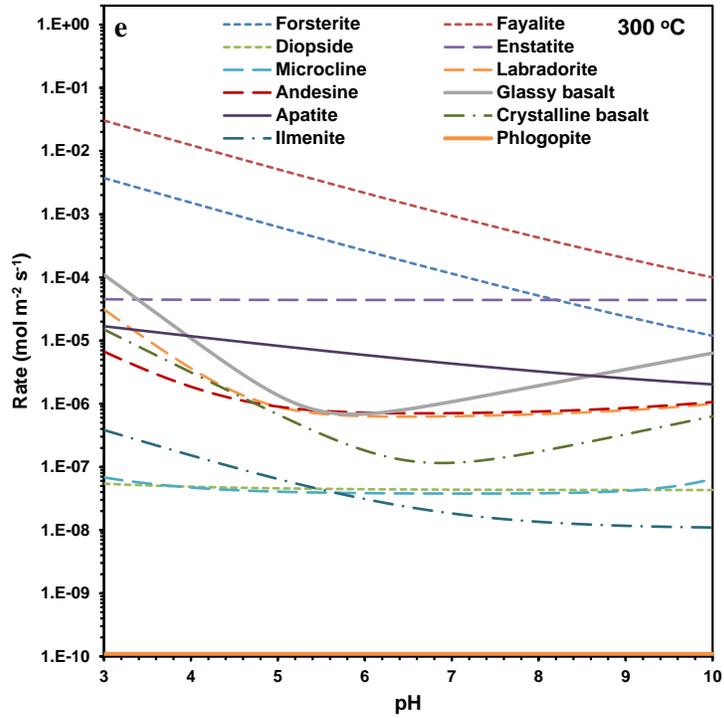

Figure C1. Range of primary mineral dissolution rates for basalts. (a) T = 25 °C; (b) T = 50 °C; (c) T = 100 °C; (d) T = 200 °C; (e) T = 300 °C.

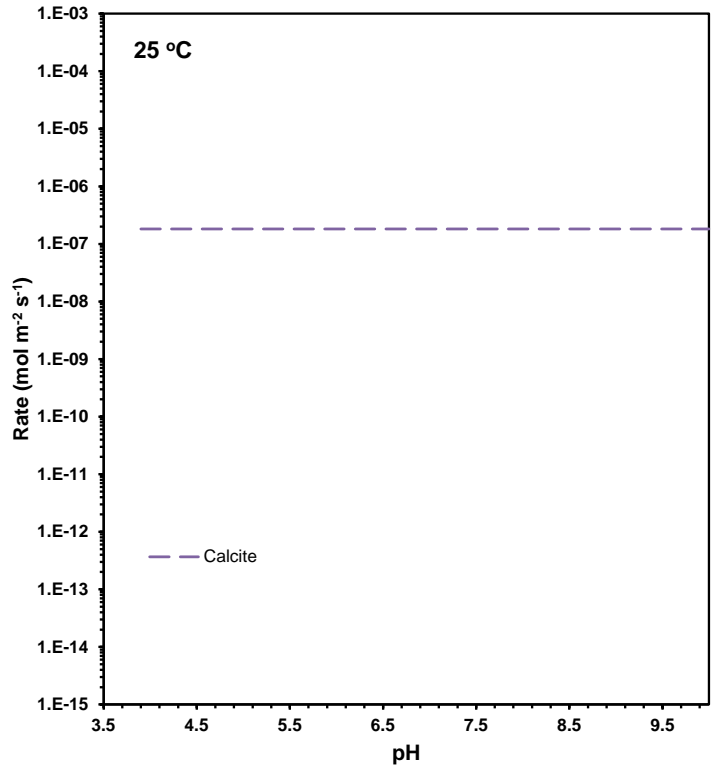
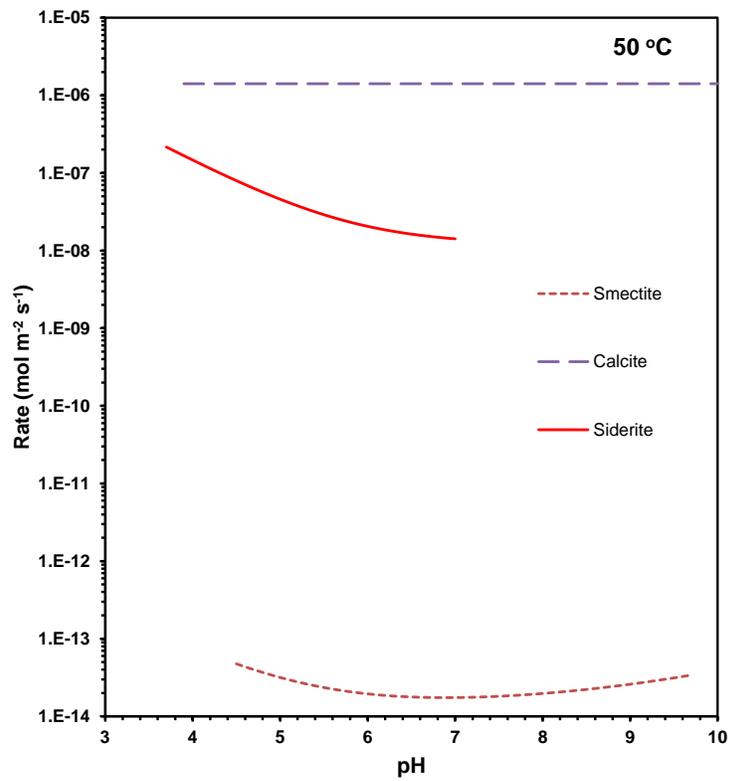

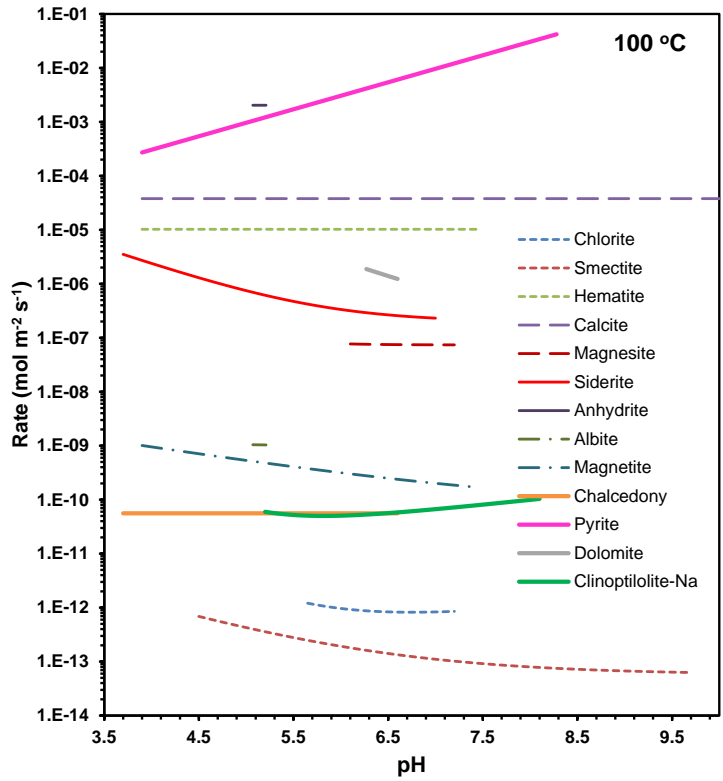
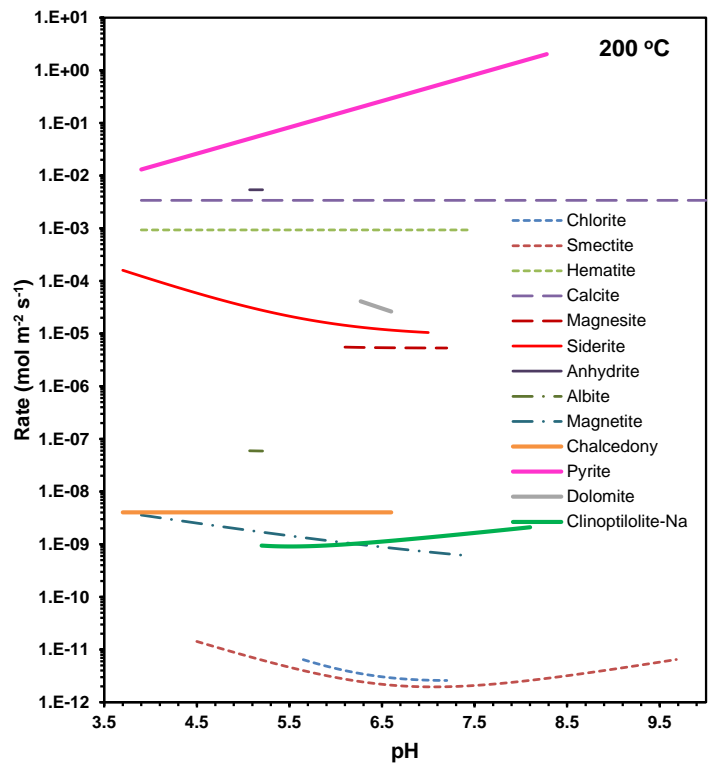

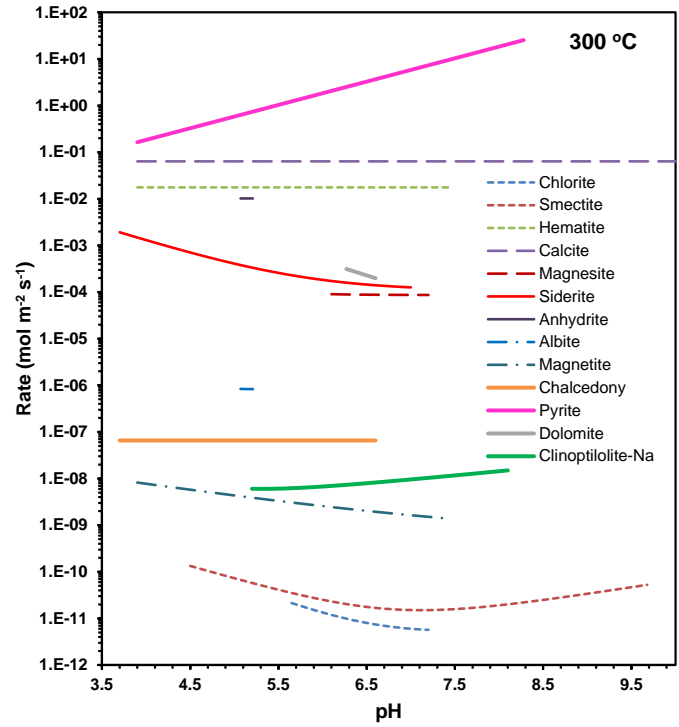

Figure C2. Range of secondary mineral reaction rates for basalts. (a) T = 25 °C; (b) T = 50 °C; (c) T = 100 °C; (d) T = 200 °C; (e) T = 300 °C.